\documentclass[twocolumn]{aastex631}

\newcommand{\KHZ}{\citetalias{hainline2024a}}
\newcommand{\RJADES}{\citetalias{Robertson2026}}
\newcommand{\JJADES}{\citetalias{Johnson2026}}

\newcommand{\Nsamp}{2081}     
\newcommand{\NsampGS}{1597}     
\newcommand{\NsampGN}{484}     
\newcommand{\Nzspec}{123}     
\newcommand{\NrejGS}{249}
\newcommand{\NrejGN}{132}

\begin{document}
\defcitealias{hainline2024a}{H24} 
\defcitealias{Robertson2026}{R26} 
\defcitealias{Johnson2026}{J26}

\title{JWST Advanced Deep Extragalactic Survey (JADES) Data Release 5: Photometrically Selected Galaxy Candidates at $z > 8$}

\author[0000-0003-4565-8239,sname='North America']{Kevin N.\ Hainline}
\affiliation{Steward Observatory, University of Arizona, 933 N. Cherry Avenue, Tucson, AZ 85721, USA}

\author[0000-0002-2929-3121]{Daniel J.\ Eisenstein}
\affiliation{Center for Astrophysics $|$ Harvard \& Smithsonian, 60 Garden St., Cambridge MA 02138 USA}

\author[0000-0003-1432-7744]{Lily Whitler}
\affiliation{Kavli Institute for Cosmology, University of Cambridge, Madingley Road, Cambridge, CB3 0HA, UK.}
\affiliation{Cavendish Laboratory - Astrophysics Group, University of Cambridge, 19 JJ Thomson Avenue, Cambridge, CB3 0HE, UK.}

\author[0000-0002-4271-0364]{Brant Robertson}
\affiliation{Department of Astronomy and Astrophysics, University of California, Santa Cruz, 1156 High Street, Santa Cruz, CA 95064, USA}

\author[0000-0002-9280-7594]{Benjamin D.\ Johnson}
\affiliation{Center for Astrophysics $|$ Harvard \& Smithsonian, 60 Garden St., Cambridge MA 02138 USA}

\author[0000-0002-6780-2441]{Peter Jakobsen}
\affiliation{Cosmic Dawn Center (DAWN), Copenhagen, Denmark}
\affiliation{Niels Bohr Institute, University of Copenhagen, Jagtvej 128, DK-2200, Copenhagen, Denmark}

\author[0000-0001-8630-2031]{D\'avid Pusk\'as}
\affiliation{Kavli Institute for Cosmology, University of Cambridge, Madingley Road, Cambridge, CB3 0HA, UK.}
\affiliation{Cavendish Laboratory - Astrophysics Group, University of Cambridge, 19 JJ Thomson Avenue, Cambridge, CB3 0HE, UK.}

\author[0000-0002-8224-4505]{Sandro Tacchella}
\affiliation{Kavli Institute for Cosmology, University of Cambridge, Madingley Road, Cambridge, CB3 0HA, UK.}
\affiliation{Cavendish Laboratory - Astrophysics Group, University of Cambridge, 19 JJ Thomson Avenue, Cambridge, CB3 0HE, UK.}

\author[0000-0003-4337-6211,sname='North America']{Jakob M.\ Helton}
\affiliation{The Department of Astronomy \& Astrophysics, The Pennsylvania State University, 525 Davey Lab, University Park, PA 16802}

\author[0000-0002-8876-5248]{Zihao Wu}
\affiliation{Center for Astrophysics $|$ Harvard \& Smithsonian, 60 Garden St., Cambridge MA 02138 USA}

\author[0000-0001-7997-1640]{Santiago Arribas}
\affiliation{Centro de Astrobiolog\'ia (CAB), CSIC–INTA, Cra. de Ajalvir Km.~4, 28850- Torrej\'on de Ardoz, Madrid, Spain}

\author[0000-0003-0215-1104]{William M.\ Baker}
\affiliation{DARK, Niels Bohr Institute, University of Copenhagen, Jagtvej 155A, DK-2200 Copenhagen, Denmark}

\author[0000-0002-8651-9879]{Andrew J.\ Bunker}
\affiliation{Department of Physics, University of Oxford, Denys Wilkinson Building, Keble Road, Oxford OX1 3RH, UK}

\author[0000-0002-0450-7306]{Alex J.\ Cameron}
\affiliation{Cosmic Dawn Center (DAWN), Copenhagen, Denmark}
\affiliation{Niels Bohr Institute, University of Copenhagen, Jagtvej 128, DK-2200, Copenhagen, Denmark}

\author[0000-0002-6719-380X]{Stefano Carniani}
\affiliation{Scuola Normale Superiore, Piazza dei Cavalieri 7, I-56126 Pisa, Italy}

\author[0000-0001-6301-3667]{Courtney Carreira}
\affiliation{Department of Astronomy and Astrophysics, University of California, Santa Cruz, 1156 High Street, Santa Cruz, CA 95064, USA}

\author[0000-0003-3458-2275]{Stephane Charlot}
\affiliation{Sorbonne Universit\'e, CNRS, UMR 7095, Institut d'Astrophysique de Paris, 98 bis bd Arago, 75014 Paris, France}

\author[0000-0002-7636-0534]{Jacopo Chevallard}
\affiliation{Department of Physics, University of Oxford, Denys Wilkinson Building, Keble Road, Oxford OX1 3RH, UK}

\author[0000-0002-9551-0534]{Emma Curtis-Lake}
\affiliation{Centre for Astrophysics Research, Department of Physics, Astronomy and Mathematics, University of Hertfordshire, Hatfield AL10 9AB, UK}

\author[0000-0003-2388-8172]{Francesco D'Eugenio}
\affiliation{Kavli Institute for Cosmology, University of Cambridge, Madingley Road, Cambridge, CB3 0HA, UK.}
\affiliation{Cavendish Laboratory - Astrophysics Group, University of Cambridge, 19 JJ Thomson Avenue, Cambridge, CB3 0HE, UK.}

\author[0009-0009-8105-4564]{Qiao Duan}
\affiliation{Kavli Institute for Cosmology, University of Cambridge, Madingley Road, Cambridge, CB3 0HA, UK.}
\affiliation{Cavendish Laboratory - Astrophysics Group, University of Cambridge, 19 JJ Thomson Avenue, Cambridge, CB3 0HE, UK.}

\author[0000-0003-1344-9475]{Eiichi Egami}
\affiliation{Steward Observatory, University of Arizona, 933 N. Cherry Avenue, Tucson, AZ 85721, USA}

\author[0000-0002-8543-761X]{Ryan Hausen}
\affiliation{Department of Physics and Astronomy, The Johns Hopkins University, 3400 N. Charles St., Baltimore, MD 21218}

\author[0000-0001-7673-2257]{Zhiyuan Ji}
\affiliation{Steward Observatory, University of Arizona, 933 N. Cherry Avenue, Tucson, AZ 85721, USA}

\author[0000-0002-3642-2446]{Tobias J.\ Looser}
\affiliation{Center for Astrophysics $|$ Harvard \& Smithsonian, 60 Garden St., Cambridge MA 02138 USA}

\author[0000-0002-4985-3819]{Roberto Maiolino}
\affiliation{Kavli Institute for Cosmology, University of Cambridge, Madingley Road, Cambridge, CB3 0HA, UK.}
\affiliation{Cavendish Laboratory - Astrophysics Group, University of Cambridge, 19 JJ Thomson Avenue, Cambridge, CB3 0HE, UK.}
\affiliation{Department of Physics and Astronomy, University College London, Gower Street, London WC1E 6BT, UK}

\author[0009-0006-4365-2246]{Petra Mengistu}
\affiliation{Department of Astronomy and Astrophysics, University of California, Santa Cruz, 1156 High Street, Santa Cruz, CA 95064, USA}

\author[0000-0003-4528-5639]{Pablo G. P\'erez-Gonz\'alez}
\affiliation{Centro de Astrobiolog\'ia (CAB), CSIC–INTA, Cra. de Ajalvir Km.~4, 28850- Torrej\'on de Ardoz, Madrid, Spain}

\author[0000-0002-7893-6170]{Marcia Rieke}
\affiliation{Steward Observatory, University of Arizona, 933 N. Cherry Avenue, Tucson, AZ 85721, USA}

\author[0000-0002-5104-8245]{Pierluigi Rinaldi}
\affiliation{Space Telescope Science Institute, 3700 San Martin Drive, Baltimore, Maryland 21218, USA}

\author[0000-0002-4622-6617]{Fengwu Sun}
\affiliation{Center for Astrophysics $|$ Harvard \& Smithsonian, 60 Garden St., Cambridge MA 02138 USA}

\author[0000-0002-9081-2111]{James A.\ A.\ Trussler}
\affiliation{Center for Astrophysics $|$ Harvard \& Smithsonian, 60 Garden St., Cambridge MA 02138 USA}

\author[0000-0003-4891-0794]{Hannah \"Ubler}
\affiliation{Max-Planck-Institut f\"ur extraterrestrische Physik (MPE), Gie{\ss}enbachstra{\ss}e 1, 85748 Garching, Germany}

\author[0000-0003-2919-7495]{Christina C. Williams}
\affiliation{NSF-DOE Vera C. Rubin Observatory/NSF NOIRLab, 950 N. Cherry Ave., Tucson, AZ 85719, USA}

\author[0000-0001-9262-9997]{Christopher N. A. Willmer}
\affiliation{Steward Observatory, University of Arizona, 933 N. Cherry Avenue, Tucson, AZ 85721, USA}

\author[0000-0002-4201-7367]{Chris Willott}
\affiliation{NRC Herzberg, 5071 West Saanich Rd, Victoria, BC V9E 2E7, Canada}

\author[0000-0002-7595-121X]{Joris Witstok}
\affiliation{Cosmic Dawn Center (DAWN), Copenhagen, Denmark}
\affiliation{Niels Bohr Institute, University of Copenhagen, Jagtvej 128, DK-2200, Copenhagen, Denmark}

\accepted{for publication in the Astrophysical Journal, April 2026}

\begin{abstract}

We present a sample of \Nsamp{} sources selected at photometric redshift $z_{\mathrm{phot}} > 8$ across the JADES DR5 data release in GOODS-S and GOODS-N over a total area of 469 square arcmin. These sources range from $M_{\mathrm{UV}} = -22$ to $M_{\mathrm{UV}} = -16$, with 19 objects at $z_{\mathrm{phot}} > 14$. We estimate the UV slopes for the full sample from fits to the photometry and find evidence for a steepening of the relationship between the UV continuum slope and $M_{\mathrm{UV}}$ to higher redshifts, a result that differs from prior analyses of brighter samples in the literature. We provide evidence that over one quarter of our sources have evidence for being morphologically extended, with many galaxies showing multiple bright knots or clumps even out to $z \sim 13 - 14$, an indication of how galaxies at Cosmic Dawn are growing and evolving. We discuss JADES-GN+189.15982+62.28899, a GOODS-N F200W dropout galaxy at $z_{\mathrm{phot}} \sim 15 - 18$ which has been observed spectroscopically with JWST/NIRSpec in prism mode, resulting in a very low signal-to-noise spectrum that is consistent with the photometry and rules out a number of low-redshift solutions for the source. Finally, we use a subsample of \Nzspec{} objects in our sample with spectroscopic redshifts to explore the usage of alternate fitting templates and a prescription for Ly-$\alpha$ damping wing absorption, finding that both produce significant improvements to the estimated photometric redshifts.

\end{abstract}


\section{Introduction} \label{sec:intro}

In the four years since the launch of the James Webb Space Telescope (JWST), our understanding of galaxy formation during the first six hundred million years after the Big Bang have come into focus. This period includes the creation of the first stars and galaxies at ``Cosmic Dawn,'' the formation of the earliest galaxy groups and clusters, and the beginning of the process of cosmic reionization. Our understanding of these early galaxies has been made possible by the unprecedented sensitivity and infrared wavelength coverage of the instruments on board JWST, specifically the imager NIRCam \citep{rieke2005, rieke2023} and spectrograph NIRSpec \citep{jakobsen2022}. Deep extragalactic surveys done with NIRCam have resulted in the identification of thousands of galaxy candidates at a redshift $z > 8$ \citep[][among many others]{castellano2022, harikane2023, finkelstein2022, adams2023, robertson2023, finkelstein2023, perezgonzalez2023, atek2023, atek2023b, whitler2023, yan2023, hainline2024a, robertson2024, finkelstein2024, weaver2024, weibel2025}. Several hundred of these distant sources have been subsequently observed with NIRSpec, confirming their redshifts and revealing their rest-UV and optical properties \citep[][and others]{bunker2023, deugenio2024, curtislake2025, scholtz2025, degraaff2025, price2025, donnan2026}. 

One of the most comprehensive programs to be undertaken thus far by the observatory is the JWST Advanced Deep Extragalactic Survey \citep[JADES; ][]{rieke2020, bunker2020, eisenstein2023, rieke2023b}, which consists of over 800 hours of imaging and spectroscopy targeting the well-studied Great Observatories Origins Deep Survey \citep[GOODS; ][]{giavalisco2004} fields in the southern (``GOODS-S'') and northern (``GOODS-N'') hemispheres. Over the last twenty years these fields have been the focus of intense study, having been observed to unprecedented depth using telescopes spanning the electromagnetic spectrum. JADES, a Guaranteed Time Observation project from the JWST NIRCam and NIRSpec science teams, has successfully been leveraged to observe a significant number of the earliest galaxies seen to date at $z > 10$, starting with photometric identification of the sources \citep{robertson2023, hainline2024a, robertson2024} and then subsequent NIRSpec spectroscopic confirmation \citep{bunker2023a, curtislake2023, deugenio2024a, hainline2024c, witstok2025}. Most notably, the extremely UV-bright galaxy JADES-GS-z14-0 ($M_{\mathrm{UV}} = -20.81$), first identified in \citet[][hereafter \KHZ{}]{hainline2024a} and \citet{robertson2024} in the ultra-deep imaging taken as part of the JADES Origins Field \citep{eisenstein2023b}, was spectroscopically confirmed to have a redshift of $z = 14.18$ \citep{carniani2024}, a value confirmed through observations done from the ground with ALMA \citep{schouws2025, carniani2025}, and with the JWST instrument MIRI \citep[][]{helton2025, helton2026}. This galaxy, and the recently discovered MoM-z14 $z = 14.44$ \citep{naidu2025} represent a type of bright, powerful source not predicted to be observed at these redshifts before the launch of JWST.

High-redshift galaxy samples, both spectroscopic and photometric, have been used to understand the formation of the first stars and galaxies. By carefully constructing luminosity functions from these sources, multiple authors have shown that there is an observed excess of UV-bright sources as compared to many pre-JWST expectations \citep{mason2015, yung2019, behroozi2020, rosdahl2022, wilkins2023}, with a slow decline with redshift in galaxy number densities at $M_{\mathrm{UV}} < -20$ \citep[][]{adams2023, castellano2022, donnan2024, finkelstein2024, robertson2024, willott2024, whitler2025, weibel2025}, a result in line with some pre-JWST predictions \citep{mcleod2016}. This result has been observed across multiple sightlines and independent extragalactic surveys. Analysis presented in \citet{whitler2025} demonstrated further evidence for a higher number density of galaxies at lower UV luminosities from observations of JADES high-redshift dropout samples. Much has been written about these surprising luminosity function observations, suggesting that perhaps galaxies had higher star formation efficiency in the early Universe \citep{dekel2023, harikane2023, ceverino2024, liz2024, feldmann2025}, or that their star formation proceeded in a more ``bursty'' mode \citep{mason2023, mirocha2023, shen2023, sun2023, gelli2024, kravtsov2024}. Larger samples of high-redshift sources are needed to fully understand the intricacies of high-redshift galaxy formation and evolution, ideally along differing sightlines to overcome cosmic variance \citep[see][]{weibel2025b}. 

Galaxies at increasingly higher redshifts have also been observed with bluer UV slopes ($\beta$, parameterized by $f_\lambda \propto \lambda^\beta$), with an evolutionary relationship with the source rest-frame UV absolute magnitude (hereafter $M_{\mathrm{UV}}$) and $\beta$ indicating that the first galaxies suffered from limited or no dust attenuation \citep{cullen2023, atek2023, topping2024, cullen2024, saxena2024, dottorini2025}. These slopes, which can reach values of $\beta < -2.5$ \citep[even beyond the most extreme slopes $\beta_{\mathrm{int}} \sim -2.6$ observed locally, ][]{chisolm2022} have helped us understand the origin of dust in the early Universe. More specifically, the UV slope provides insight into the age of a galaxy's stellar population, its metallicity, the contribution from two-photon nebular continuum emission, or even active galactic nuclei (AGNs). The contribution to the UV flux from the nebular continuum has come into focus owing to the recent discovery of a population of nebular-continuum-emission-dominated galaxies \citep{cameron2024, trussler2025}. Notably, there is a disagreement in the literature on the evolution of the relationship between $\beta$ and $M_{\mathrm{UV}}$ as a function of redshift, likely driven by a lack of sources at the highest redshifts at faint $M_{\mathrm{UV}}$ values, but also the existence of high-redshift sources with relatively reddened UV slopes \citep{mitsuhashi2025, donnan2025}, including JADES-GS-z14-0, which has been observed spectroscopically with $\beta = -2.2$ \citep{carniani2024}. The full JADES Data Release 5 (``DR5'') photometry offers a perfect opportunity to expand the exploration of the evolution of the UV slope to intrinsically fainter sources, given the unique depth of the full survey. 

The population of galaxies in the literature at $z > 8$ with spectroscopic redshifts have additionally demonstrated a pressing issue: the estimated photometric redshifts are often overpredicted \citep{arrabalharo2023b, fujimoto2023, hainline2024a, finkelstein2024, willott2024, hainline2024c}, which has been attributed to both damping wings in the IGM \citep{umeda2024, chen2024a, chen2024b, keating2024, mason2025} and the presence of strong damped Lyman-$\alpha$ absorption (DLA) weakening the observed strength of a sharp Lyman-$\alpha$ break \citep{heintz2024, heintz2025}, although care must be taken when attributing this entirely to one cause \citep{huberty2025}. This effect is seen even for galaxies observed spectroscopically, where redshifts derived from the Lyman-$\alpha$ break alone can be biased high when compared to what is measured using emission line detections \citep{heintz2024, deugenio2024a, hainline2024c, carniani2024}. At the same time, given the very blue UV slopes observed in ultra-high-redshift galaxies, multiple authors have produced alternate template sets for fitting these sources and assisting in the derivation of accurate photometric redshifts \citep{larson2023b, steinhardt2023, luberto2025}. As accurate photometric redshifts are vital for deriving luminosity functions and exploring other evolutionary trends in the early Universe, multiple authors have explored how template choice, as well as filter availability and depth affect derived photometric redshifts \citep{clausen2025, adams2025}. In \citet{asada2025}, the authors derive a straightforward prescription for this circumgalactic medium (CGM) absorption as a function of redshift from observations of JWST galaxies as part of the CAnadian NIRISS Unbiased Cluster Survey \citep[CANUCS; ][]{willott2022}. Given the large number of existing campaigns that have derived spectroscopic redshifts for sources in GOODS-S and GOODS-N \citep[see ][hereafter \RJADES{}, and Pusk\'as et al. in prep for more details]{Robertson2026}, these fields are ideal to continue this exploration of how templates and DLAs affect photometric redshift recovery. 

In this paper, we expand on the early JADES work from \KHZ{} to explore the full DR5 JADES dataset in GOODS-S and GOODS-N. Now with almost four times the survey area surveyed in \KHZ{}, the GOODS fields represent the current best regions for exploring Cosmic Dawn given the number of filters, observational depth, survey cadence (to remove transients), and ancillary data coverage. We select and visually inspect a sample of over 2000 galaxies at $z_{\mathrm{phot}} > 8$, and find 19 sources at $z_{\mathrm{phot}} > 14$. We discuss the source selection and properties and estimate $M_{\mathrm{UV}}$ and $\beta$ based on photometric or spectroscopic redshifts. We explore the evolution of $\beta$ as a function of redshift, and observe a evolving slope of the $\beta$ vs. $M_{\mathrm{UV}}$ relationship to higher redshift. We examine the morphologies of these high-redshift sources, finding that $\sim25\%$ of galaxies at this redshift have evidence for being significantly extended, including those with multiple separate knots, a fraction that is relatively constant with increasing redshift. Finally, we compare photometric and spectroscopic redshifts for a subsample, and discuss the usage of alternate template sets, as well as the \citet{asada2025} prescription for CGM absorption. 

Throughout this paper, we assume the \citet{planck2020} cosmology with
$H_0 = 67.4$ km s$^{-1}$ Mpc$^{-1}$, $\Omega_M = 0.315$, and $\Omega_\Lambda = 0.685$. All magnitudes are provided using the AB magnitude system \citep{oke1974, oke1983}.

\section{JADES Survey Photometry} \label{sec:survey}

The JADES DR5 dataset is a combination of over 800 total hours of photometry and spectroscopy from the NIRCam and NIRSpec Guaranteed Time Observing (GTO) teams along with data from numerous other imaging programs. The survey combines JWST/NIRCam and JWST/MIRI imaging with NIRSpec spectroscopy focused on the GOODS-S (R.A. = 53.126 deg, decl. = -27.802 deg) and GOODS-N (R.A. = 189.229, decl. = +62.238 deg) extragalactic fields \citep{giavalisco2004}. A complete description of the imaging observations taken and assembled as part of JADES DR5, including the programs from which they were selected, the data reduction, mosaic production, and photometric measurements, is provided in \citep[][hereafter \JJADES{}]{Johnson2026} and \RJADES{}. We note that the spectroscopic counterpart of the JADES survey was recently presented in \citet{curtislake2025} and \citet{scholtz2025}. Here, we briefly describe the photometric catalogs from which we selected our $z > 8$ candidates. 

\begin{figure*}[ht!]
  \centering
  \includegraphics[width=0.8\linewidth]{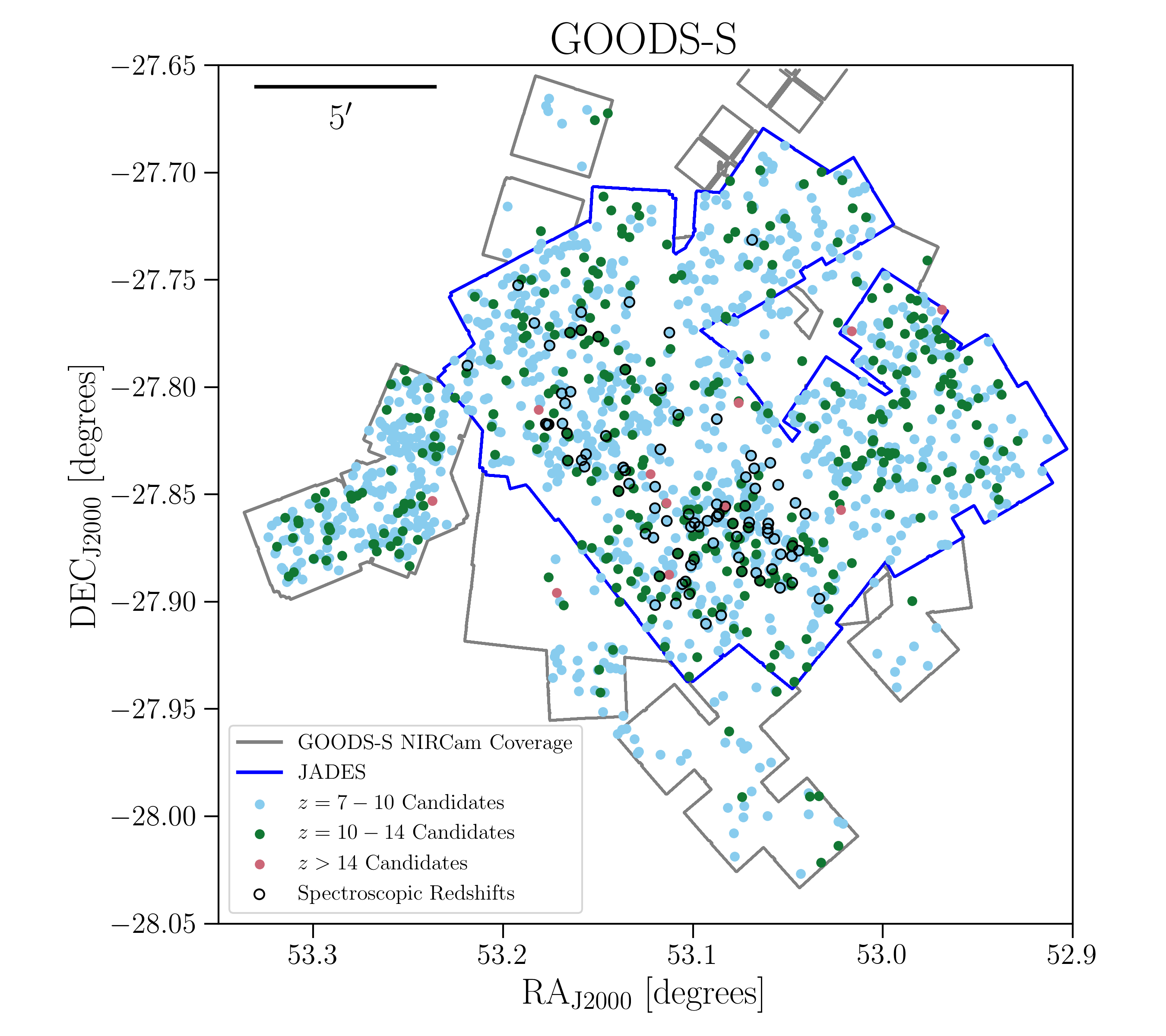}
  \caption{Footprint showing the \NsampGS{} GOODS-S galaxies and candidates at $z > 8$ in our sample. We show a 5$^{\prime}$ scale bar in each panel, and in blue, the JADES-specific survey area, while we plot ancillary JWST/NIRCam data that we search over as part of JADES DR5 with a grey outline. The points are colored in bins of redshift, as shown in the legend in each panel. We indicate those sources with spectroscopic redshifts with black outlines. The highest density of sources are found in regions with increased observational depth, as described in the text.}
  \label{fig:GS_footprint}
\end{figure*}

\begin{figure*}[t!]
  \centering
  \includegraphics[width=0.8\linewidth]{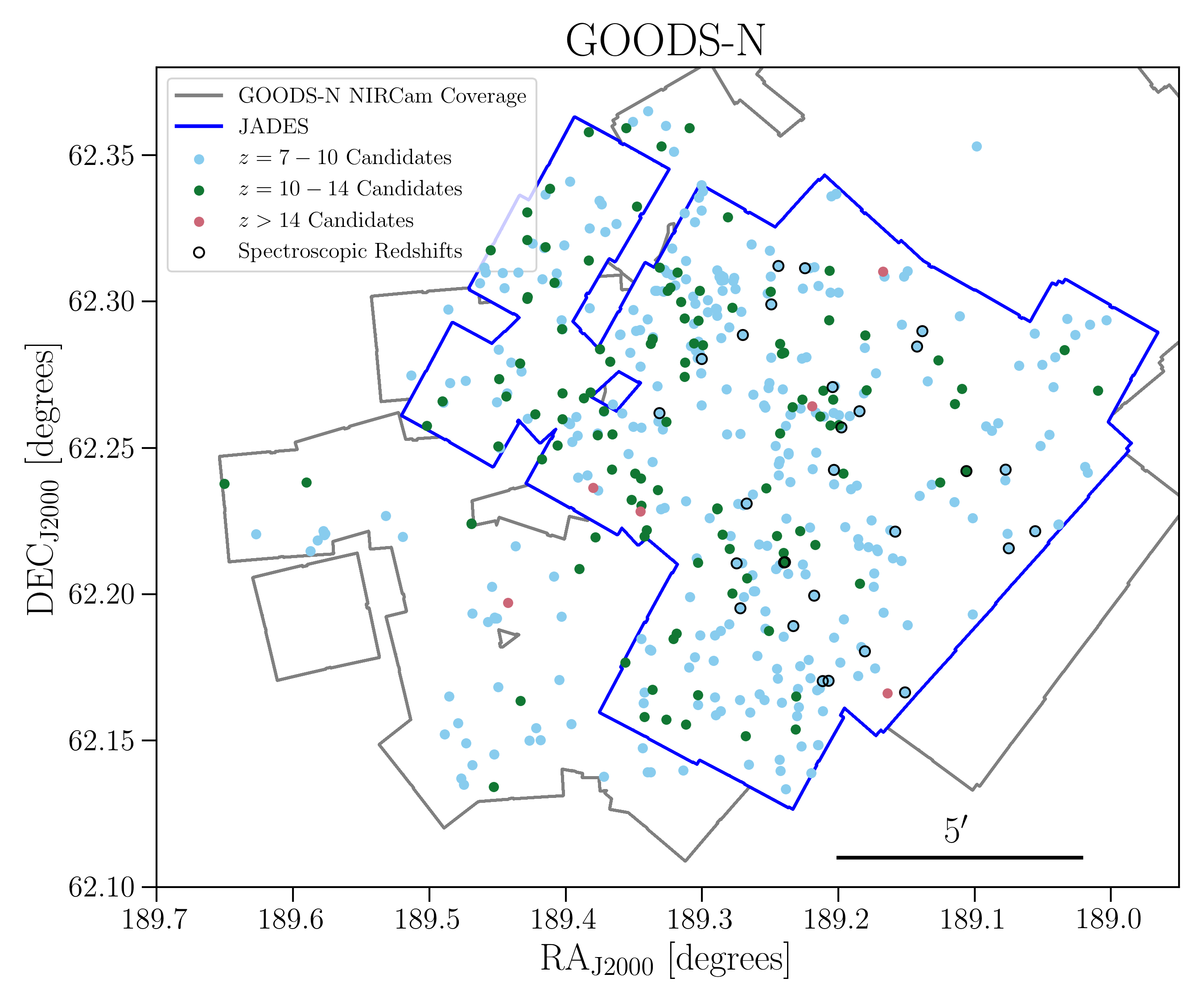}
  \caption{Footprint showing the \NsampGN{} GOODS-N galaxies and candidates at $z > 8$ in our sample, with colors and lines as in Figure \ref{fig:GS_footprint}.}
  \label{fig:GN_footprint}
\end{figure*}

\subsection{JADES Observations and Photometry} \label{sec:observations}

To assemble a sample of galaxy candidates at $z > 8$, we explored the full JADES DR5 catalogs from across GOODS-S and GOODS-N. As discussed in \JJADES{} and \RJADES{}, the final GOODS-S footprint has sub-regions observed with both JWST NIRCam wide filters (F070W, F090W, F115W, F150W, F200W, F277W, F356W, and F444W) and medium filters (F162M, F182M, F210M, F250M, F300M, F335M, F410M, F430M, F460M, and F480M), making this the most comprehensive extragalactic deep field observed thus far by JWST. In GOODS-N, the footprint we explored was observed with fewer JWST NIRCam wide filters (F090W, F115W, F150W, F200W, F277W, F356W, and F444W) and medium filters (F182M, F210M, F335M and F410M). While there are also MIRI observations taken with across the JADES footprint \citep{alberts2026}, we choose not to include these data in our fitting due to a lack of observational depth for observing the faint sources in our sample \citep[See ][for a discussion of the MIRI photometry for the small number of sources in JADES at $z > 8$]{helton2025b}. The JADES GTO data was supplemented by NIRCam imaging from multiple programs, many taken in parallel: JWST Extragalactic Medium-band Survey \citep[JEMS, 1963, PIs Williams, Tacchella, Maseda,][]{williams2023}, Parallel wide-Area Nircam Observations to Reveal And Measure the Invisible Cosmos \citep[PANORAMIC, 2514, PI Williams,][]{williams2025}, Bias-free Extragalactic Analysis for Cosmic Origins with NIRCam \citep[BEACON, PID 3990, PI Morishita,][]{morishita2025}, the Slitless Areal Pure-Parallel High-Redshift Emission Survey \citep[SAPPHIRES, PID 6434, PI Egami,][]{sun2025} and the Public Observation Pure Parallel Infrared Emission-Line Survey \citep[POPPIES, PID 5398, PI Kartaltepe,][]{kartaltepe2024}. JADES DR5 also includes image data from other GTO and GO programs, including the JADES Origins Field \citep[JOF, PID 3215, PI Eisenstein,][]{eisenstein2023b}, the MIRI Deep Imaging Survey \citep[MIDIS, PID 1283, PI {\"O}stlin,][]{ostlin2025}, the First Reionization Epoch Spectroscopically Complete Observations \citep[FRESCO, PID 1895, PI Oesch,][]{oesch2023}, the Next Generation Deep Extragalactic Exploratory Public \citep[NGDEEP, PID 3215, PIs: Finkelstein, Papovich, Pirzkal,][]{bagley2024} survey, the Complete NIRCam Grism Redshift Survey (CONGRESS, PID 3577, PI Egami, Sun et al. in prep), Observing All phases of StochastIc Star formation (OASIS, PID 5997, PI Looser, Looser et al. in prep), and Cycle 2 Director Discretionary Program 6541 (PI Egami) among others, see \JJADES{} for a full list.

We supplement these near-IR JWST/NIRCam data with observations from the Hubble Space Telescope ACS instrument. We used the HST/ACS mosaics from the Hubble Legacy Fields (HLF) v2.0 for GOODS-S and v2.5 for GOODS-N \citep{illingworth2013, illingworth2017, whitaker2019}, and utilize data from these surveys in the ACS F435W, F606W, F775W, F814W, and F850LP filters. 

The full sample of JADES sources were selected following the methodology outlined in \RJADES{}, and were selected from signal-to-noise (SNR) ``detection images'' created from a stack of the NIRCam long-wavelength mosaics using the F277W, F335M, F356W, F410M, F430M, F444W, and F460M filters. In GOODS-S, imaging with the F480M filter was included when available. Photometry was extracted for each source found in these detection images, and for the purposes of this study, we utilize forced photometry extracted using 0.2$^{\prime\prime}$ diameter circular apertures (which we will refer to as ``CIRC1'' fluxes, to match what is provided in the public DR5 JADES catalogs) for selecting and fitting the galaxies. For measuring photometric redshifts, we use aperture corrected fluxes extracted from images that are at their native resolution and have not been convolved to a single instrumental resolution, however, for estimating $\beta$ and $M_{\mathrm{UV}}$, we fit to fluxes for the sources measured from images convolved to the F444W PSF. The uncertainties that we use in this study were estimated as a combination, in quadrature, of the Poisson noise from the source and the noise measured using random apertures placed throughout the mosaics. We apply aperture corrections appropriate for point sources using empirical HST/ACS and JWST/NIRCam point-spread functions. A full table of the photometric depths across the GOODS-S and GOODS-N mosaics, in AB magnitudes for each filter, is found in \RJADES{}.  

\section{Selecting Galaxies Beyond Redshift 8} \label{sec:selection}

To select high-redshift galaxy candidates across the assembled GOODS-S and GOODS-N survey areas, we utilize a modified method and criteria from what was presented in \KHZ{}. We start by fitting the circular aperture fluxes for each source in the JADES photometric catalogs using \texttt{eazy-py}, the python implementation of the photometric redshift code \texttt{EAZY} \citep{brammer2008}. \texttt{EAZY} combines user-defined galaxy templates and compares the template-derived photometry to the observed photometry for each source using $\chi^2$ minimization to find the best fits. The primary output of \texttt{EAZY} is the $\chi^2$ as a function of redshift for all of the fits which can be used to find the redshift at the overall $\chi_{\mathrm{min}}^2$ (which we adopt as our fiducial value, referred to as $z_a$ throughout) as well as the probability of an individual source being at a given redshift using the input templates $P(z) = \exp{[-\chi^2(z) / 2]}$, normalized such that $\int P(z)dz = 1.0$. 

We ran \texttt{EAZY} on the full GOODS-S and GOODS-N photometric catalogs using a set of 16 templates provided in \KHZ{} (in Section \ref{fig:spec_z_vs_redshift_alttemplates} we explore the usage of other templates designed to fit ultra high-redshift galaxies), and we allow the code to combine all of the templates when fitting each galaxy. We explored a redshift range of $z = 0.01 - 22.0$ with a step size of $\Delta z = 0.01$. We set an error floor on the photometry of 5$\%$, and furthermore employ the \texttt{EAZY} template error file ``template error.v2.0.zfourge.'' Given the uncertainties in high-redshift luminosity functions, we choose to not adopt an apparent magnitude prior. In our fits, we also used a set of photometric offsets to account for the differences between the HST and NIRCam fluxes, the derivation of which is discussed in \RJADES{}.  

Our $z > 8$ galaxy selection criteria are straightforward: 

\begin{enumerate}
    \item We require that the redshift corresponding to the $\chi^2$ minimum is greater than 8: $z_a > 8$. 
    \item The SNR in at least two photometric bands (F150W, F200W, F277W, F335M, F356W, F410M, or F444W) must be greater than 5, from photometry derived using 0.2'' diameter apertures. 
    \item The difference in overall minimum $\chi^2$ and the $\chi^2$ restricted for fits at $z < 7$ must be greater than 4: $\Delta \chi^2 = \chi_{min,z < 7}^2 - \chi_{min}^2 > 4$. 
\end{enumerate}

These selection criteria are notably different from \KHZ{} as we do not explicitly require that the total integrated probability of a galaxy being at $z > 7$ be above 70$\%$: $\int_{7}^{22} P(z)dz > 0.7$. We find that the photometric redshift minimum and $\Delta \chi^2 > 4$ restrictions will generally lead to this being true; there are no sources in GOODS-S or GOODS-N with $\int_{7}^{22} P(z)dz < 0.7$ in our initial sample of high-redshift candidates. We do not make any restrictions based on the position of each galaxy with respect to a neighbor, which differs from the selection process outlined in \KHZ{}. Finally, unlike \KHZ{}, we do not provide the information or IDs for those sources with $\Delta \chi^2 < 4$, as there is significantly less confidence that these sources are at high redshift. 

This initial selection results in a sample of 1843 sources in GOODS-S and 607 sources in GOODS-N. We performed a round of visual inspection to reject non-astrophysical data artifacts (extended diffraction spikes from stars, hot pixels from cosmic rays), extended low-redshift, dusty star forming galaxies, sources associated with the edges of local, bright galaxies, sources associated with transients and supernovae \citep[see discussion in][]{decoursey2025}, saturated bright stars, brown dwarfs \citep{hainline2024b, hainline2025b}, and sources that show flux in a NIRCam photometric band shortward of the Lyman-$\alpha$ break at the \texttt{EAZY} photometric redshift. We removed \NrejGS{} sources in GOODS-S ($\sim 13 \%$) and \NrejGN{} sources in GOODS-N ($\sim 21 \%$) in our visual inspection. The higher rejection fraction in GOODS-N is likely a result of both the larger number of bright stars in the field with extended diffraction spikes being misidentified as high-redshift galaxies as well as from the regions in this area with fewer stacked images resulting in more data artifacts. We discuss a sample of four notable rejected sources which have unexplained spectral energy distributions (SED) in Section \ref{sec:notable_rejected_candidates}. 

\section{Results} \label{sec:results}

The final sample of $z > 8$ galaxy candidates across the JADES DR5 area we present in this study consists of \Nsamp{} sources (\NsampGS{} sources in GOODS-S and \NsampGN{} sources in GOODS-N). We provide the full sample of sources in a catalog that can be found online\footnote{DOI: 10.5281/zenodo.18306484}, and the columns of this output catalog are given in Table \ref{tab:catalog_columns}. In summary, we provide the JADES ID (both the catalog ID number and a name for the source: ``-GS'' or ``-GN'' followed by the R.A. and decl. values in decimal degrees), the internal JADES DR5 catalog ID, the source R.A., decl., circular aperture photometry, \texttt{EAZY} output parameters, and derived properties for each source, including $M_{\mathrm{UV}}$ and UV slope $\beta$. For objects found within the JADES survey area\footnote{We consider JADES affiliated programs those with PIDs 1180, 1181, 1210, 1286, 1287, 3215, 4540, 5997, 6541, and 3577.}, we also append "JADES-" to the front of the source name, while we keep this blank for sources found in portions of the DR5 footprint outside of JADES. We provide spectroscopic redshifts for the \Nzspec{} objects in our sample which have been confirmed spectroscopically, as described in Section \ref{sec:spectroscopic_redshifts}. We also indicate, for each source, the JWST dataset(s) from where the source was selected using the JWST PID. 

\begin{deluxetable*}{llcl}
\tabletypesize{\footnotesize}
\tablecolumns{8}
\tablewidth{0pt}
\tablecaption{Column Descriptions in the $z > 8$ Source Output Table \label{tab:catalog_columns}}
\tablehead{
\colhead{HDU} & \colhead{Column Name} &  \colhead{format}  &  \colhead{Description}}
\startdata
& \texttt{ID} & \texttt{int} & JADES DR5 Catalog ID \\
& \texttt{RA} & \texttt{float} & Right Ascension \\
& \texttt{DEC} & \texttt{float} & Declination \\
& \texttt{Source\_Name} & \texttt{string} & JADES DR5 Source Name \\
& \texttt{z\_spec} & \texttt{float} & Spectroscopic Redshift \\
& \texttt{EAZY\_z\_a} & \texttt{float} & \texttt{EAZY} Photometric Redshift (at $\chi^2_{\mathrm{min}}$) \\
& \texttt{EAZY\_z\_a\_chisq} & \texttt{float} & \texttt{EAZY} fit minimum $\chi^2$ \\
& \texttt{EAZY\_z\_a\_delta\_chisq} & \texttt{float} & \texttt{EAZY} $\chi^2_{z < 7} - \chi^2_{\mathrm{min}}$ \\
& \texttt{EAZY\_z160} & \texttt{float} & \texttt{EAZY} 16th Percentile Photometric Redshift \\
& \texttt{EAZY\_z500} & \texttt{float} & \texttt{EAZY} 50th Percentile Photometric Redshift \\
& \texttt{EAZY\_z840} & \texttt{float} & \texttt{EAZY} 84th Percentile Photometric Redshift \\
\texttt{PROPERTIES} & \texttt{N\_SNR\_gt\_5} & \texttt{int} & Number of primary NIRCam filters with SNR $> 5$ \\
& \texttt{MUV} & \texttt{float} & UV Magnitude \\
& \texttt{MUV\_err} & \texttt{float} & UV Magnitude 1$\sigma$ Uncertainty \\
& \texttt{beta} & \texttt{float} & UV Slope $\beta$ \\
& \texttt{beta\_err} & \texttt{float} & UV $\beta$ 1$\sigma$ Uncertainty \\
& \texttt{object\_PIDs} & \texttt{int} & Source Photometry JWST PIDs\\
& \texttt{FILTER\_flux\_nJy}$^a$ & \texttt{float} & JWST/NIRCam Photometry \\
& \texttt{FILTER\_flux\_err\_nJy}$^a$ & \texttt{float} & JWST/NIRCam 1$\sigma$ Photometric Uncertainty \\
& \texttt{Survey\_Area} & \texttt{string} & Survey Area (GOODS-S or GOODS-N) \\
\hline
& \texttt{z\_a\_TEMPLATE}$^b$ & \texttt{float} & \texttt{EAZY} Photo-$z$ ($\chi^2_{\mathrm{min}}$) \\
& \texttt{z\_a\_chisq\_TEMPLATE}$^b$ & \texttt{float} & \texttt{EAZY} fit minimum $\chi^2$ \\
& \texttt{z\_160\_TEMPLATE}$^b$ & \texttt{float} & \texttt{EAZY} 16th Percentile Photo-$z$ \\
& \texttt{z\_500\_TEMPLATE}$^b$ & \texttt{float} & \texttt{EAZY} 50th Percentile Photo-$z$ \\
\texttt{ALTERNATE\_TEMPLATES} & \texttt{z\_840\_TEMPLATE} & \texttt{float} & \texttt{EAZY} 84th Percentile Photo-$z$ \\
& \texttt{z\_a\_TEMPLATE\_asada}$^b$ & \texttt{float} & \texttt{EAZY} Photo-$z$ (at minimum $\chi^2$), with \citet{asada2025} CGM \\
& \texttt{z\_a\_chisq\_TEMPLATE\_asada}$^b$ & \texttt{float} & \texttt{EAZY} fit minimum $\chi^2$, with \citet{asada2025} CGM \\
& \texttt{z\_160\_TEMPLATE\_asada}$^b$ & \texttt{float} & \texttt{EAZY} 16th Percentile Photo-$z$, with \citet{asada2025} CGM \\
& \texttt{z\_500\_TEMPLATE\_asada}$^b$ & \texttt{float} & \texttt{EAZY} 50h Percentile Photo-$z$, with \citet{asada2025} CGM \\
& \texttt{z\_840\_TEMPLATE\_asada}$^b$ & \texttt{float} & \texttt{EAZY} 84th Percentile Photo-$z$, with \citet{asada2025} CGM \\
\hline 
& \texttt{F277W\_ext\_ratio\_0p10} & \texttt{float} & Flux ratio between source and PSF at 0.10$^{\prime\prime}$, F277W \\
& \texttt{F277W\_ext\_ratio\_0p15} & \texttt{float} & Flux ratio between source and PSF at 0.15$^{\prime\prime}$, F277W \\
\texttt{MORPHOLOGY} & \texttt{F277W\_ext\_ratio\_0p20} & \texttt{float} & Flux ratio between source and PSF at 0.20$^{\prime\prime}$, F277W \\
& \texttt{F444W\_ext\_ratio\_0p10} & \texttt{float} & Flux ratio between source and PSF at 0.10$^{\prime\prime}$, F444W \\
& \texttt{F444W\_ext\_ratio\_0p15} & \texttt{float} & Flux ratio between source and PSF at 0.15$^{\prime\prime}$, F444W \\
& \texttt{F444W\_ext\_ratio\_0p20} & \texttt{float} & Flux ratio between source and PSF at 0.20$^{\prime\prime}$, F444W \\
\hline 
& \texttt{Hainline\_et\_al\_2023\_chisq\_gt\_4\_cat\_flag} & \texttt{bool} & Boolean if this source was included in \KHZ{} as a ``\texttt{chisq\_gt\_4}'' source\\
& \texttt{Hainline\_et\_al\_2023\_chisq\_lt\_4\_cat\_flag} & \texttt{bool} & Boolean if this source was included in \KHZ{} as a ``\texttt{chisq\_lt\_4}'' source\\
& \texttt{Hainline\_et\_al\_2023\_flag\_bn\_2\_cat\_flag} & \texttt{bool} & Boolean if this source was included in \KHZ{} as a ``\texttt{flag\_bn\_2}'' source\\
& \texttt{Austin\_et\_al\_2023\_cat\_flag} & \texttt{bool} & Boolean if this source was included \citet{austin2023}.\\
\texttt{OTHER\_CATALOGS} & \texttt{Leung\_et\_al\_2023\_cat\_flag} & \texttt{bool} & Boolean if this source was included \citet{leung2023}\\
& \texttt{Robertson\_et\_al\_2024\_cat\_flag} & \texttt{bool} & Boolean if this source was included \citet{robertson2024}\\
& \texttt{Whitler\_et\_al\_2025\_cat\_flag} & \texttt{bool} & Boolean if this source was included \citet{whitler2025}\\
& \texttt{Weibel\_et\_al\_2025\_cat\_flag} & \texttt{bool} & Boolean if this source was included \citet{weibel2025}\\
& \texttt{First\_Reference\_Bibcode} & \texttt{string} & Earliest reference Bibcode for source\\
\enddata
\tablecomments{a) Filters include F435W, F606W, F775W, F814W, and F850LP (from HST/ACS), as well as F070W, F090W, F115W, F150W, F162M, F182M, F200W, F210M, F250M, F277W, F300M, F335M, F356W, F410M, F430M, F444W, F460M, and F480M filters (from JWST/NIRCam) b) Templates include: \texttt{tweak\_fsps} and \texttt{agn\_blue\_sfhz} (built-in \texttt{EAZY} templates), \texttt{larson} from \citet{larson2023b}, \texttt{fsps\_45k} and \texttt{fsps\_60k} from \citet{steinhardt2023}, and \texttt{ares\_no\_lya} and \texttt{ares\_with\_lya} from \citet{luberto2025}.}
\end{deluxetable*}

We plot the full GOODS-S and GOODS-N footprints along with our sources in Figures \ref{fig:GS_footprint} and \ref{fig:GN_footprint}. On both these figures we plot outlines for both JADES data, as well as the full footprint of the DR5 GOODS-S and GOODS-N imaging datasets from which we selected our sources. In each panel, we color the points by their photometric or spectroscopic redshift. We find that, as expected, the deeper portions of the footprints contain a higher density of galaxies at higher redshifts, in particular the JOF in the southern part of GOODS-S and in the NGDEEP area to the east of the GOODS-S footprint\footnote{This is shown clearly with NIRCam filter depth maps presented in Section 4.2 of \JJADES{}.}. In both fields there are regions within the JADES DR5 footprint, specifically in some of the regions where NIRCam was used in parallel, where we find no candidates owing to the lack of specific filter coverage.

For each source in our sample, we use the photometric or spectroscopic redshift to estimate $M_{\mathrm{UV}}$ and $\beta$. As discussed previously, to estimate these values, we use the CIRC1 fluxes measured from images convolved to the F444W PSF. We followed the approach from \citet{topping2024}, where we fit a \citet{calzetti1994} power law ($f_{\lambda} \propto \lambda^{\beta}$) to the photometry spanning the rest-frame UV for each object. For objects at $z = 8 - 9$, we fit to photometry in the F150W, F200W, and F277W filters, for objects at $z = 9 - 13.2$, we used the F200W, F277W, and F356W filters, and for sources at $z > 13.2$, we used the F277W, F356W, and F444W filters. While the exact wavelength range that this procedure spans is different depending on the source redshift, the average range is $1650$\AA$- 2900$\AA$\,$ rest-frame, which will not be affected by the presence of Lyman-$\alpha$ emission or Lyman damping. The slope of the power-law fit we report as $\beta$, and we calculated the rest-frame magnitude through a synthetic top-hat transmission filter between 1400 and 1600 Angstroms (rest-frame) for this model fit. In a small fraction of sources (14 sources in GOODS-S, and 66 sources in GOODS-N), lack of available photometry sampling the rest-frame UV prevented us performing this fit and estimating values. We report both $M_{\mathrm{UV}}$ and $\beta$ for each source in our sample in our final table, along with uncertainties estimated from the fits. 

\begin{figure*}[t]
  \centering
  \includegraphics[width=0.95\linewidth]{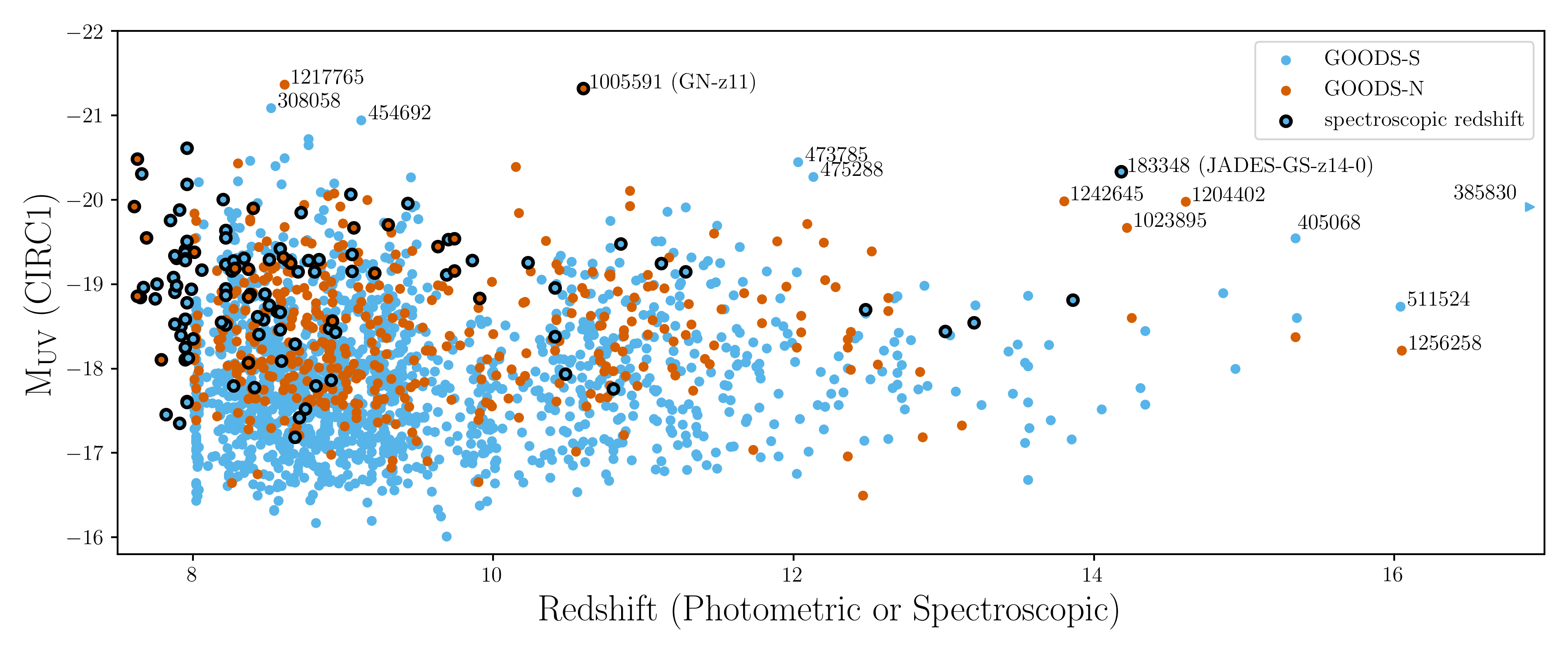}\
  \caption{UV magnitude plotted against photometric (solid color points) and spectroscopic (black outlined points) for our \NsampGS{} GOODS-S (blue) and \NsampGN{} GOODS-N (red) sources. We highlight some of the brighter source IDs along the top of the Figure. The UV magnitudes were estimated using convolved CIRC1 photometry, as described in Section \ref{sec:results}. The observed lack of sources at $z \sim 10$ and $z \sim 13$ is primarily due to the impact of the gaps between the F115W, F150W, and F200W filters when measuring photometric redshifts, as discussed in Appendix A of \KHZ{}.}
  \label{fig:MUV_vs_redshift}
\end{figure*}

In Figure \ref{fig:MUV_vs_redshift} we plot the estimated $M_{\mathrm{UV}}$ against the photometric or spectroscopic redshifts for our sources. We plot the GOODS-S sources in blue, and the GOODS-N sources in red, and if a source has a spectroscopic redshift (see Section \ref{sec:spectroscopic_redshifts}), we use a thick black outline. We also highlight some of the brighter sources, including GN-z11 \citep[``1005591'', ][]{oesch2016, bunker2023a} and JADES-GS-z14-0 \citep[``183348'', ][]{hainline2024a, robertson2024, carniani2024}. The lower density of sources with photometric redshifts observed at $z \sim 10$ (and to a lesser extent, $z \sim 13$) is due to the gap between the F115W and F150W photometric filter curves and our use of minimum-$\chi^2$ redshifts from \texttt{EAZY}, as discussed in detail in Appendix A of \KHZ{}.

From this figure, we can see that the depth of the GOODS-S observational data allows for us to detect a population of intrinsically fainter sources at $M_{\mathrm{UV}} > -17$, primarily at $z < 12$. For both GOODS-S and GOODS-N, the distribution of sources at the faint end, along the bottom of the figure, demonstrates the observational depths of these programs, while we also observe that the distribution at the bright end of the figure shows the growth of galaxies over cosmic time. It is notable that the detection limits of the deepest portions in the assembled JADES DR5 footprint are such that finding higher redshift galaxies would require significantly deeper observations, or finding additional rare bright sources like the ones indicated with labels. It does appear, however, that the density of sources drops significantly at $z > 15$, even with the depth of this dataset.

\subsection{Comparison to Literature Sources}\label{sec:comparisonliterature}

Given the importance of the GOODS-S and GOODS-N extragalactic fields, there have been multiple literature studies exploring samples of galaxies from these survey areas in the redshift range we consider. In this Section we discuss the results from cross-matching our current sources with both pre- and post-JWST samples of high-redshift galaxies. Prior to the launch of JWST, a number of the sources in this sample were identified using both HST and the Spitzer space telescope, as discussed in Section 4.7 of \KHZ{} (we refer the reader to Table 4 in that paper for a detailed breakdown of the references for the initial detection of a number of very bright high-redshift sources in GOODS-S and GOODS-N). We followed the procedure described there, searching the literature to match against sources in our current sample, and we report the earliest citation for those sources with matches in our final data table, as described in Table \ref{tab:catalog_columns}. While we only report the earliest paper describing each matched source, the full set of literature catalogs we searched included \citet{bunker2009, lorenzoni2011, bouwens2011, yan2012, ellis2013, mclure2013, oesch2013, schenker2013, oesch2014, bouwens2015, finkelstein2015, harikane2016, oesch2018, bouwens2021, robertson2023, donnan2023b, tacchella2023} and \citet{tang2025}. Separately, we also cross-matched our sources against those presented in \KHZ{}, to the dropout and photometric redshift selected samples in \citet{whitler2025}, \citet{robertson2024}, the NGDEEP sources discussed in \citet{austin2023} and \citet{leung2023}, and the PANORAMIC dropout sources discussed in \citet{weibel2025}. 

As this paper is building upon earlier results from \KHZ{}, our sample has a significant overlap with the candidates presented in that study, although we probe a larger area. In GOODS-S, 401 sources from our current sample were also presented in the \KHZ{} $\Delta \chi^2 > 4$ sample (75\% of the sample in that earlier paper), and in GOODS-N, 127 sources (70\%) are in our updated sample. In addition, with the updated photometry in JADES DR5, an additional 27 objects in GOODS-S and 10 objects in GOODS-N that were presented in the $\Delta \chi^2 < 4$ sample from \KHZ{} are in our final sample. Similarly, as we do not use a nearby neighbor flag as was done in \KHZ{}, an additional 22 sources in GOODS-S and 7 sources in GOODS-N that were flagged in that study are in our current sample.

For the $\Delta \chi^2 > 4$ sources in \KHZ{} that do not appear in the sample presented in this study, many of the objects (55\% of these sources in GOODS-S and 51\% in GOODS-N) have updated photometric redshifts $z_{\mathrm{a}} < 8$, with most being at $7 < z_{\mathrm{a}} < 8$. However, the vast majority of the sources from \KHZ{} that do not appear in this study have an updated $\Delta \chi^2 < 4$ (83\% of these sources in GOODS-S and 82\% in GOODS-N) or in a few rare cases, because they are blended with other sources in the updated segmentation map. 

In \citet{whitler2025}, the authors present a sample of F115W ($z \sim 8.5-12$), F150W ($z \sim 12-16$), and F200W ($z \sim 16-22.5$) dropout galaxies from GOODS-S and GOODS-N JADES observations and use these sources to construct UV luminosity functions. We compare our current sample to theirs, with the caveat that dropout selection will miss galaxies in very specific redshift ranges where the Lyman-$\alpha$ break pushes sources to red colors outside of standard selection boxes \citep[see discussions in][]{hainline2024a, whitler2025}. The \citet{whitler2025} sample has 235 sources in GOODS-S and 137 sources in GOODS-N, and the sample presented here includes 145 (62\%) and 84 (61\%) of the \citet{whitler2025} sources respectively. The majority of the sources that are in the \citet{whitler2025} sample, but not our own, are fit with \texttt{EAZY} at $z_{\mathrm{a}} < 8$ (58\% in GOODS-S and 77\% in GOODS-N), while of the remaining at $z_{\mathrm{a}} > 8$ the majority of the galaxies that are not in our current sample have an updated $\Delta \chi^2 < 4$ (32 sources, or 86\% in GOODS-S, and 9 sources, or 82\% in GOODS-N). 

In \citet{robertson2024}, a sample of galaxies with photometric redshifts putting them at $z > 11$ is presented from within the extremely-deep GOODS-S JOF \citep{eisenstein2023b}. We matched these sources to our current sample, and find that of the 14 objects provided in that paper, 8 appear in the current sample (57\%), with the remaining six having $\Delta \chi^2 < 4$, or in the case of one source (ID 55733), an updated photometric redshift of $z_a = 3.37$. 

A sample of 37 sources at $z_{\mathrm{phot}} > 5.9$ was presented for the NGDEEP field \citep{bagley2024} in GOODS-S in \citet{austin2023}. We matched these sources to our catalog and find that 15 appear in the current sample. For these sources, our photometric redshift estimates broadly agree, although two sources at $z_{\mathrm{phot}} = 7.4 - 8.0$ appear in our sample at $z_a > 8$. The highest redshift source from their sample that appears in the current sample is NGD-z12a, which they estimate to be at $z_{\mathrm{phot}} = 12.1$, and has $z_a = 12.7$ in our sample. Notably, the highest-redshift ``robust'' source in their sample, NGD-z15a, does not appear in our sample (it is matched to JADES ID 510779) as we observe a detection (SNR $= 3.82$) of the source in the F150W filter, likely indicating the source is at $z < 5$ with a strong Balmer break. 

In \citet{leung2023}, the authors present a sample of 38 NGDEEP sources with photometric redshifts $z_{\mathrm{phot}} > 9$. We cross matched each of these sources to our catalog, finding that 30 of the sources are found in our current sample, with a similar strong agreement in photometric redshift. The remaining sources are either at $z_a < 8$, or at $z_a > 8$ but with $\Delta \chi^2 < 4$. One source in their sample, ``NGDEEP 13406'' (RA: 53.240619 DEC: -27.862122 and a $z_{\mathrm{phot}} = 9.31$ in that paper), is cross-matched to JADES ID 468805 ($z_a = 2.6$), but the true source from this study may be found slightly northwest of 468805, and not properly segmented in the JADES reduction. This study also includes the \citet{austin2023} source NGD-z15a (as ``NGDEEP 1369'') previously mentioned, which we reject. 

Finally, in \citet{weibel2025}, the authors assembled a sample of F115W, F150W, and F200W dropouts from across a number of publicly available JWST NIRCam survey datasets, including JADES, while including pure parallel imaging from PANORAMIC \citep{williams2025}. They employ a slightly different SNR limit than we do in this paper, requiring SNR $ > 8$ for the detection band redward of the Lyman-$\alpha$ break and then SNR $ > 3$ for the next redder band, such that they will return intrinsically brighter sources than we do across the same footprint. We compare the sources assembled in this paper with our current sample, while focusing on the F115W and F150W dropouts (as the authors do not report any F200W dropouts in either GOODS-S or GOODS-N). Combining these two samples, there are 47 high-redshift galaxies and candidates (33 in GOODS-S and 14 in GOODS-N), and all but six (which are all in GOODS-S) are also included in the $z > 8$ sample in this study. Three of those six sources are reported by \texttt{EAZY} at $z_a < 8$, and of the remaining galaxies, two have $\Delta \chi^2 < 4$, and the final source (their ID 32068, and ID 164615 in the JADES catalog) has a detection in F090W at SNR $= 3.47$, and was removed from our selection during visual inspection. 

We indicate which of the \KHZ{}, \citet{whitler2025}, \citet{robertson2024}, \citet{austin2023}, \citet{leung2023}, and \citet{weibel2025} sources appear in our sample with a separate column in our final output table. As discussed above, we also include a column for matches between our current sample and those presented in literature studies with the earliest reference. After exploring the literature, 1346 (65\% of the total sample) of the sources (1057 in GOODS-S and 289 in GOODS-N) are reported in this study for the first time. 

\subsection{Spectroscopic Redshifts}\label{sec:spectroscopic_redshifts}

As the GOODS-S and GOODS-N regions have been the target of numerous spectroscopic campaigns in the past several decades, there are many sources in our sample with confirmed spectroscopic redshifts, as plotted in Figure \ref{fig:MUV_vs_redshift}. In GOODS-S, 95 of our sources have spectroscopic redshifts, and in GOODS-N, 28 of our sources have spectroscopic redshifts, for a total of \Nzspec{} sources. These redshifts originate from multiple surveys, primarily done with either JWST/NIRSpec or JWST/NIRCam Wide Field Slitless Spectroscopy (WFSS), although there are 7 spectroscopic redshifts obtained using data from the MUSE WIDE survey (M. Maseda, private communication). In GOODS-S, the NIRSpec catalogs we drew from were a part of the publicly available JADES DR4 catalogs \citep{curtislake2025, scholtz2025}, and for some ultra-high-redshift sources, we include additional redshift measurements in the literature for GN-z11 \citep{bunker2023a}, JADES-GS-z11-0 \citep{hainline2024c,witstok2025b}, JADES-GS-z12-0 \citep{deugenio2024a}, JADES-GS-z13-1-LA \citep{witstok2025}, JADES-GS-z13-0 \citep{hainline2024c}, JADES-GS-z14-1 \citep{wu2025}, and JADES-GS-z14-0 \citep{carniani2024}.

We also include NIRSpec observations from the JOF \citep[PID 3215, ][]{eisenstein2023b}, OASIS (PID 5997, T. Looser, private communication), DarkHorse \citep{deugenio2025}, data taken as part of DDT spectroscopic follow-up of transients \citep[PID 6541, ][]{decoursey2025}, and data taken as part of a spectroscopic program targeting quiescent galaxies \citep[PID 2198, ][]{barrufet2024}. In GOODS-N, we also found a source with a spectroscopic confirmation (JADES-GN+189.22441+62.31137, JADES ID 1095809, $z_{\mathrm{spec}} = 7.61$) taken as part of a program targeting reionization galaxies \citep[PID 1871, ][]{chisolm2024}. We additionally supplemented these data with redshifts from NIRCam WFSS observations taken as part of JOF (PID 4540, PI: Eisenstein), as well as the publicly available FRESCO \citep[PID 1895,][]{oesch2023} and CONGRESS \citep[PID 3577,][]{egami2023} programs from F. Sun (private communication).

As would be expected, and as can be seen in Figure \ref{fig:MUV_vs_redshift}, the bulk of the spectroscopic redshifts are for brighter sources at $z < 10$. A small number of the spectroscopic redshifts for our sources selected with $z_{\mathrm{phot}} > 8$ are at $z_{\mathrm{spec}} = 7.5 - 8$, and we choose to include these sources to demonstrate the uncertainties in photometric redshift selection. In Section \ref{sec:spec_z_vs_phot_z} we will explore a comparison of the spectroscopic to photometric redshifts for our sample, and we will discuss the usage of alternate \texttt{EAZY} templates in fitting, as well as the effects of IGM and Lyman damping absorption on the overall photometric redshifts. 

\subsection{The Highest Redshift Candidates in GOODS-S and GOODS-N}\label{sec:highest_redshift_sources}

\begin{figure*}[t!]
  \centering
  \includegraphics[width=0.32\linewidth]{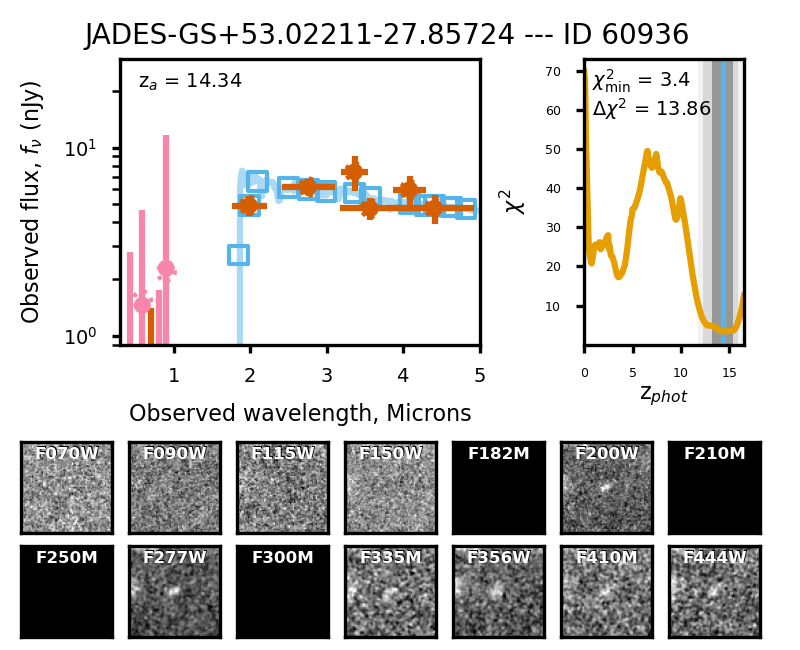}\
  \includegraphics[width=0.32\linewidth]{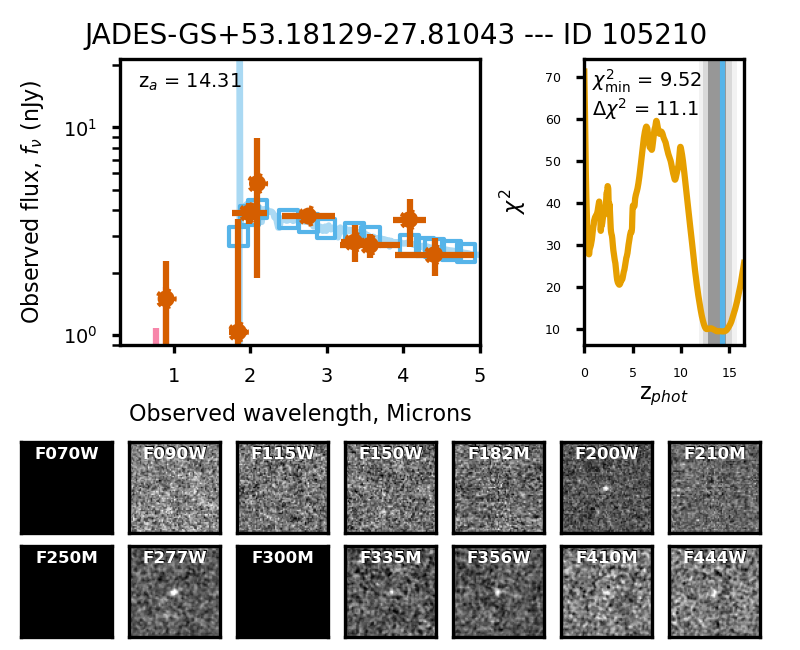}\
  \includegraphics[width=0.32\linewidth]{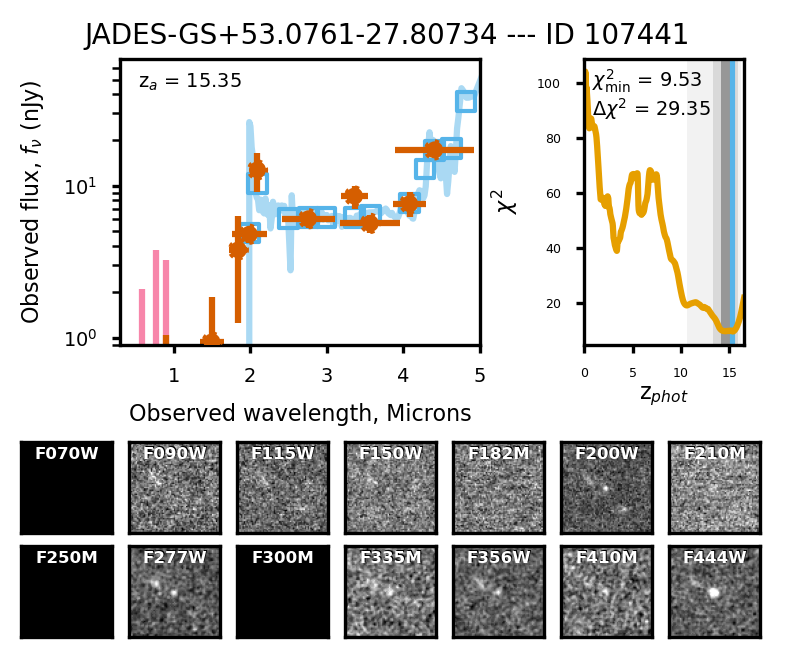}\
  \includegraphics[width=0.32\linewidth]{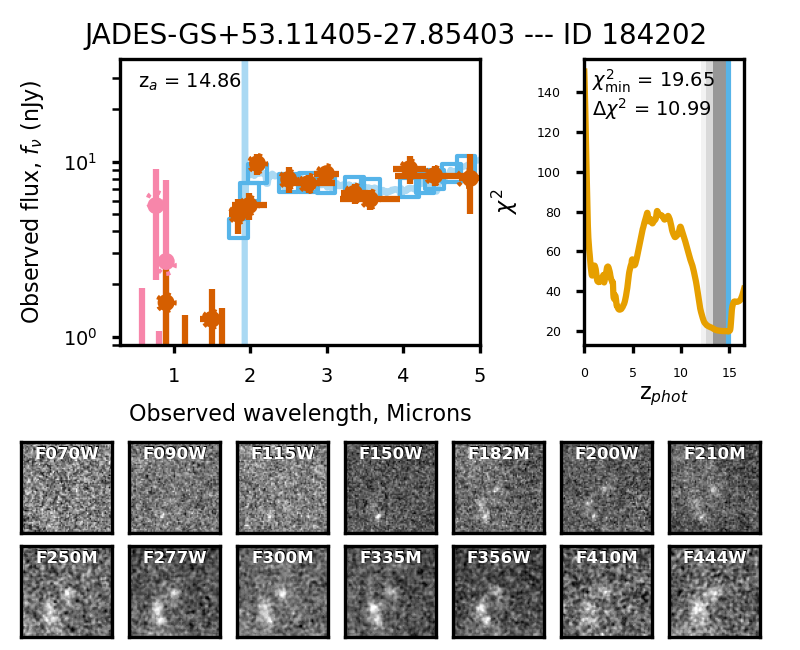}\
  \includegraphics[width=0.32\linewidth]{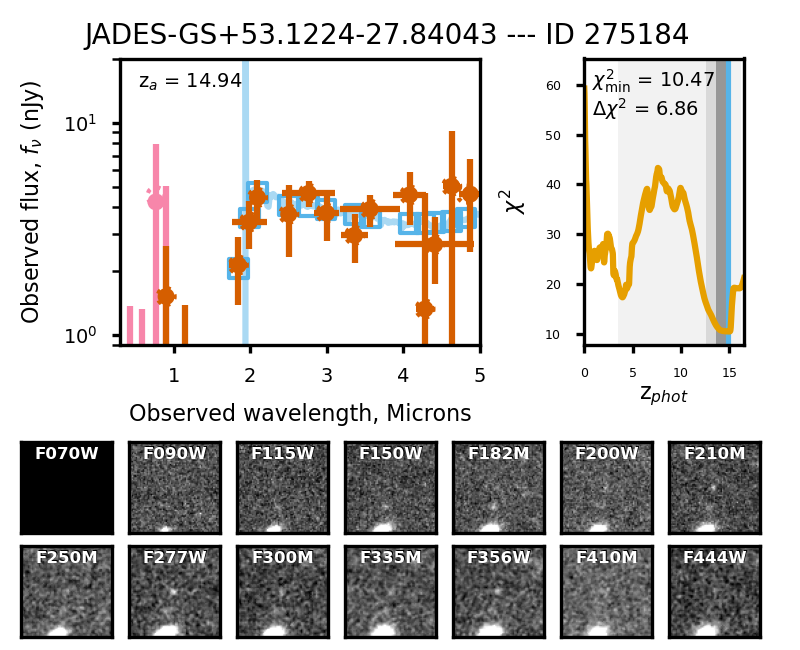}\
  \includegraphics[width=0.32\linewidth]{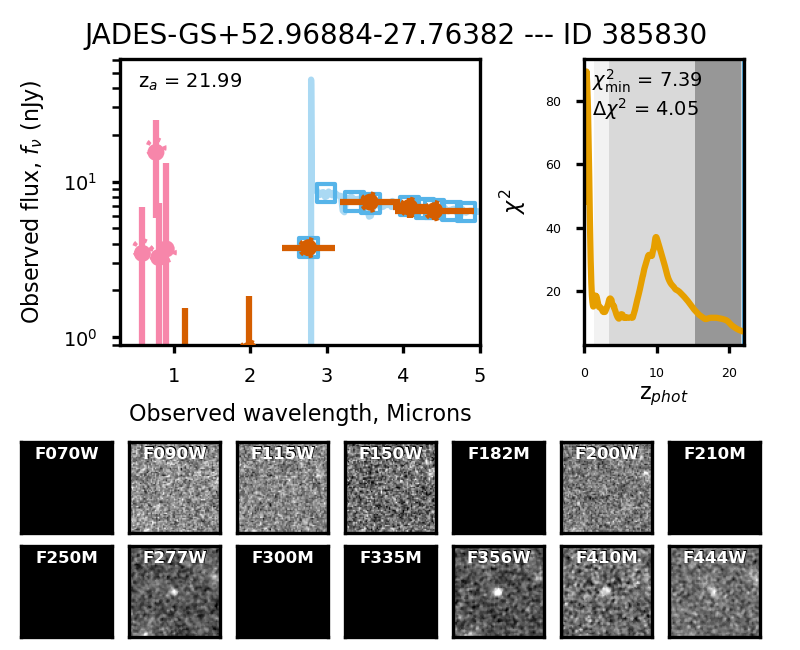}\
  \includegraphics[width=0.32\linewidth]{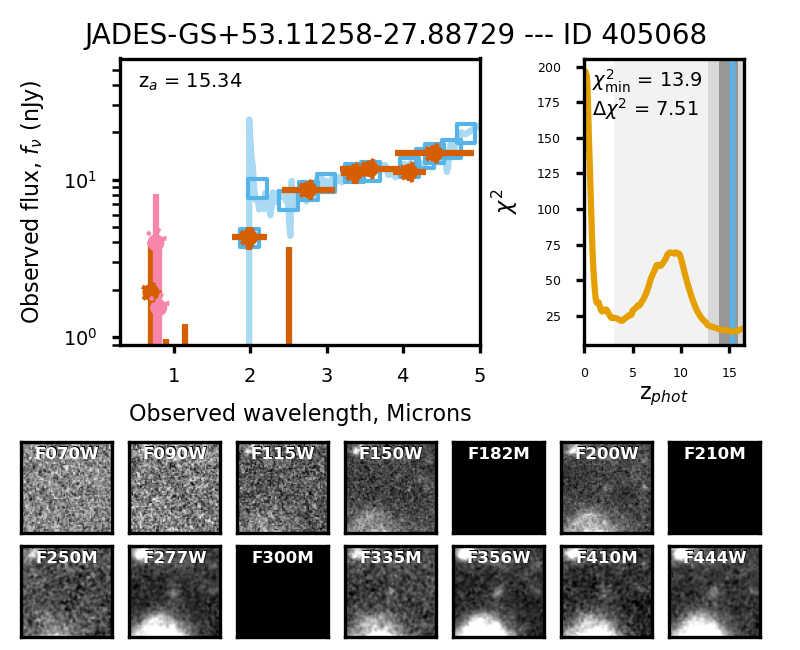}\
  \includegraphics[width=0.32\linewidth]{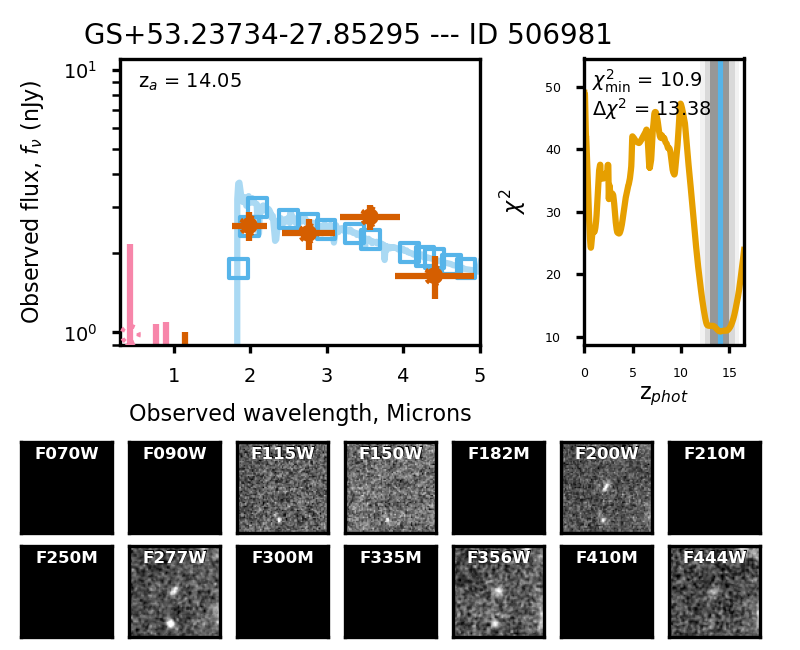}\
  \includegraphics[width=0.32\linewidth]{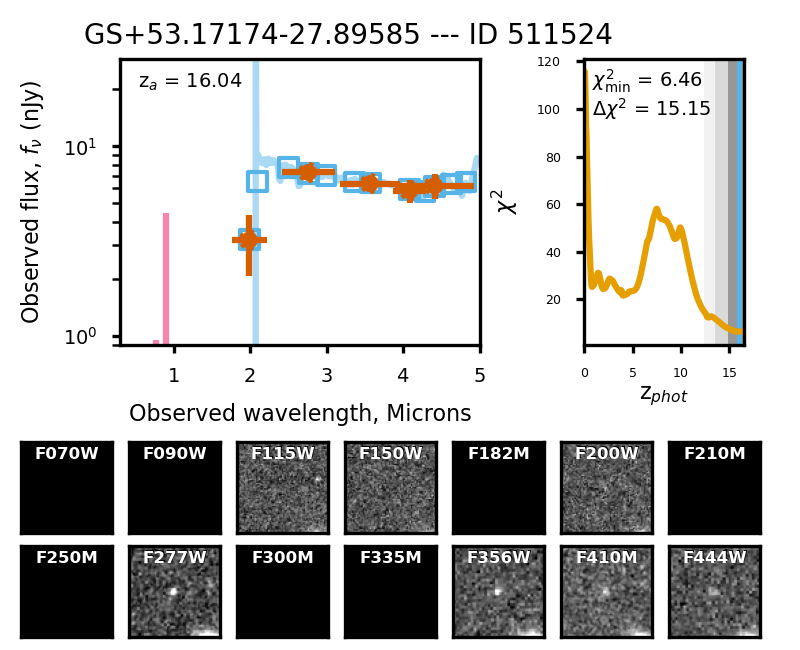}\
  \includegraphics[width=0.32\linewidth]{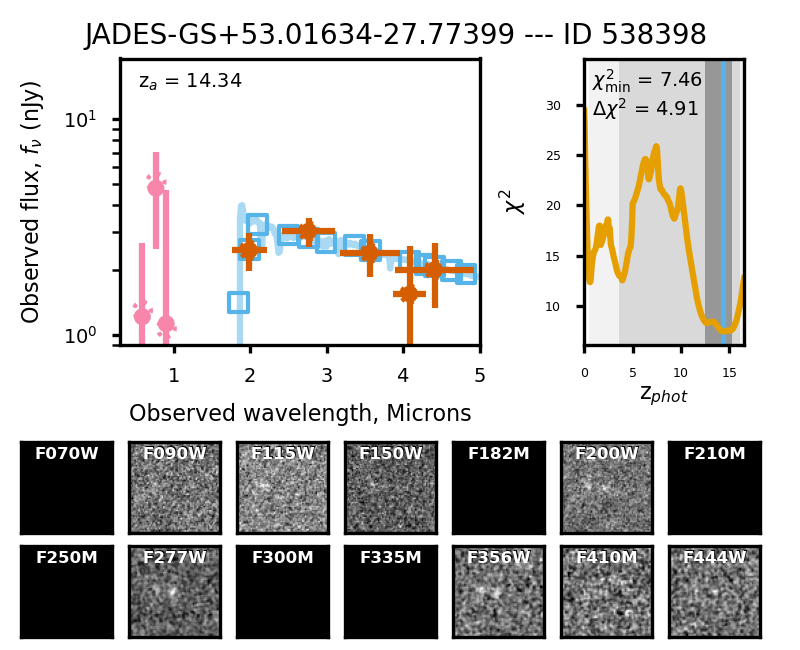}\  
  \caption{SEDs, \texttt{EAZY} $\chi^2$ surfaces, and JWST/NIRCam thumbnails for the GOODS-S galaxies at $z_a > 14$ in our sample. For each source, we show the HST and NIRCam photometry in the top left panel with pink and red points respectively, and the \texttt{EAZY} fit in blue. In the top right panel, we plot \texttt{EAZY} $\chi^2$ as a function of redshift with an orange line, the $z_a$ value with a vertical blue line, and the 1$\sigma$ (2$\sigma$) redshift uncertainty with a dark grey band (light grey band). Underneath, we plot $2.0^{\prime\prime} \times 2.0^{\prime\prime}$ NIRCam thumbnails, for those filters without coverage, the thumbnail is black.}
  \label{fig:gs_high_z_galaxies}
\end{figure*}

\begin{figure*}[t!]
  \centering
  \includegraphics[width=0.32\linewidth]{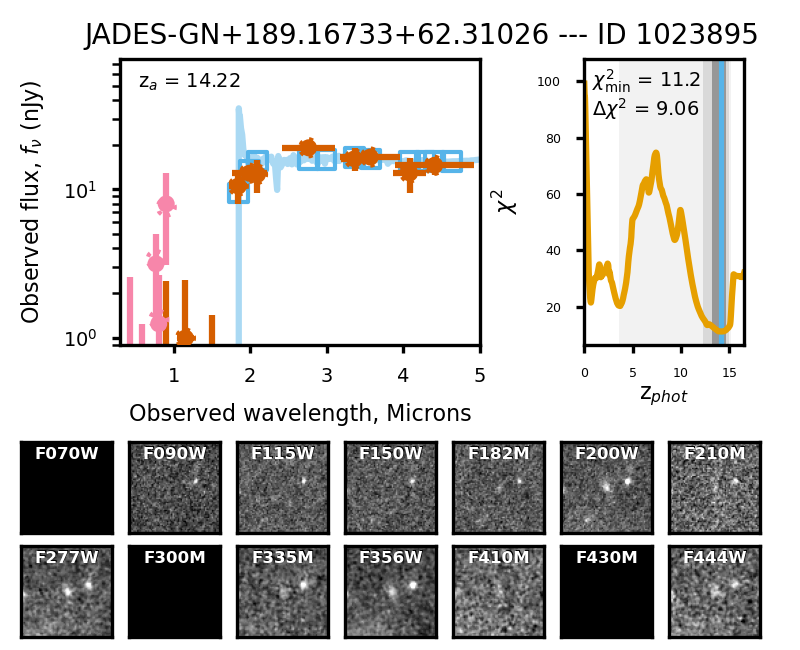}\
  \includegraphics[width=0.32\linewidth]{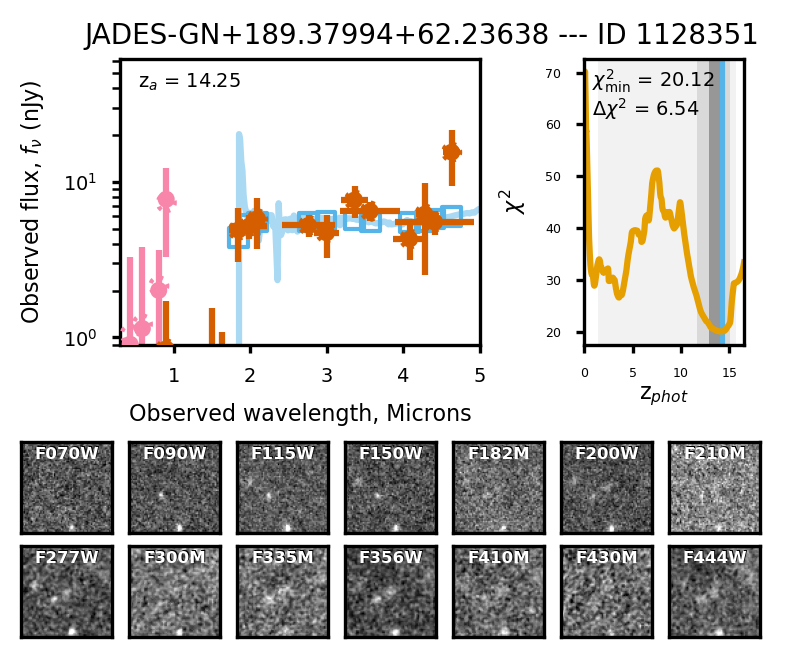}\
  \includegraphics[width=0.32\linewidth]{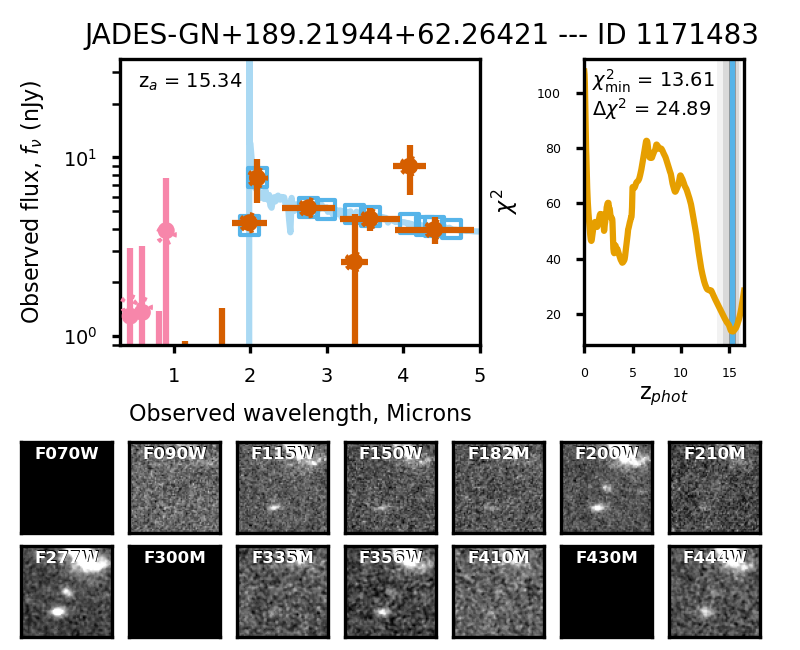}\
  \includegraphics[width=0.32\linewidth]{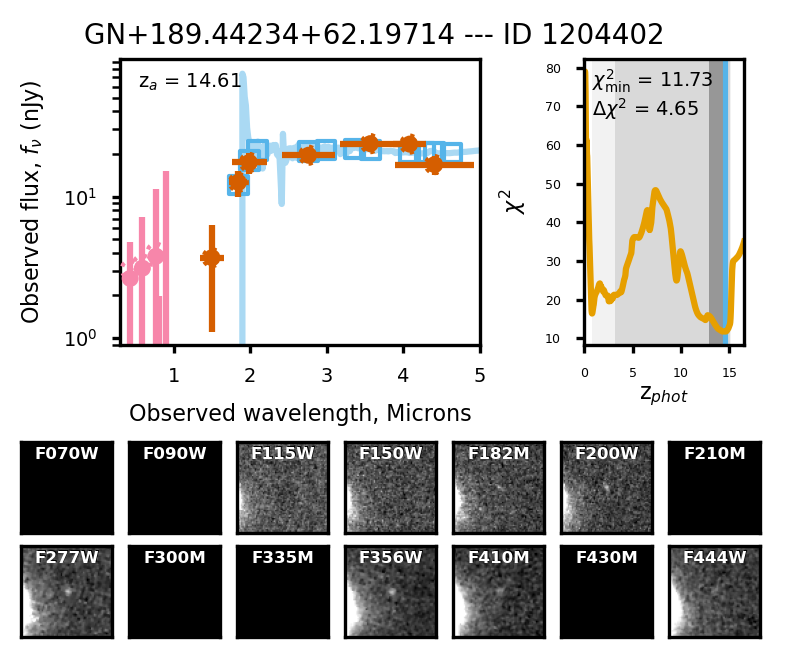}\
  \includegraphics[width=0.32\linewidth]{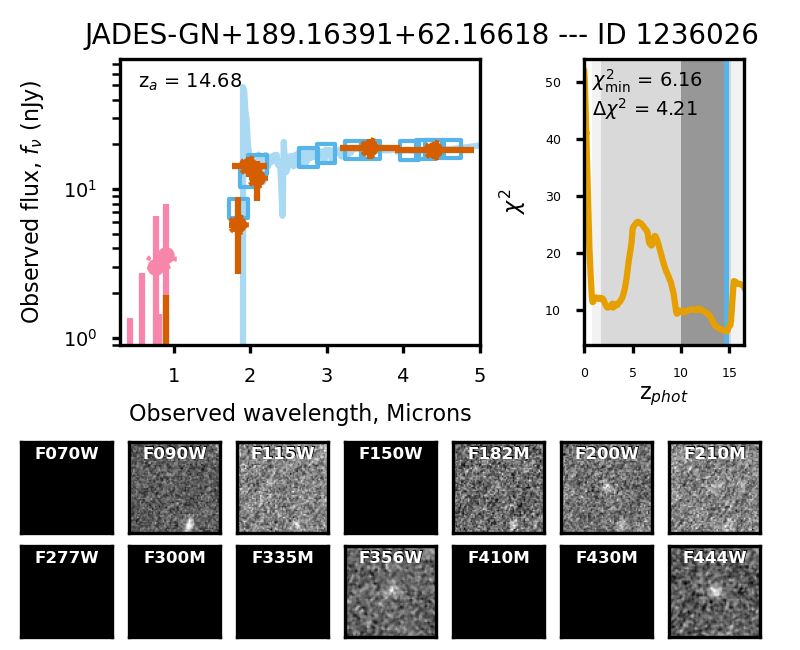}\
  \includegraphics[width=0.32\linewidth]{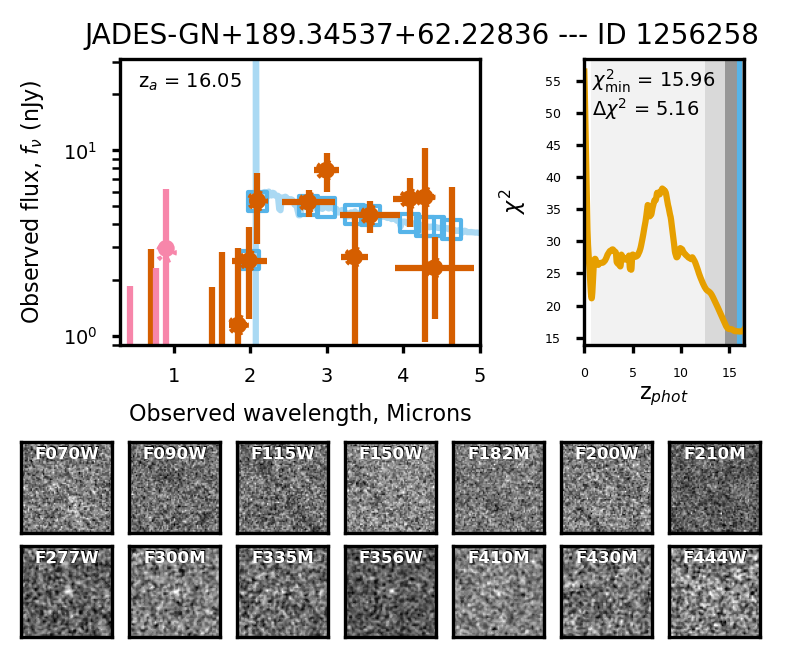}\
  \caption{GOODS-N galaxies at $z_a > 14$ in our sample. The colors, lines, and symbols are as in Figure \ref{fig:gs_high_z_galaxies}.}
  \label{fig:gn_high_z_galaxies}
\end{figure*}

There are 19 objects in our full sample with \texttt{EAZY} $z_a > 14$ (13 in GOODS-S and 6 in GOODS-N). Three of these objects, JADES-GS-z14-0, JADES-GS-z14-1, and JADES-GS-z13-0 have been spectroscopically confirmed, and are described further in \citet{curtislake2023,hainline2024c}, \citet{carniani2024} and \citet{wu2025}. We will discuss the remaining 16 sources in this section, and we plot their SEDs in Figures \ref{fig:gs_high_z_galaxies} and \ref{fig:gn_high_z_galaxies}, along with thumbnails for the objects and the $\chi^2$ surface, indicating the minimum $\chi^2$ and $\Delta \chi^2$ values. 

\textbf{JADES-GS+53.02211-27.85724} (ID 60936, $z_a = 14.34$) This F150W dropout galaxy, which was included in the sample presented in \KHZ{}, appears to be extended with two knots in both the F200W and F277W filters, with no significant detection in four NIRCam filters, F070W, F090W, F115W, and F150W. As is the case for many of these sources in this section, the exact photometric redshift is driven largely by the observed flux for the filter containing the Lyman-$\alpha$ break, which for this source is F200W. Given the uncertainty on F200W, the 1$\sigma$ uncertainty on the redshift could range between $z = 13.19 - 14.60$. 

\textbf{JADES-GS+53.18129-27.81043} (ID 105210, $z_a = 14.31$) This is another F150W or F182M dropout galaxy, also from the \KHZ{} sample, with a significant (SNR $>$ 5) detection in F200W, F277W, and F356W, and a very blue UV slope ($\beta = -2.89 \pm 0.41$).

\textbf{JADES-GS+53.0761-27.80734} (ID 107441, $z_a = 15.35$) This F182M dropout is notable because of the relatively bright F444W ($17.2 \pm 1.0$ nJy) and F210M ($12.7 \pm 3.6$ nJy) fluxes, as compared to the F277W flux ($17.2 \pm 1.0$ nJy). While there appears to be F150W flux in the thumbnail, there is high flux uncertainty ($0.9 \pm 0.9$ nJy). To the northwest of the source is a faint extended source seen in F090W, F115W, and F150W, which was included as part of the same source in the JADES DR5 segmentation map, but is likely a foreground object. 

\textbf{JADES-GS+53.11405-27.85403} (ID 184202, $z_a = 14.86$) This source is somewhat extended, with the brighter core only visible at 2.5 - 3.0$\mu$m. It is within 0.5$^{\prime\prime}$ of another faint source, ID 184201, that has a photometric redshift of $z_a = 11.91$, and resides within $\sim 2^{\prime\prime}$ of a much larger, dusty galaxy (ID 183722) at $z_a = 3.28$. If 184201 and 184202 were associated with extended emission from 183722 and dusty, then the Balmer break would be expected to lie at $\sim 1.5 - 1.7\mu$m, where we see the potential Lyman-$\alpha$ break for these sources, given the photometric uncertainty. In support of this hypothesis, Sun et al. (submitted) presents evidence for CGM dust in a GOODS-S foreground galaxy cluster obscuring a $z \sim 4.3$ background galaxy and leading it to be mistakenly identified at $z > 16$ (see Section 4.6 in their paper). 

\textbf{JADES-GS+53.1224-27.840427} (ID 275184, $z_a = 14.94$) This F150W or F182M dropout galaxy has multiple detections, although with only $2-3\sigma$ detections at $> 4\mu$m. There is a lower redshift galaxy within $\sim 1^{\prime\prime}$ (ID 80874, seen in the bottom of the thumbnail), but the photometric redshift for this source, $z_a = 1.7$, does not coincide with the secondary minimum observed for 275184. 

\textbf{JADES-GS+52.96884-27.76382} (ID 385830, $z_a = 21.99$) This is the highest redshift candidate in the whole sample, an F200W dropout (F200W SNR = 0.85) near the western edge of the GOODS-S mosaic. The redshift is very uncertain, being driven largely by the red F277W $-$ F356W color ($f_{\mathrm{F356W}} / f_{\mathrm{F277W}} = 1.96 \pm 0.27$), and it may be that this source is at lower redshift with strong emission lines boosting the fluxes in the detected filters \citep[a so-called ``Schr\"{o}dinger's Galaxy,'' see][]{naidu2022,arrabalharo2023}. As a result, for the \texttt{EAZY} fit, the 1$\sigma$ uncertainty range on the redshift is $z = 15.26 - 22$. 

\textbf{JADES-GS+53.11258-27.88729} (ID 405068, $z_a = 15.34$) This galaxy is located between two bright, red galaxies at $z \sim 2.7$ (IDs 163806 and 405056 in the JADES DR5 catalog) and could be associated with these sources, given the red UV slope ($\beta = -1.13 \pm 0.24$). It is possible this source is at a lower redshift and is being heavily obscured by dust in the CGM in these more nearby sources along the line of sight (see Sun et al. submitted for discussion of similar sources). 

\textbf{GS+53.23734-27.85295} (ID 506981, $z_a = 14.05$) This object features an extended morphology ($0.3-0.4^{\prime\prime}$, 1.0 - 1.3 kpc at $z = 14$) in both the F200W and F277W photometry. The exact redshift of the source is dependent on the F200W flux, and as a result the 1$\sigma$ range on the redshift could be $z_{\mathrm{phot}} = 13.0 - 14.9$.

\textbf{GS+53.17174-27.89585} (ID 511524, $z_a = 16.04$) This very exciting, compact galaxy  lies in a region covered by both the PANORAMIC \citep{williams2025} GOODS-S coverage, as well as JWST PID 2198 \citep{barrufet2024}. It has a very red F200W $-$ F277W color ($f_{\mathrm{F277W}} / f_{\mathrm{F200W}} = 2.24 \pm 0.83$), with no detection (SNR $< 1$) in F115W, F150W, or the HST ACS filters. With a F277W SNR of 11.6, this source represents the brightest galaxy across GOODS-S at this photometric redshift ($M_{\mathrm{UV}} = -18.7 \pm 0.1$), albeit with a 1$\sigma$ range of $z_{\mathrm{phot}} =14.9 - 16.8$. Object 511524 does not appear in the catalog of $z = 9 - 17$ galaxies from PANORAMIC presented in \citet{weibel2025}, possibly because the F200W $-$ F277W color is too red to satisfy the color selection for F150W dropouts, or due to differing photometric measurements from the PANORAMIC data. 

\textbf{JADES-GS+53.01634-27.7739} (ID 538398, $z_a = 14.34$) This relatively faint F150W dropout is only significantly detected with SNR $>5$ in F200W and F277W. It is within 0.5$^{\prime\prime}$ of another faint source observed at $\sim 1\mu$m (ID 546753) at $z_a = 1.1$, a redshift that is inconsistent with the $\chi^2$ minima observed for 538398.

\textbf{JADES-GN+189.16733+62.31026} (ID 1023895, $z_a = 14.22$) This F150W dropout, which appears in the sample from \KHZ{}, is slightly resolved, extending northeast to south-west, as seen in F200W and F356W most notably. It is northwest of a large extended source (ID 1035590) at $z_a = 1.18$ but does not appear to be associated. The photometric redshift is largely driven by the F182M flux, and the 1$\sigma$ range on the redshift is $z_{\mathrm{phot}} = 13.1 - 14.6$. 

\textbf{JADES-GN+189.37994+62.23638} (ID 1128351, $z_a = 14.25$) This faint object is unresolved, with a fairly red UV slope unusual for sources at this redshift ($\beta = -1.77 \pm 0.55$). It is immediately proximate to another, separate lower-redshift source that has flux detected only at $> 0.9$ $\mu$m, but is not selected as its own source in the JADES DR5 segmentation catalog. 

\textbf{JADES-GN+189.21944+62.26421} (ID 1171483, $z_a = 15.34$) This F182M dropout is observed to be extended in each of the filters it is observed in, and in F200W there are two distinct morphological peaks separated by $0.1^{\prime\prime}$ (0.3 kpc at $z = 15.3$). The source lies between two galaxies separated by 1.5$^{\prime\prime}$ at $z_a = 1.5 - 1.6$ (seen in the thumbnails), inconsistent with the $\chi^2$ minima for ID 1171483. Because of the lack of F182M flux (SNR $ = -1.82$), the 1$\sigma$ redshift range is fairly well constrained to lie at $z_{\mathrm{phot}} =14.9 - 16.8$.

\textbf{GN+189.44234+62.19714} (ID 1204402, $z_a = 14.61$) This galaxy is an F150W dropout northwest of a galaxy (ID 1204399) at $z_a = 1.2$. While this redshift is consistent with the secondary minimum on the $\chi^2$ surface for the source, this is largely driven by excess flux in the aperture in the F150W band from the extended nearby source. This source lies in a relatively shallow region in the GOODS-N mosaic with data from both PANORAMIC and JWST PID 2674 (PI: Arrabal-Haro). 

\textbf{JADES-GN+189.16391+62.16618} (ID 1236026, $z_a = 14.68$) This F182M dropout is in a region of the GOODS-N mosaic with NIRCam LW coverage with F356W and F444W, where it is detected at SNR $> 6$. Owing to the lack of F150W photometric coverage for the source, the F182M non-detection is driving the photometric redshift, with the full 1$\sigma$ range on the redshift being $z_{\mathrm{phot}} = 9.7 - 14.7$. 

\textbf{JADES-GN+189.34537+62.22836} (ID 1256258, $z_a = 16.05$) This faint, compact source is an F162M or F182M dropout, and it is relatively isolated from bright nearby galaxies. At $z_a > 16$, it represents the farthest GOODS-N candidate in our primary sample. The UV slope we measure is very blue at $\beta = -3.12 \pm 0.78$, although this is likely biased by the faint F444W flux. 

\subsection{The $z > 16$ Candidate JADES-GN+189.15982+62.28899}\label{sec:ID_1019411}

\begin{figure*}[]
  \centering
  \includegraphics[width=0.44\linewidth]{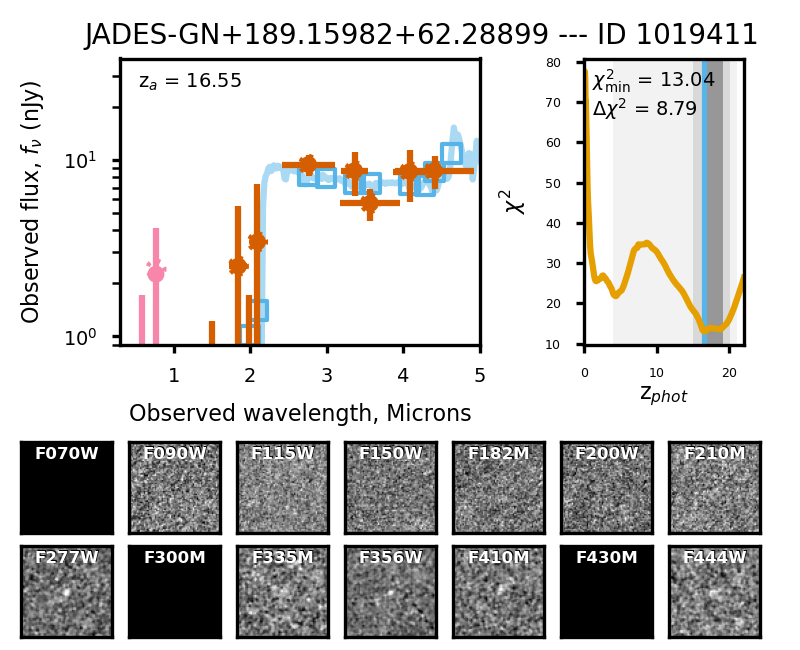}\
  \includegraphics[width=0.55\linewidth]{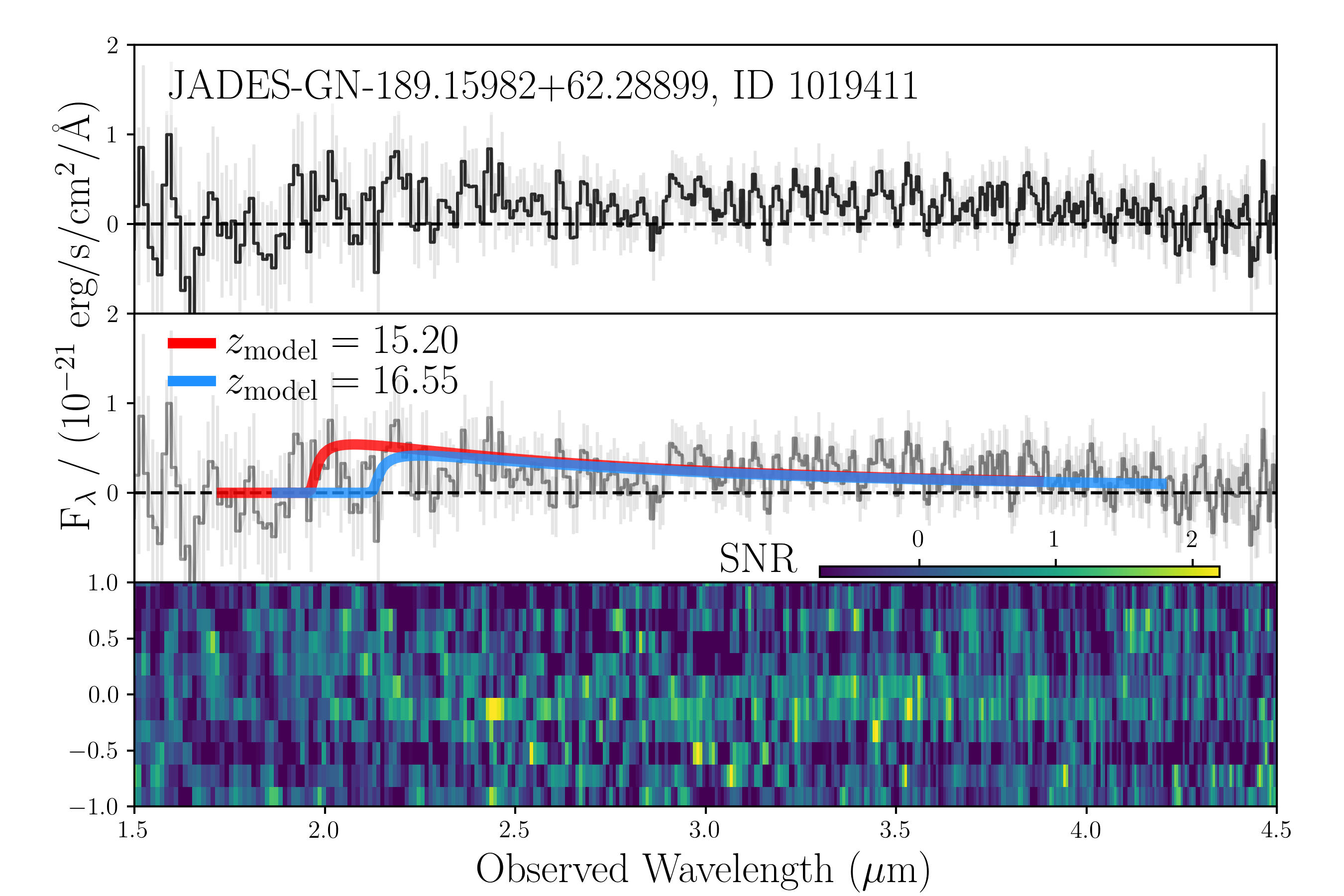}\  
  \caption{(Left) The SED of GOODS-N ID 1019411, the highest-redshift candidate in \KHZ{}, with the colors, lines, and symbols the same as in Figure \ref{fig:gs_high_z_galaxies}. (Right) The 9ksec JADES ``mediumjwst'' program NIRSpec prism spectrum of the source, in black, with the 1$\sigma$ uncertainties plotted in grey in the top panel. The 1D spectrum shows low significance flux between 2.4 - 5$\mu$m in agreement with the SED. In the middle panel, we plot a simple fit with UV slope $\beta = -2.48$ attenuated by a x$_{\mathrm{HI}}$=1 neutral damped IGM GP trough at a redshift of $z_{\mathrm{model}} = 15.20$ with a red line. For comparison, we plot a model spectrum at $z = 16.55$, the \texttt{EAZY} photometric redshift, with a blue line, which better agrees with the observed spectrum. In the bottom panel we plot the 2D spectrum SNR for the source. 
  }
  \label{fig:ID_1019411_SED_spectrum}
\end{figure*}

The highest-redshift candidate in \KHZ{}, JADES-GN+189.15982+62.28899 (JADES ID 1019411), does not appear in our primary sample, as it does not satisfy the SNR requirement with the updated photometry in JADES DR5. We plot the SED for the source in the left panel of Figure \ref{fig:ID_1019411_SED_spectrum}, where the CIRC1 F200W flux is $0.8 \pm 0.9$ nJy (SNR = 0.8), the F277W flux is $9.4 \pm 1.3$ nJy (SNR = 7.2) and $f_{\mathrm{F200W}}/ f_{\mathrm{F277W}} = 0.085 \pm 0.095$, demonstrating the strength of the observed break in the observed SED at 2$\mu$m. We fit the source with \texttt{EAZY}, where we adopted the CGM absorption prescription from \citet{asada2025}, as discussed in Section \ref{sec:spec_z_vs_phot_z}, and the resulting $z_a = 16.55$ (a 1$\sigma$ uncertainty range of $z_{\mathrm{phot}} = 16.4 - 19.0$). We plot this fit in the top left panel, and the $\chi^2$ surface in the top right panel of Figure \ref{fig:ID_1019411_SED_spectrum}. From this redshift, we estimate $M_{\mathrm{UV}} = -19.22 \pm 0.18$, and $\beta = -1.59 \pm 0.56$, although this slope is uncertain owing to the noisy photometry. We fit the NIRCam photometry for the source using the stellar population inference code \texttt{Prospector} \citep{johnson2021}, and recover $z_{\mathrm{phot}} = 16.4^{+0.6}_{-0.4}$, in agreement with the \texttt{EAZY} redshift. 

With possible F200W dropouts, there are a number of ways in which these sources can be low-redshift interlopers mimicking the SED of a high-redshift galaxy. Strong optical emission lines in dusty galaxies at $z = 4 - 5$ have been observed to boost the flux in multiple bands, which, along with a strong Balmer break, can produce an observed photometric SED that can be mistaken for an ultra-high-redshift dropout \citep{naidu2022, arrabalharo2023, weibel2025}. Low-mass quiescent galaxies at $z \sim 2 - 5$ \citep{sandles2023, baker2025b} can also have SEDs that can be mistaken for high-redshift galaxies. Quiescence within low-mass galaxies is often associated with environmental effects \citep[e.g., ][]{alberts2024b}. Related to this, it is important for us to note that GOODS-N contains two of the richest spectroscopically confirmed overdensities yet discovered with JWST \citep{lin2025}. These two protoclusters are at $z \approx 4.4$ and $z \approx 5.2$, hosting $98$ and $144$ $\mathrm{H}\alpha$ emitting galaxies, respectively. These structures are known to contain numerous dusty star-forming galaxies \citep[e.g., HDF850.1; ][]{sunf2024}, Balmer break galaxies \citep[e.g., ][]{helton2024}, and AGNs/LRDs \citep{lin2026}. 

Another possible interloper type for high-redshift galaxies are transients, like potential supernovae, as discussed in \citet{decoursey2025} and \citet{castellano2025}. A few sources in \KHZ{} (AT2022aews, known as JADES-GS+53.12692-27.7910, and AT2022aevk, known as JADES-GS+53.20055-27.784) were found to be transient sources in \citet{decoursey2025}, and further photometric observations showed that they had faded from their first detection. The spectrum for JADES-GN+189.15982+62.28899 was observed on 2023-05-05, and the imaging for the source was only from 2023-02, only three months prior, likely not enough time to fully rule out a transient event as the source of the object. 

JADES-GN+189.15982+62.28899 was observed for $\approx 9$ ksec with the JWST/NIRSpec prism as part of the JADES ``\texttt{medium\_jwst}'' program, and released in the recent JADES DR4 spectroscopic release \citep[see][ for further information about this program, as well as the data reduction]{curtislake2025, scholtz2025}. We plot this spectrum in the right panel of Figure \ref{fig:ID_1019411_SED_spectrum}. The spectrum is quite noisy, but at very low significance we observe flux concurrent with what would be expected from the SED: a lack of flux at $< 2.0 \mu$m, and a flat spectrum that fades to higher wavelengths, where the detector is less sensitive. 

To further explore this spectrum, we used a similar analysis as presented in Appendix A of \citet{hainline2024c}, where we attempt to statistically explore the spectrum to determine a redshift, and find that the best-fitting redshift (at low significance) is $z \sim 15.2$. We plot a fit of the form $f_\lambda \propto \lambda^\beta$ at this redshift in the right, middle panel of Figure \ref{fig:ID_1019411_SED_spectrum} with a red line, with a slope $\beta = -2.48$ and with $x_{\mathrm{HI}} = 1$ neutral damped intergalactic medium absorption to guide the eye. The exact redshift is highly dependent on the significance of the flux at 2.0 - 2.4$\mu$m, in agreement with the photometric solution. We consider this result to be too low of a significance to merit claiming a true spectroscopic redshift has been measured, and the fit predicts flux in excess of what is observed in the F200W photometry. To help explore alternate redshift solutions, we additionally plot a model at $z = 16.55$ (the photometric redshift from \texttt{EAZY} when including the CGM DLA prescription from \citet{asada2025}) with a blue line. This model spectrum is in excellent agreement to the low SNR spectrum.

From the observed spectrum, we measure 3$\sigma$ upper limits on potential line fluxes of $< 5\times10^{-19}$ erg s$^{-1}$ cm$^{-2}$ at $2\mu$m $< \lambda < 3\mu$m, and $< 3\times10^{-19}$ erg s$^{-1}$ cm$^{-2}$ at $\lambda > 3\mu$m. These limits provide evidence against the possibility that the source is at low redshift (e.g. $z = 3 - 5$); if we adopt an observed continuum flux of 7 nJy at 2.7 $\mu$m for the source from the SED, this flux limit would correspond to only a boost of $\sim$ 1.8 nJy ($\Delta \mathrm{mag} < 0.25$). From the spectrum, it is unlikely that the observed NIRCam photometric fluxes are boosted by strong emission lines like H$\alpha$ or the [\ion{O}{3}]$\lambda4959,5007$ + H$\beta$. For comparison, the spectrum for the $z = 4.9$ source CEERS 93316 presented in \citet{arrabalharo2023} included an [\ion{O}{3}]$\lambda5007$ emission line with a flux of $75.5\times10^{-19}$ erg s$^{-1}$ cm$^{-2}$ and H$\alpha$ with flux $53.8\times10^{-19}$ erg s$^{-1}$ cm$^{-2}$, lines which boosted the NIRCam F277W, F356W, F410M, and F444W fluxes and mimicked the SED of a $z \sim 16$ galaxy. 

One alternative is presented by comparing JADES-GN+189.15982+62.28899 to CEERS\_15937 in \citet{castellano2025}, a source with a Balmer break and no emission lines at $z_{\mathrm{spec}} = 4.91$. The spectrum presented for CEERS\_15937 in \citet{castellano2025} is similar to what we observe in the right panel of Figure \ref{fig:ID_1019411_SED_spectrum}, although the Balmer break in CEERS\_15937 is not nearly as sharp as what is seen in the spectrum. The extremely red $z \sim 5$ quiescent sources in \citet{baker2025b} have similar spectra, although with more significant flux at $< 2$ microns than JADES-GN+189.15982+62.28899. In Eisenstein et al. (in prep), the authors look at the full sample of JADES sources with strong Balmer breaks to explore the frequency of this type of interloper as a function of observational depth. 

Given the observed flux and limited exposure time for the JADES-GN+189.15982+62.28899 spectroscopy, it is difficult to make any concrete claims about the source and its spectroscopic redshift. Clearly, given the importance in finding objects at $z > 14.5$ for our understanding of Cosmic Dawn, it would be of interest to further observe this intriguing source. As shown in Section \ref{sec:highest_redshift_sources}, we find numerous $z > 14$ sources within GOODS-N that would be worth focusing on with a targeted NIRSpec MSA campaign. 

\subsection{Notable Rejected Candidates in GOODS-S and GOODS-N}\label{sec:notable_rejected_candidates}

\begin{figure*}[t!]
  \centering
  \includegraphics[width=0.48\linewidth]{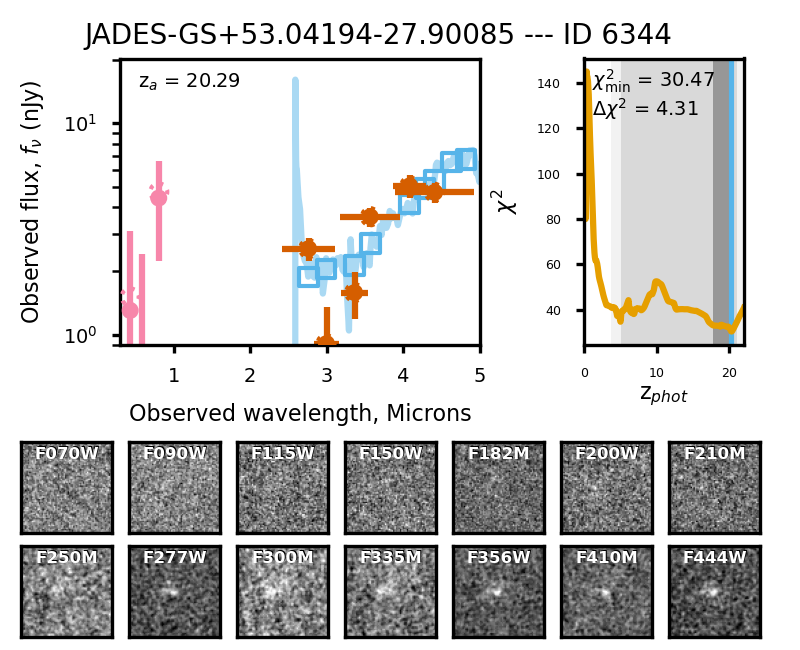}\
  \includegraphics[width=0.48\linewidth]{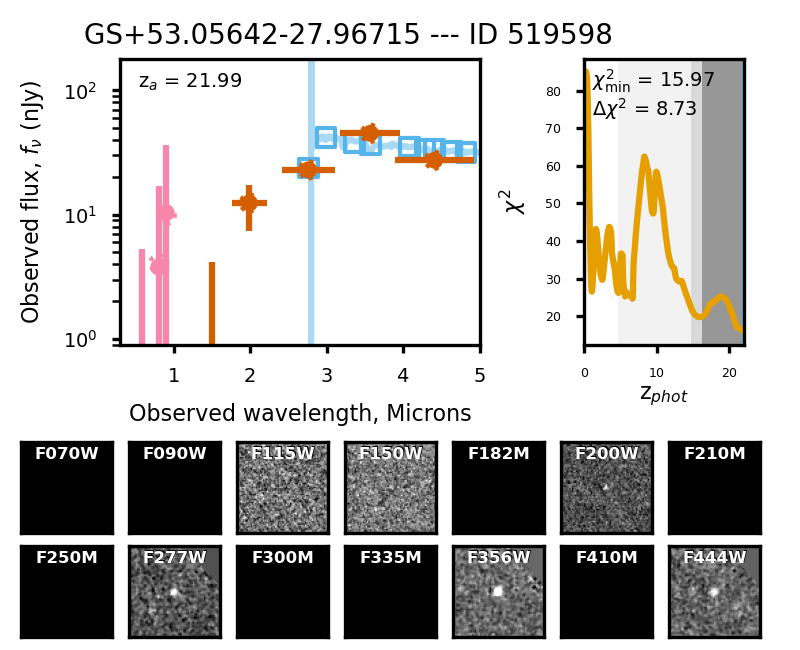}\
  \includegraphics[width=0.48\linewidth]{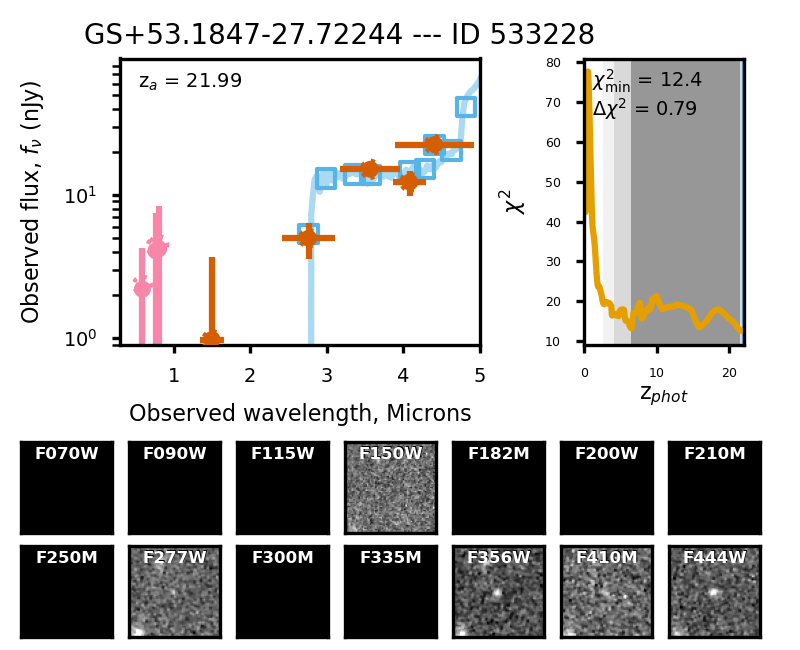}\
  \includegraphics[width=0.48\linewidth]{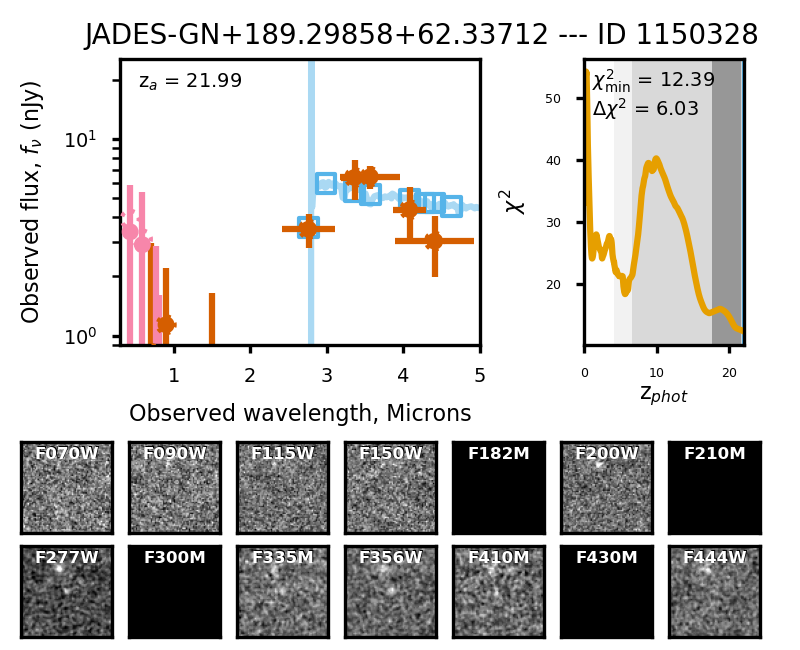}\
  \caption{SEDs, \texttt{EAZY} $\chi^2$ surfaces, and JWST/NIRCam thumbnails for the GOODS-S and GOODS-N sources that were rejected from the primary sample, but have notable SEDs. The panels and lines are the same as in Figure \ref{fig:gs_high_z_galaxies}.}
  \label{fig:rejected_candidates}
\end{figure*}

During the visual inspection of the $z > 8$ candidates discussed in Section \ref{sec:selection}, we found four sources across the GOODS-S (three objects) and GOODS-N (one object) that, while we rejected them from our primary sample, we wanted to highlight given their interesting SEDs. We plot their SEDS, thumbnails, and \texttt{EAZY} fits in Figure \ref{fig:rejected_candidates}. All four sources are isolated and separate from brighter, low-redshift galaxies that we might associate with each object. We will discuss each of the four in turn in this section. 

\textbf{JADES-GS+53.04194-27.90085} (ID 6344, $z_a = 20.29$) This source is observed at SNR $>$ 5 in F277W, F356W, and F444W and is quite red, with no observed flux at $\lambda < 2.5\mu$m. The red color is not consistent with what would be expected from a galaxy at this redshift, and the low F300M ($0.9 \pm 0.4$ nJy) and F335M ($1.6 \pm 0.4$ nJy) fluxes additionally cast doubt on the fit for this source. While there is a star (JADES ID 159416) within $5^{\prime\prime}$ of the source, this source does not coincide with a diffraction artifact from that object. The stark difference in fluxes between F300M, F335M and the F277W, F356W, F410M, and F444W filters could indicate that this is a transient source. 

\textbf{GS+53.05642-27.96715} (ID 519598, $z_a = 21.99$) The minimum $\chi^2$ solution from \texttt{EAZY} for this object does not take into account the F200W detection (SNR = 2.5). This object sits at the edge of the mosaic near the southern portion of the full assembled GOODS-S footprint, and the significant F277W (SNR = 10.6), and F356W (SNR = 22.0) detections for the object supports the observed turnover, likely indicating that the object is at lower redshift with emission line boosting.

\textbf{GS+53.1847-27.72244} (ID 533228, $z_a = 21.99$) This object is only detected in the NIRCam LW filters, and as it sits in a gap in the mosaic, there is only one relatively shallow NIRCam SW filter image covering the galaxy, in F150W (SNR = 0.36). The redshift is primarily driven by the F277W flux ($5.0 \pm 1.4$ nJy) and the red $f_{\mathrm{F356W}} / f_{\mathrm{F277W}} = 1.0 \pm 0.3$. The F444W flux ($22.5 \pm 2.0$ nJy) also raises suspicion. 

\textbf{JADES-GN+189.29858+62.33712} (ID 1150328, $z_a = 21.99$) This object is quite faint, and again, the photometric redshift is driven by the F277W photometry ($3.5 \pm 0.7$ nJy). The F444W flux (SNR = 2.88) for this source, if it were at this redshift, would perhaps indicate an extremely blue UV slope.

\subsection{UV Slope Evolution}\label{sec:uv_slopes}

The UV slope of a galaxy provides insight into both recent star formation, nebular emission, and the properties of dust in galaxies, and as such, exploring its evolution for the first galaxies is of significant importance. In \citet{topping2024}, the authors explored the evolution of the UV slope $\beta$ with $M_{\mathrm{UV}}$ for dropout-selected galaxies in an early JADES release (only using data taken between 2022 October and 2023 February) across a smaller area and with less observational depth than what we have now with JADES DR5. With our updated sample, we can explore the evolution with higher significance, especially at $z > 11$, where the \citeauthor{topping2024} sample only had 24 sources ($M_{\mathrm{UV, median}} = -18.61^{+0.81}_{-0.91}$), while our sample includes 254 sources at $z > 11$, across a wider range of $M_{\mathrm{UV}}$. As a reminder, both $M_{\mathrm{UV}}$ and $\beta$ were estimated from photometry measured using mosaics convolved to the F444W PSF to avoid effects with a changing PSF as a function of wavelength. 

\begin{figure*}[t!]
  \centering
  \includegraphics[width=0.95\linewidth]{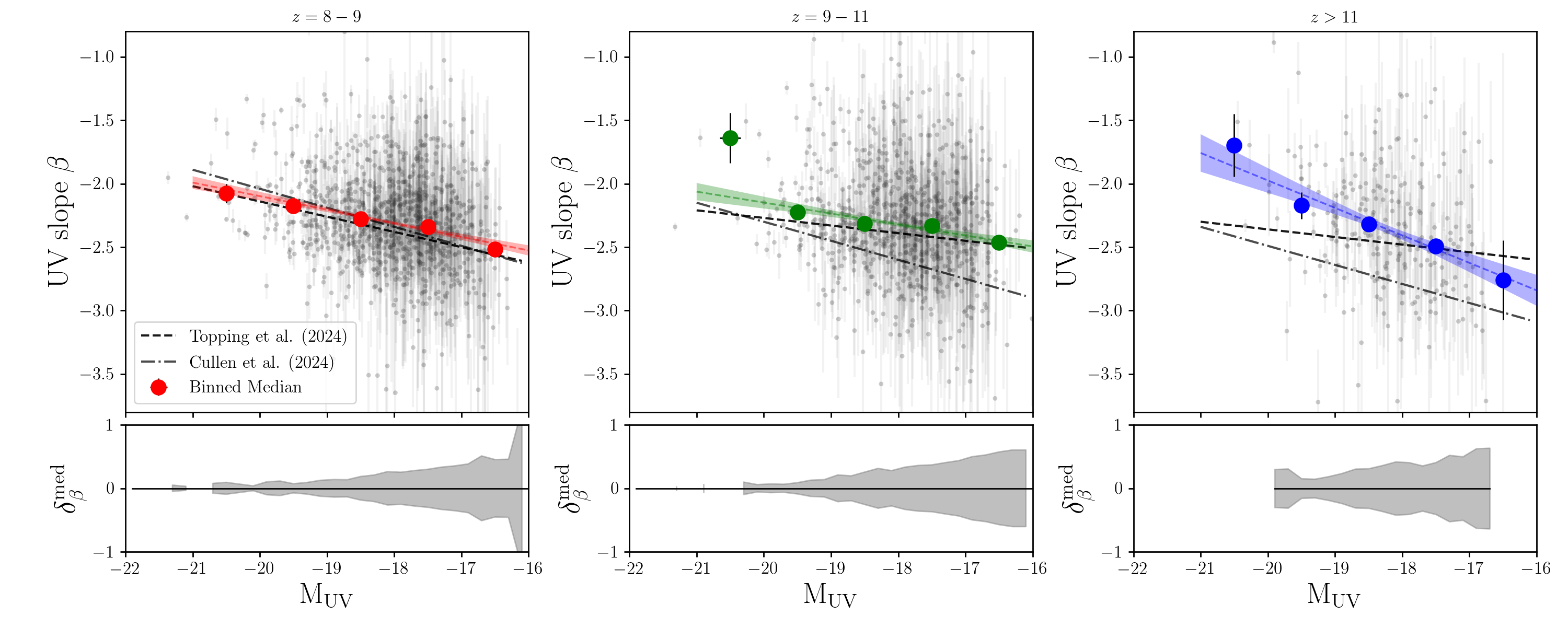}\
  \caption{Evolution of the UV slope $\beta$ vs. $M_{\mathrm{UV}}$ for our high-redshift sources in three bins, (Left) $z = 8 - 9$, (Middle) $z = 9 - 11$, and (Right) $z > 11$. We measure $\beta$ following the method outlined in \citet{topping2024}, and present the individual measurements with grey points, with the median uncertainty on $\beta$ shown in grey in the bottom panel for each plot. We plot a binned median with larger colored points, and a linear fit to each set of median values with a colored line, with the shaded region representing the 1$\sigma$ uncertainty from the fit. The binned values are provided in Table \ref{tab:median_uv_slopes}, and the evolutionary fit parameters are provided in Table \ref{tab:best_fit_slope_evolution_parameters}. In all three panels, we additionally include the results for similar redshift ranges for the samples presented in \citet{topping2024} (dashed line) and \citet{cullen2024} (dot-dashed line) for comparison with our own sample of high-redshift galaxies.
  }
  \label{fig:beta_vs_muv}
\end{figure*}

\begin{deluxetable}{lcccc}
\tabletypesize{\footnotesize}
\tablecolumns{8}
\tablewidth{0pt}
\tablecaption{Median UV Slopes For Our Sources Binned by Redshift and UV Magnitude \label{tab:median_uv_slopes}}
\tablehead{
\colhead{Redshift Limits} &  \colhead{N}  &   \colhead{$z_{\mathrm{med}}$}  &  \colhead{$M_{\mathrm{UV}}^{\mathrm{med}}$} & \colhead{$\beta_{\mathrm{med}}$}}
\startdata
$z = 8 - 9$ &  19 &  8.55 & $-20.22 \pm 0.05$ & $-2.08 \pm 0.07$\\
            & 132 &  8.44 & $-19.29 \pm 0.02$ & $-2.17 \pm 0.03$\\
            & 338 &  8.50 & $-18.40 \pm 0.02$ & $-2.28 \pm 0.02$\\
            & 412 &  8.55 & $-17.56 \pm 0.01$ & $-2.34 \pm 0.02$\\
            &  90 &  8.54 & $-16.78 \pm 0.02$ & $-2.52 \pm 0.06$\\
$z = 9 - 11$ &  5 &  9.45 & $-20.27 \pm 0.16$ & $-1.64 \pm 0.20$\\
            &  81 &  9.46 & $-19.27 \pm 0.03$ & $-2.23 \pm 0.04$\\
            & 248 &  9.55 & $-18.41 \pm 0.02$ & $-2.31 \pm 0.03$\\
            & 345 &  9.46 & $-17.56 \pm 0.02$ & $-2.33 \pm 0.03$\\
            &  82 &  9.56 & $-16.83 \pm 0.02$ & $-2.46 \pm 0.05$\\
   $z > 11$ &   3 & 12.13 & $-20.34 \pm 0.05$ & $-1.70 \pm 0.25$\\
            &  33 & 11.48 & $-19.38 \pm 0.05$ & $-2.17 \pm 0.11$\\
            & 111 & 11.75 & $-18.41 \pm 0.03$ & $-2.32 \pm 0.04$\\
            &  85 & 11.77 & $-17.57 \pm 0.03$ & $-2.49 \pm 0.06$\\
            &  14 & 11.46 & $-16.84 \pm 0.04$ & $-2.76 \pm 0.31$\\
\enddata
\end{deluxetable}

We split our sources into three bins of redshift: $8 < z <  9$, $9 < z < 11$, and $z > 11$, and for each redshift interval, we binned the sources by $M_{\mathrm{UV}}$ (with $\Delta M_{\mathrm{UV}} = 1.0$) and calculated the median $M_{\mathrm{UV}}$ values in each bin, along with the standard deviation of the mean for the sources in each bin. We provide these values in Table \ref{tab:median_uv_slopes}. We then performed a linear fit to the median $M_{\mathrm{UV}}$ values as a function of $\beta$ in each redshift bin. 

In Figure \ref{fig:beta_vs_muv}, we plot the sources in each redshift bin, the median uncertainties on $\beta$, and the fits to the binned median $\beta$ values. In Table \ref{tab:best_fit_slope_evolution_parameters}, we report the median redshifts, best-fit slopes and $\beta_{0}$ values (the value for $\beta$ at $M_{\mathrm{UV}} = -19$ for each fit) for our redshift bins. There is a large spread in uncertainties on both $\beta$ (shown with error bars), and $M_{\mathrm{UV}}$ (shown in the bottom panels), with the uncertainty increasing at higher redshifts. We see a gradual steepening of the slope of the $M_{\mathrm{UV}}$ vs. $\beta$ relationship, most significantly observed in the $z > 11$ bin, with consistent values for $\beta_{0}$ within the uncertainties. Notably, we measure a significantly non-zero slope in each redshift bin, in tension with other surveys which primarily sample brighter sources ($M_{\mathrm{UV}} < -18$), such as those in the JADES spectroscopic sample from \citet{saxena2024} and the photometrically selected sources in the NIRCam pure-parallel BEACON sample from \citet{zhang2026}. Indeed, it is the UV faint sources at $M_{\mathrm{UV}} > -18$ that are systematically bluer in our sample.

\begin{deluxetable}{lccc}
\tabletypesize{\footnotesize}
\tablecolumns{8}
\tablewidth{0pt}
\tablecaption{Best-fit Linear Relation Showing UV Slope Evolution \label{tab:best_fit_slope_evolution_parameters}}
\tablehead{
\colhead{Redshift Limits} &  \colhead{$z_{\mathrm{med}}$}  &  \colhead{d$\beta$/dM$_{\mathrm{UV}}$} & \colhead{$\beta_{0}$}}
\startdata
$z = 8 - 9$  & 8.51 & $-0.10 \pm 0.02$ & $-2.21 \pm 0.02$\\
$z = 9 - 11$ & 9.48 & $-0.09 \pm 0.02$ & $-2.24 \pm 0.03$\\
    $z > 11$ & 11.68 & $-0.22 \pm 0.05$ & $-2.19 \pm 0.05$\\
\enddata
\end{deluxetable}

In \citet{topping2024}, the authors provide estimates of the evolution of the slope of the $M_{\mathrm{UV}}$ vs. $\beta$ relationship and the $\beta_0$ values for a sample of F090W dropouts at $6.5 \lesssim z \lesssim 8.5$, F115W dropouts at $8.5 \lesssim z \lesssim 11$, and F150W dropouts at $11 \lesssim z \lesssim 14$. For the F090W dropouts, our estimated $M_{\mathrm{UV}}$ vs. $\beta$ slope and $\beta_0$ values are in agreement with the \citet{topping2024} values, although we note that there is likely a larger number of sources at $z < 8$ in that study compared to our own due to the use of dropout selection. For the F115W dropouts, the $M_{\mathrm{UV}}$ vs.\ $\beta$ slopes agree, although the normalization, $\beta_0$, is larger in our fits than what is provided in \citeauthor{topping2024}. Given that the F150W dropout sample in \citet{topping2024} spans one $M_{\mathrm{UV}}$ bin, the authors of this study fix the $M_{\mathrm{UV}}$ vs.\  $\beta$ slope to what they measure for the F115W dropouts. We observe a significantly steeper $M_{\mathrm{UV}}$ vs.\ $\beta$ slope, driven by the sources at $-18 < M_{\mathrm{UV}} < -17$. 

In \citet{cullen2024}, the authors assemble a sample of 172 galaxies at $8 < z < 16$ from multiple JWST NIRCam imaging and COSMOS/UltraVISTA ground-based near-infrared surveys. From this sample, they measure $\beta$ by modeling each SED in a Bayesian framework that includes all of the available photometry and accounts for potential neutral hydrogen in the IGM. In the paper, they confirmed that their results are comparable to the method we employ that was also used in \citet{topping2024}, albeit with a median offset of $\Delta \beta < +0.04$. Our results for the evolution of the $M_{\mathrm{UV}}$ vs $\beta$ relation differ in that they find a slightly steeper slope ($d \beta / M_{\mathrm{UV}} = -0.15 \pm 0.03$) at $7.5 < z < 10$ than what we calculate for our sources, but in agreement within the uncertainties. They adopt this $M_{\mathrm{UV}}$ vs.\ $\beta$ slope for their redshift bins at $z > 10$, and while we find agreement at $z > 11$, the $M_{\mathrm{UV}}$ vs. $\beta$ slope at $z = 9 - 11$ is slightly flatter in our sample than what they observe. 

The largest discrepancy between our results and what is presented in \citet{cullen2024} arises by comparing $\beta_0$. These authors find a strong redshift evolution of $\beta_0$ that we do not observe, reaching $-2.6$ at $11 < z < 12$. It should be noted that our sample includes a significant population of sources with $M_{\mathrm{UV}} > -18$ that are not seen in the \citeauthor{cullen2024} sample, especially at $z > 10$, and these faint sources we observe at these redshifts have a range of UV slopes which raise the $\beta_0$ values we estimate from our fits. As can be seen from the error bars on $\beta$ plotted in the figure, there are significant measurement uncertainties for these sources, especially given that we are fitting only three photometric bands. While photometric redshift uncertainties will also play a part, they should be fairly robust given the large redshift bins we use. The \citeauthor{cullen2024} sample occupies a blue and very bright portion of $M_{\mathrm{UV}}$ vs.\ $\beta$ space not well-represented in our sample, likely arising because of their assembling sources from across a large sample of surveys, many significantly shallower than JADES, such that their highest redshift objects trace the population of more rare, UV bright galaxies that may not be found in the same numbers within the GOODS-S and GOODS-N footprints.

\subsection{Extended Source Morphologies}\label{sec:extended}

In \KHZ{}, the authors presented evidence that galaxies at $z > 8$ can have significantly extended and often clumpy morphologies \citep[but see, as a counter-example, the ultra-compact GN-z11,][]{tacchella2023}, suggesting these sources are growing across multiple regions of star formation that may be merging. The radial surface brightness profile for JADES-GS-z14-0 has flux out to 1 kpc from the galaxy center, and the source has a half-light radius of 260 pc \citep{carniani2024}. These results join a growing number of observations of high redshift galaxies being seen with clustered star formation, with estimates of the fraction of bright ($M_{\mathrm{UV}} \sim -22$) galaxies with clumps at $z \sim 7$ being as high as 66\% \citep{adamo2024, harikane2025, fujimoto2025}. These clumps, which may arise from galaxy mergers \citep{dimatteo2008} or violent disk instabilities \citep{dekel2009, dekel2013}, may allow for dense star formation that can punch holes in the galaxy ISM, allowing for the escape of ionizing photons \citep[see discussion in][]{gazagnes2025}.

To explore which sources have evidence for being extended, we looked at the NIRCam F277W thumbnails for each object (in the cases where F277W was not available, we used images with the F444W filter). We first masked nearby objects using the JADES DR5 GOODS-S and GOODS-N segmentation maps, making sure to exclude from the masking any object in our $z > 8$ candidate list given that some extended, clumpy sources in our catalog have IDs for individual knots, and we wanted to include them if they are part of the same larger galaxy structure. For each source in our catalog we then calculated a radial profile centered on the object, estimating the flux for the object in annuli one pixel wide which we used to interpolate and find the radial profile values at 0.10$^{\prime\prime}$, 0.15$^{\prime\prime}$, and 0.20$^{\prime\prime}$. We additionally performed this estimation on the F277W (and F444W) mosaic PSF (the derivation of this PSF is discussed in \JJADES{}) from which aperture corrections were estimated, so that we could calculate the ratio between the source radial profile and that derived from the PSF. To assess which sources are extended, we sought to find those where the source profile was significantly wider than the PSF out to 0.20$^{\prime\prime}$, so we flagged those sources where the ratio between the source profile and the PSF at 0.10$^{\prime\prime}$ was larger than 2, the ratio at 0.15$^{\prime\prime}$ was larger than 3, and the ratio at 0.20$^{\prime\prime}$ was larger than 4. Because this ratio could be noisy at low SNR, we also required a SNR in F277W greater than 5. For those sources without F277W coverage, we performed the same analysis on the F444W images, but given the difference in PSF between 2.7$\mu$m and 4.44$\mu$m, we update the minimum ratios to 1, 2, and 3 at 0.10$^{\prime\prime}$, 0.15$^{\prime\prime}$, and 0.20$^{\prime\prime}$, and we required the F444W SNR to be greater than 5. 

The result of this analysis indicates that $\sim$27\% of the sources in GOODS-S and $\sim 28$\% of the sources in GOODS-N have evidence for being extended. The fraction of morphologically extended sources in our sample only mildly decreases to higher redshift: 30\% at $z = 8 - 9$, 25\% at $z = 9 - 10$, 22\% at $z = 10 - 11$, and then 25\% at $z > 11$, although these sources are fainter, and so the actual fraction is more uncertain. The median $M_{\mathrm{UV}}$ for the extended sources ($M_{\mathrm{UV, median}} = -18.$2) is only slightly lower (i.e. brighter) than the full sample ($M_{\mathrm{UV, median}} = -18.0$), but with a broad standard deviation for each distribution ($\sim$0.8). The extended sources are only slightly redder, as measured by F200W $-$ F277W color, with $\Delta(m_{\mathrm{F200W}} - m_{\mathrm{F277W}}) = 0.06$, with a large spread in the values for both samples (the median F277W $-$ F444W color between the full sample and the extended subsample is the same, however). 

\begin{figure*}[]
  \centering
  \includegraphics[width=0.49\linewidth]{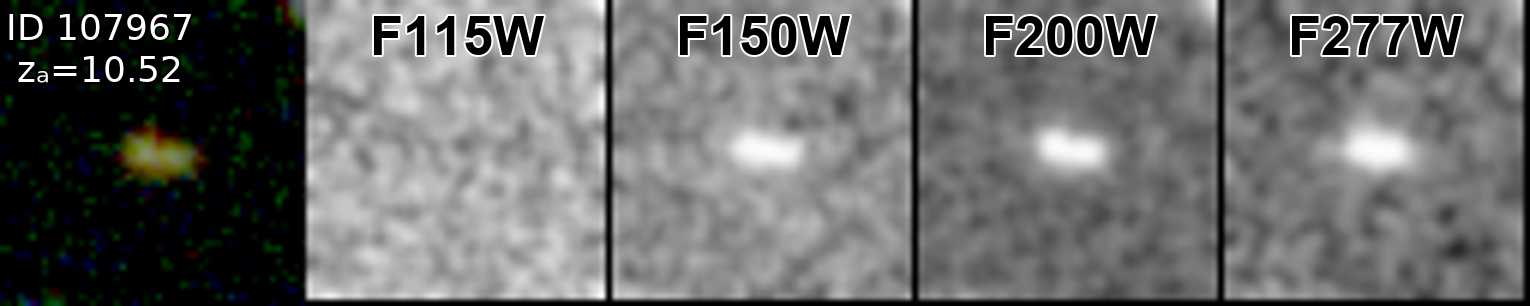}\  
  \includegraphics[width=0.49\linewidth]{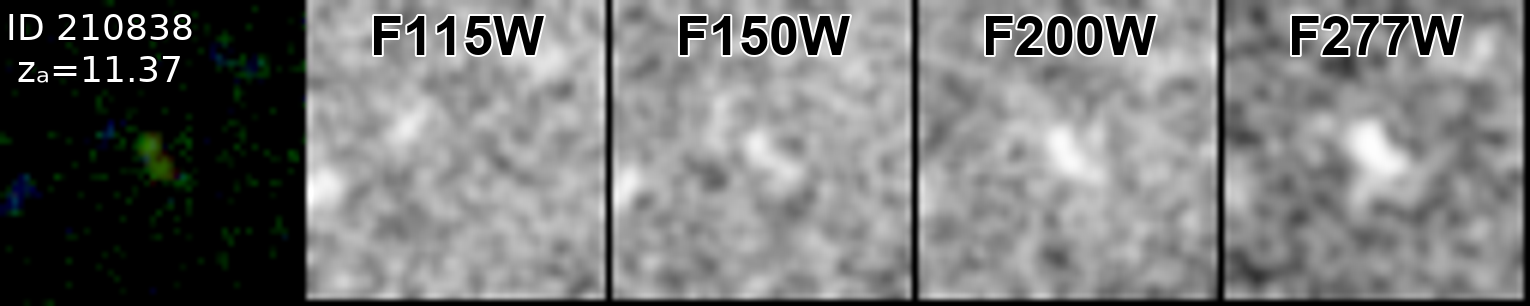}\
  \includegraphics[width=0.49\linewidth]{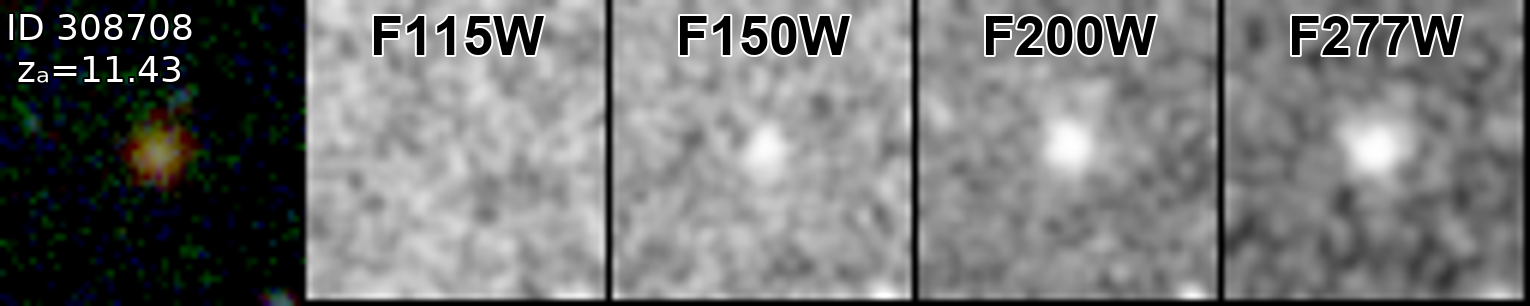}\  
  \includegraphics[width=0.49\linewidth]{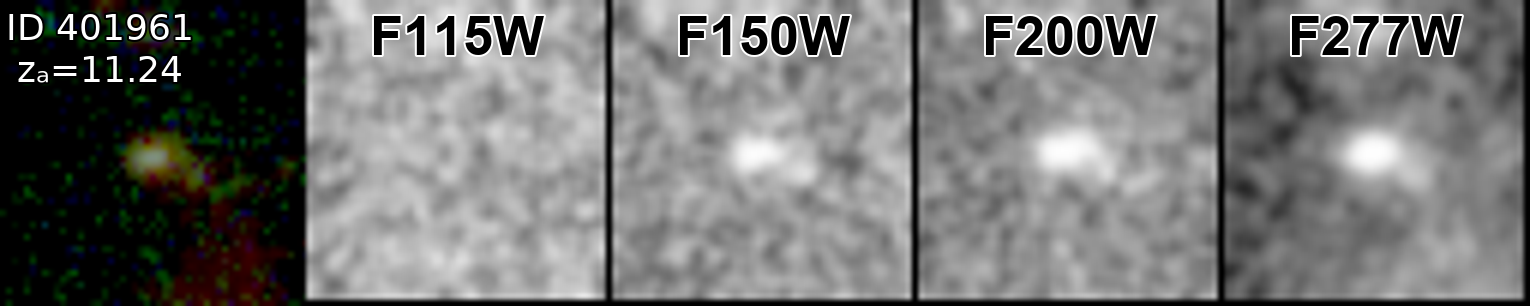}\  
  \includegraphics[width=0.49\linewidth]{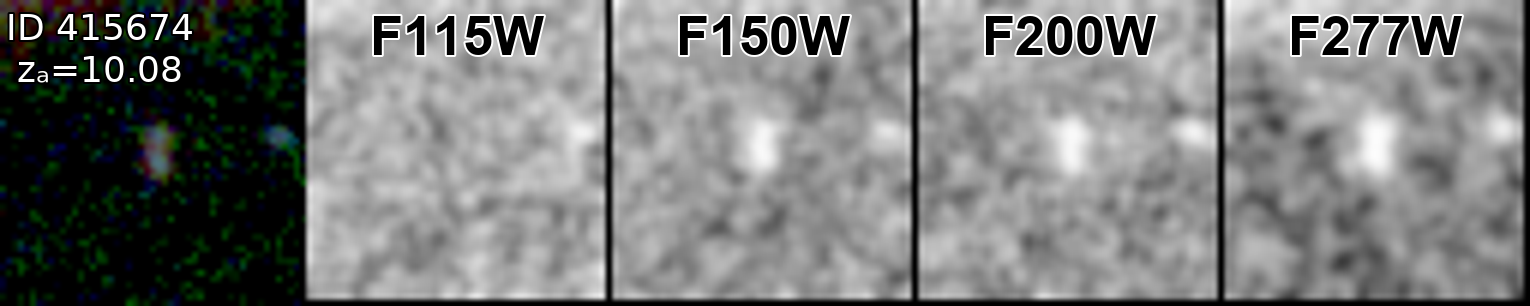}\   
  \includegraphics[width=0.49\linewidth]{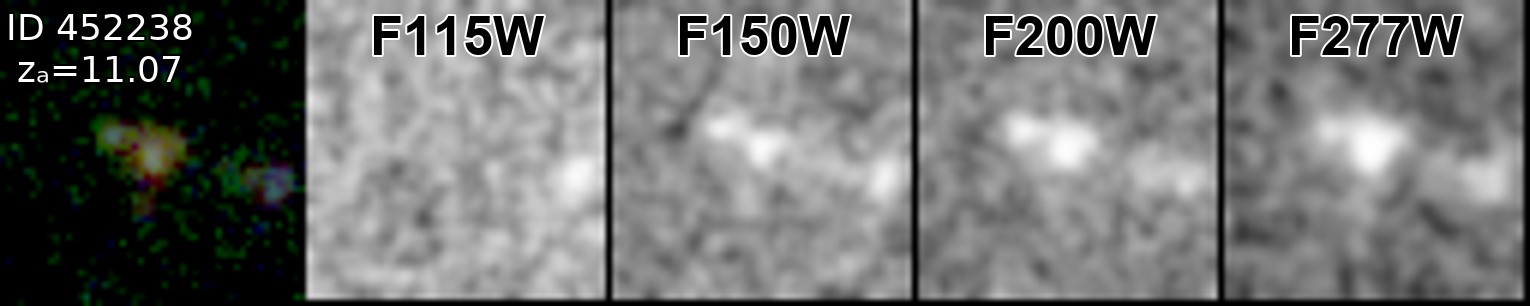}\   
  \includegraphics[width=0.49\linewidth]{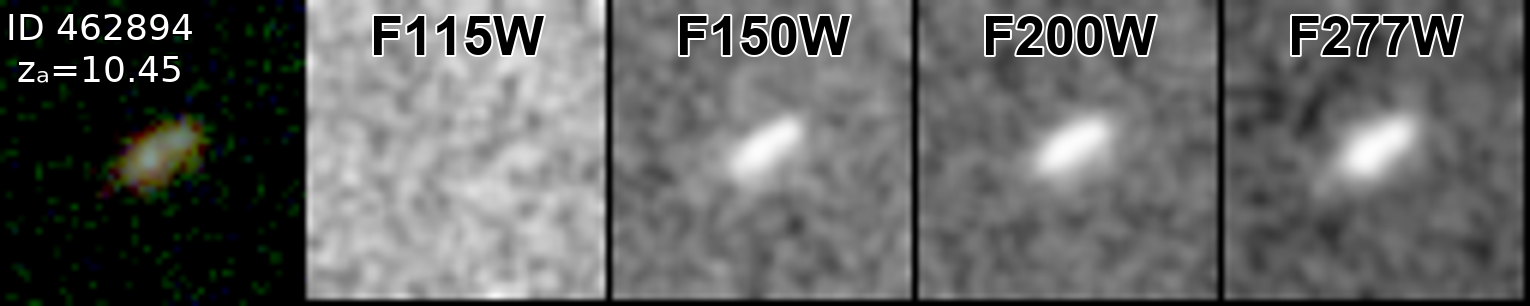}\  
  \includegraphics[width=0.49\linewidth]{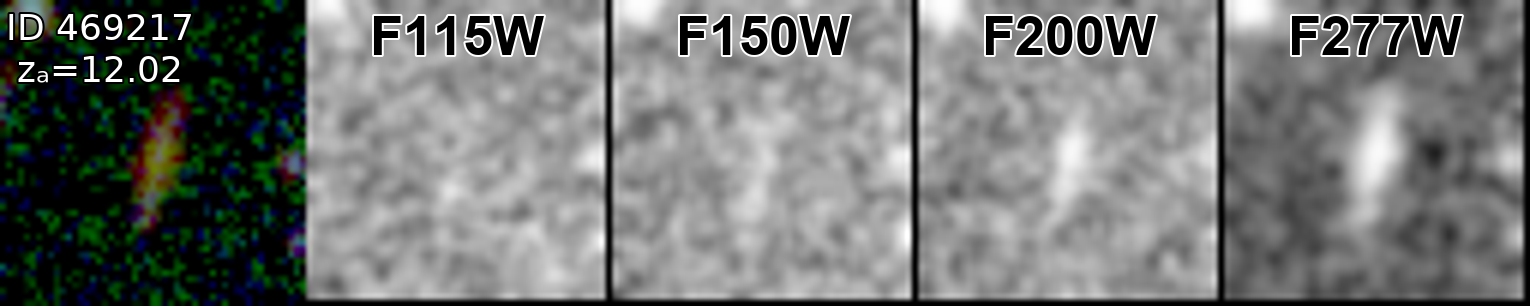}\  
  \includegraphics[width=0.49\linewidth]{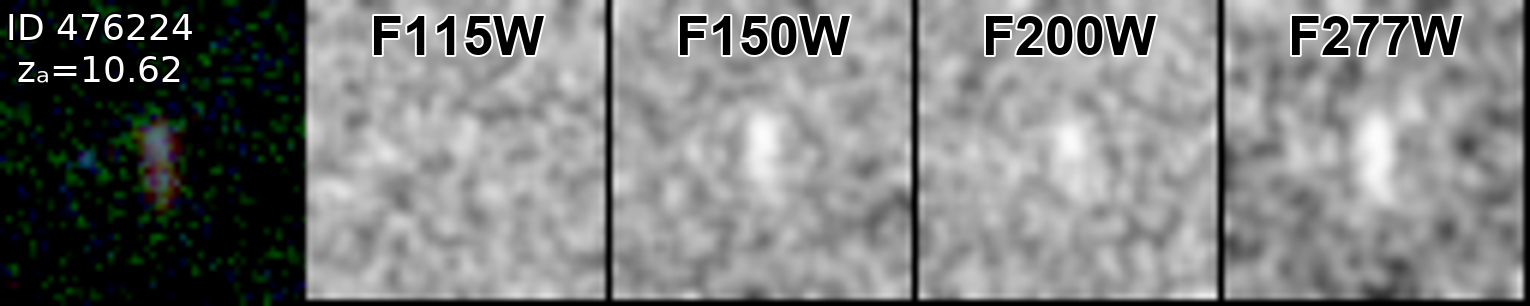}\  
  \includegraphics[width=0.49\linewidth]{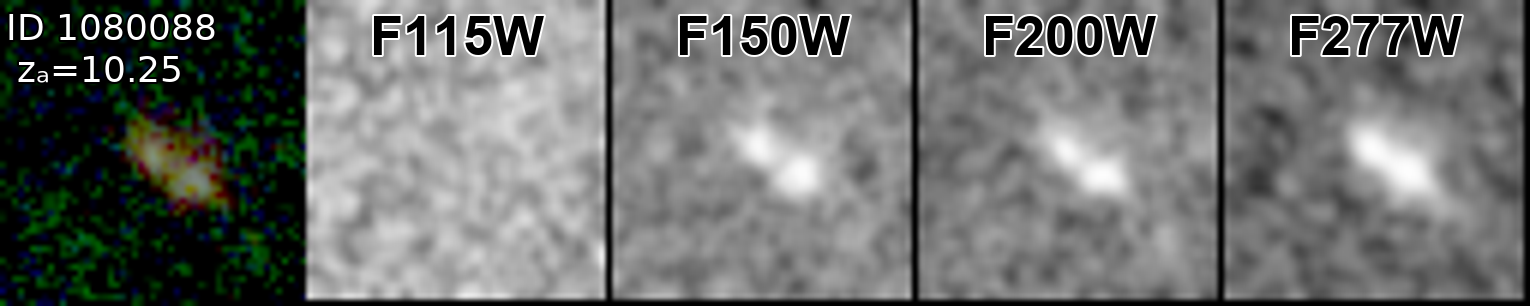}\  
  \includegraphics[width=0.49\linewidth]{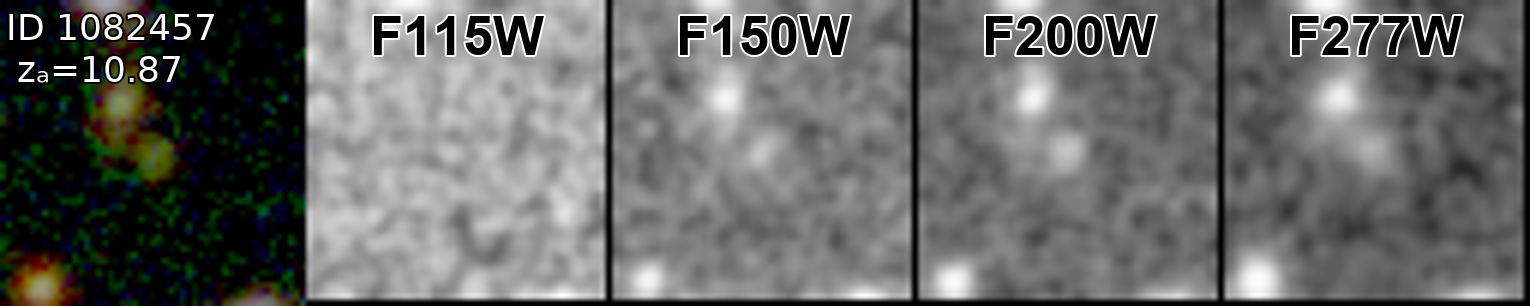}\  
  \includegraphics[width=0.49\linewidth]{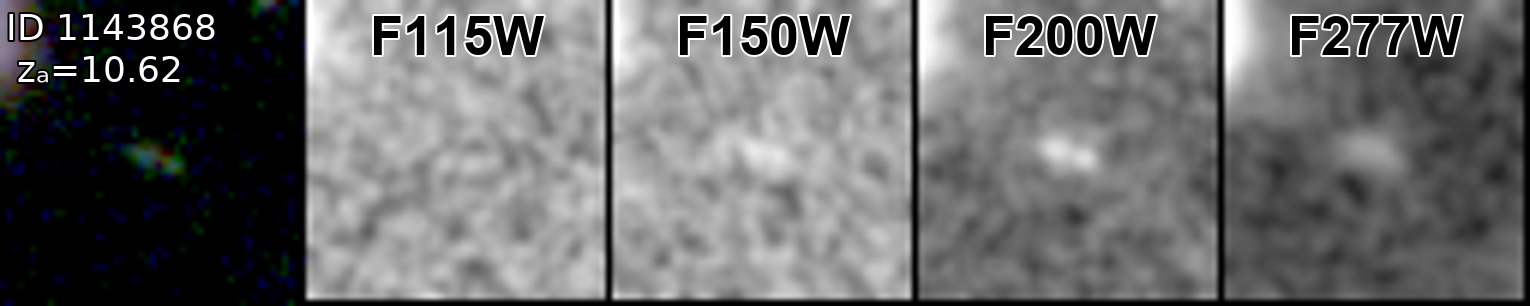}\  
  \includegraphics[width=0.49\linewidth]{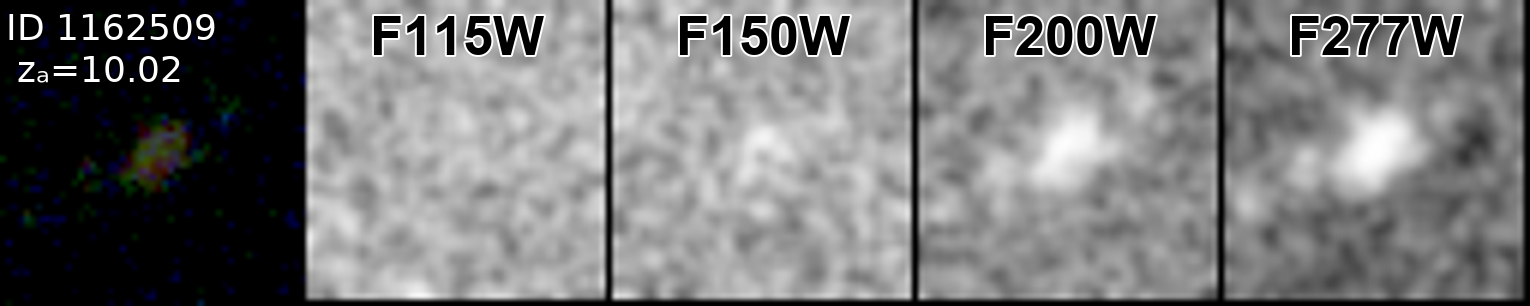}\
  \includegraphics[width=0.49\linewidth]{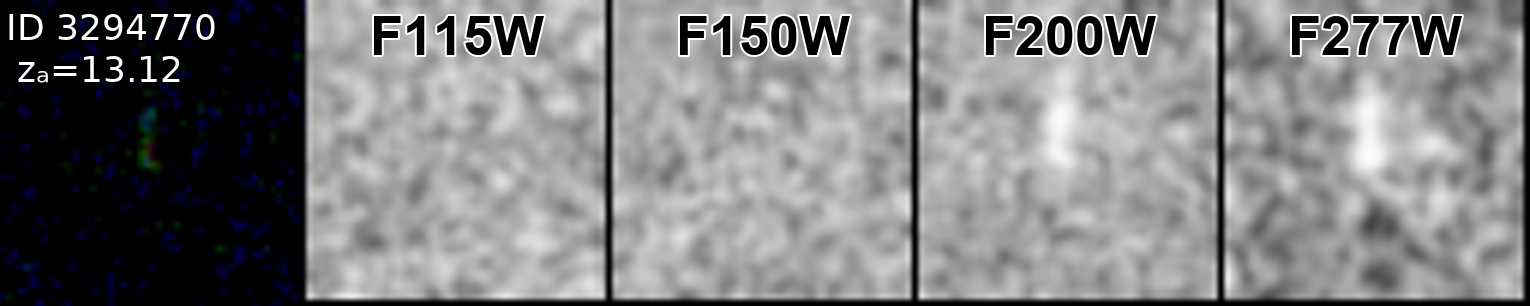}\
  \caption{Thumbnails for a set of 14 example extended sources at $z > 10$. Each thumbnail is $1.5^{\prime\prime} \times 1.5^{\prime\prime}$, which at $z = 10$ (12) corresponds to 6.4 kpc (5.6 kpc). For each galaxy, we plot, in the left-most column, a $1.5^{\prime\prime} \times 1.5^{\prime\prime}$ RGB thumbnail, constructed using multiple combined filters: in GOODS-S, the blue image is composed of the F090W through F182M NIRCam filters, the green image is composed of F200W and F210M, and the red image is the F277W through F460M filters, and in GOODS-N this is similar, but the red image is the F277W through F444W filters. In the other columns, we plot F115W, F150W, F200W, and F277W thumbnails for each source, at their native resolution for each filter. These galaxies demonstrate the morphological diversity for sources at $z > 10$ in our sample.}
  \label{fig:extended_thumbnails}
\end{figure*}

In Figure \ref{fig:extended_thumbnails} we plot the thumbnails for a subsample of 14 extended objects at $z_a > 10$ in our sample. These objects have distinct morphologies: multiple clump sources like GOODS-N 1082457, and compact, but hazy sources like the very bright GOODS-S 308708, which has a compact core surrounded by a more extended haze to the north. We see flattened morphologies for GOODS-S 469217 at $z_a = 12.02$ ($\sim$$0.6^{\prime\prime}$ in extent, or 2.2 kpc), 476224 at $z_a = 10.62$ ($\sim$$0.5^{\prime\prime}$ in extent, or 2.0 kpc), and perhaps most intriguing, GOODS-N 3294770 at $z_a = 13.12$ ($\sim$$0.4^{\prime\prime}$ in extent, or 1.4 kpc). These values agree with the large-effective-radius tails in the distributions of the sizes of Cosmic Dawn galaxies from multiple literature studies \citep[see][and others]{morishita2024, ormerod2024, allen2025, miller2025}.

We also highlight GOODS-S ID 462894 and 452238, which are separated by $\sim$$2^{\prime\prime}$, which at the redshift of the sources, is only $8 - 9$ kpc (projected). The eastern source, ID 452238, has a color gradient, being significantly redder in the larger western clump. Importantly, these sources are a part of the large $z \sim 10$ overdensity discussed in \citet{Wu2026}, which may represent the early formation of a massive galaxy cluster. This study finds that a full third (significantly higher than we see for the full population) of the sources within this overdensity have evidence for being extended, a fact they claim is indicative of frequent interactions in the overdense galaxy environment. While a full detailed morphological analysis of these sources is outside the scope of our current paper, in our final output table  we include the radial profile ratios used to select these sources, as well as a column marking which sources have been selected as being morphologically complex or extended. 

\section{Discussion}

\subsection{Implications for the Evolution of UV Slopes at $z > 8$}\label{sec:uv_slope_discussion}

While there has been a large number of high-redshift sources selected from multiple surveys with $\beta$ and $M_{\mathrm{UV}}$ measurements, there is still no consensus about the evolution of the UV slope at Cosmic Dawn. As shown in Section \ref{sec:uv_slopes}, we do not observe a significant change in $\beta_0$ with redshift, in contrast with other studies in the literature \citep[e.g.][]{cullen2024}. Our sample contains a significant number of sources at $M_{\mathrm{UV}} > -18$, which have a large scatter in their measured $\beta$ values (albeit with larger uncertainties). However, the median $\beta$ values are bluer than the UV-brighter galaxies, indicating that they are perhaps genuinely dust-poor, low-mass systems, or they are caught in a brief, high specific-star-formation rate (sSFR) phase where the UV is being dominated by very young stars. Blue UV slopes ($\beta < -2.6$) are theoretically possible for sources with stellar ages less than 10 Myr \citep{topping2024, cullen2024, saxena2024}, although any nebular continuum will serve to redden these slopes, and may contribute to the observed scatter. For sources with such blue slopes, a weak nebular continuum may indicate high gas temperatures or a lack of nebular gas entirely in these sources. \citet{topping2024} and \citet{cullen2024} also discuss how the most extreme blue $\beta$ values may imply a non-zero escape fraction for these sources, as theorized for the $z \sim 13$ spectroscopically-confirmed galaxy JADES-GS-z13-0, where $\beta = -2.69$ \citep{hainline2024c}. 

In \citet{narayanan2025}, the authors explore the mechanisms that drive UV slope at $z > 8$ in cosmological zoom-in simulations. The conclusions they draw from their simulated galaxies can help to understand the trends observed in Figure \ref{fig:beta_vs_muv}. In the absence of nebular continuum, bursty star-formation leads to a large range in dust-unreddened UV slopes ($\beta \sim -3$ to $-2.2$). The \citet{narayanan2025} simulations indicate that the impact of the nebular continuum is only a reddening of the UV slope by $\Delta \beta = 0.2 - 0.4$. Alternately, they find that dust obscuration plays a significant role for sources at $z = 8 - 10$ owing to an increase in dust grain-grain shattering leading to smaller dust grains, which is more pronounced in more massive sources. This scenario could be what is pushing $d\beta / dM_{\mathrm{UV}}$ shallower at $z < 11$, and we do observe that our UV-brightest sources are systematically redder as a function of cosmic time. This conclusion is also supported by observations of the UV-bright JADES sources discussed in \citet{saxena2024}. Without difficult-to-obtain observations of the thermal emission in these sources, or further detections of the 2175 \AA\ attenuation feature \citep[seen out to $z = 7-8$, see][]{witstok2023, markov2025}, it will be challenging to fully explore the role of dust obscuration in these sources, which may also be reddened by nebular continua. 

\subsection{Comparing Spectroscopic Redshifts to Photometric Redshifts}\label{sec:spec_z_vs_phot_z}

In Section \ref{sec:spectroscopic_redshifts} we discussed the subset of \Nzspec{} sources with spectroscopic redshifts from our assembled final sample across GOODS-S and GOODS-N. In this section, we explore the relationship between photometric and spectroscopic redshifts at $z > 8$, and discuss both the usage of alternate \texttt{EAZY} templates for fitting these sources as well as the usage of a prescription for CGM absorption from \citet{asada2025} within \texttt{EAZY}. 

\begin{figure*}[t!]
  \centering
  \includegraphics[width=0.95\linewidth]{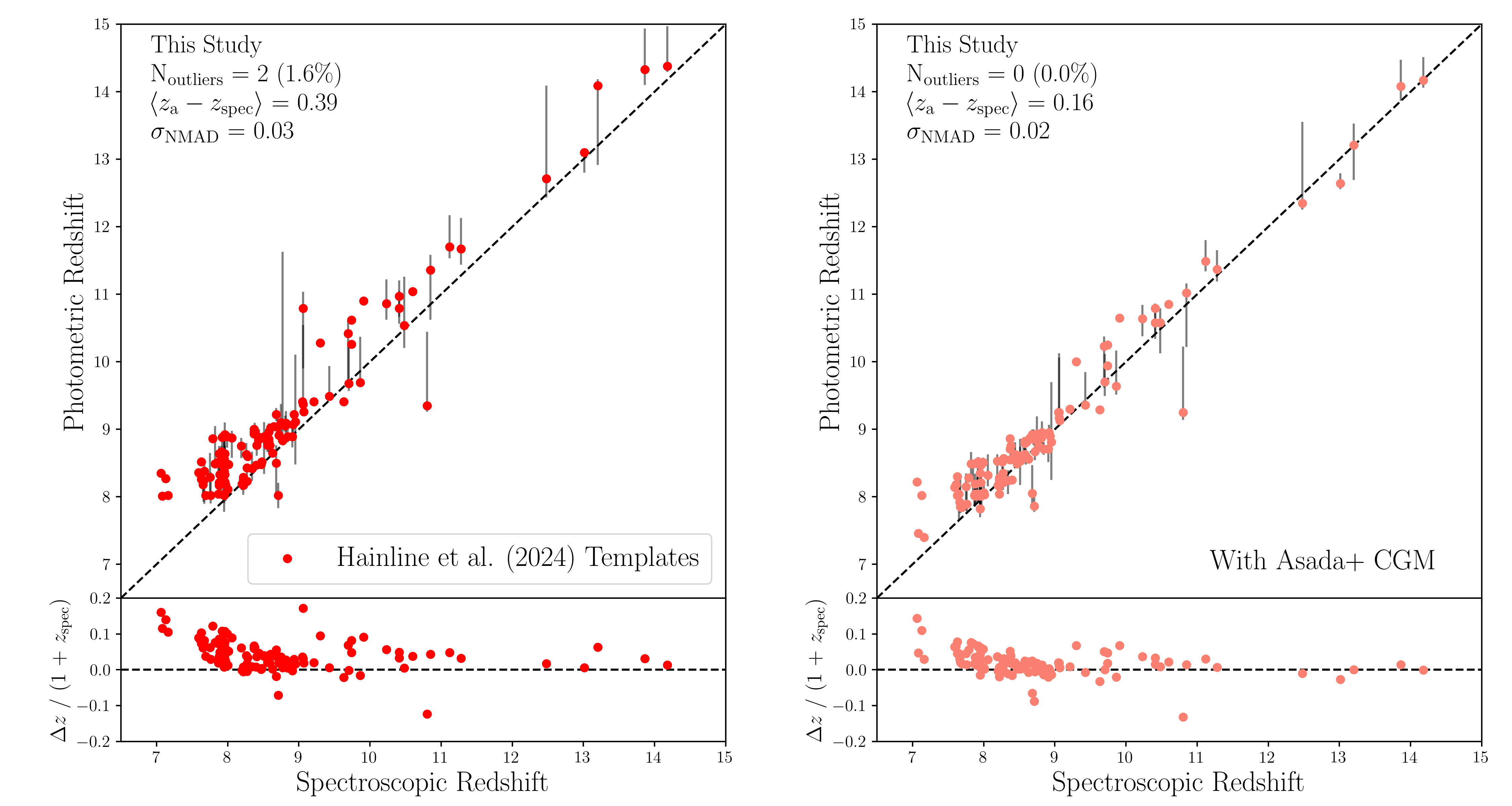}\
  \caption{\texttt{EAZY} photometric redshift plotted against spectroscopic redshift for \Nzspec{} sources in our sample. (Left) Raw photometric redshifts derived from our fits.  (Right) Photometric redshifts estimated using the \citet{asada2025} CGM evolution prescription. In both panels we provide statistics for the fits: N$_{\mathrm{outliers}}$, the number of catastrophic outliers with $|z_a - z_{\mathrm{spec}}|/(1+z_{\mathrm{spec}}) > 0.15$ (and the outlier fraction of the total number of sources), the average offset between spectroscopic and photometric redshift ($\langle \Delta z = z_a - z_{\mathrm{spec}} \rangle$), and $\sigma_{\mathrm{NMAD}}$, the normalized mean absolute deviation of the redshifts.}
  \label{fig:spec_z_vs_redshift}
\end{figure*}

In the left panel of Figure \ref{fig:spec_z_vs_redshift}, we compare our $z_a$ values to these spectroscopic redshifts as red circles. As can be seen from the residuals below the plot, the values largely agree, with only two catastrophic outliers, defined as any source with $|z_{\mathrm{spec}} - z_a|/(1+z_{\mathrm{spec}}) > 0.15$. These two sources, GOODS-S 73568 ($z_{\mathrm{spec}} = 9.06$, $z_a = 10.79$) GOODS-N 1085981 ($z_{\mathrm{spec}} = 7.06$, $z_a = 8.35$) have $\chi^2$ surfaces entirely consistent with the spectroscopic redshift. ID 73568 in GOODS-S is a F115W dropout, and the exact minimum $\chi^2$ redshift is dependent on the measured flux in the F115W band. ID 1085981 in GOODS-N also shows evidence of excess flux at 4$\mu$m, and could be a ``little red dot'' \citep{matthee2024, hainline2025a}, which may serve to push the photometric redshift higher to account for the flux with possible [\ion{O}{3}]$\lambda$5007 + H$\beta$ line emission. 

Notably, all of the photometric redshifts we measure are biased high compared to the spectroscopic redshifts ($\langle \Delta z = z_a - z_{\mathrm{spec}} \rangle = 0.39$), similar to the offsets seen in \KHZ{}, \citet{arrabalharo2023b}, \citet{fujimoto2023}, \citet{finkelstein2024}, \citet{willott2024}, and other high-redshift samples. This trend is primarily due to two effects. At $z < 9$, in Figure \ref{fig:spec_z_vs_redshift}, the bottom left panel shows that the residuals have a redshift trend. This effect arises owing to the existence of the [\ion{O}{3}]$\lambda\lambda 4959,5007$ emission line boosting the flux in the NIRCam F444W filter, which strongly influences the photometric redshift at the minimum \texttt{EAZY} $\chi^2$. For a $\chi^2$ minimization fitting approach, the model will be strongly influenced by bright, high SNR photometric fluxes, like F444W and added flux from [\ion{O}{3}]. As \texttt{EAZY} is performing template fitting, and not a full accounting of the underlying photoionization of the nebular emission in these galaxies, the potential [\ion{O}{3}] line strengths that can be modeled are limited to what can be achieved in the co-added templates. As such, if the [\ion{O}{3}] flux from a given source at $z \sim 8$ is significantly higher or lower than what can be achieved with the \texttt{EAZY} templates, the redshift will be biased to better account for the observed F444W photometric flux. For our high-redshift sample, there is uncertainty in the photometric redshifts at $z \sim 8$ as the Lyman-$\alpha$ break moves between the F090W and F115W bands, and the presence of [\ion{O}{3}] emission will result in redshifts that are biased high. For some sources in our sample, extra medium-band filter coverage from NIRCam F430M, F460M, and F480M help to better define the continuum, and the redshifts are far more accurate. At $z > 8.8$, [\ion{O}{3}] is redshifted out of F444W, which is where the redshift dependent positive residuals seen in Figure \ref{fig:spec_z_vs_redshift} are largely mitigated. 

The secondary origin of the photometric redshifts being biased high, seen most notably at $z > 9$, is the presence of both Lyman-damping wings in the IGM, and strong DLA absorption in the CGM surrounding these sources, causing the turnover in the Lyman-$\alpha$ break to smooth out in a way not accounted for by the templates and CGM prescription in \texttt{EAZY}. This effect is discussed in multiple studies, including \citet{chen2024b}, \citet{umeda2024}, \citet{heintz2024}, \citet{deugenio2024a}, \citet{heintz2025}, and \citet{hainline2024c}. At fixed hydrogen column density N$_{\mathrm{H}}$, this effect increases linearly with redshift, which leads to an increase with the photometric redshift offset at higher redshifts. In addition, this $\Delta z$ bias is larger as N$_{\mathrm{H}}$ increases. 

To further explore the strength of this effect, \citet{asada2025} derived an empirical prescription for the strength of the DLA as a function of redshift through fits to sources at $z > 6$ in the CANUCS survey. They derive a series of transmission curves, and demonstrate how DLA strength increases at higher redshift, and accounting for this effect results in a significant improvement in the photometric redshifts compared to their sample of spectroscopic redshifts. In this prescription, they assume that the column density of neutral hydrogen in the CGM ($N_{\mathrm{HI, CGM}}$) around galaxies increases significantly from $\log{(N_{\mathrm{HI, CGM}})} \sim 19.8$ cm$^{-2}$ at $z = 6$ to a plateau of $\log{(N_{\mathrm{HI, CGM}})} = 21.5$ cm$^{-2}$ at $z > 9$. 
As seen in Figure 2 in \citet{asada2025}, there is a large scatter on the actual measurements of $N_{\mathrm{HI, CGM}}$ for spectroscopically confirmed sources at $z > 6$, and we can use the spectroscopic sample we have assembled to explore the accuracy of this CGM correction. 

To that end, the \citet{asada2025} prescription was recently incorporated within \texttt{EAZY}, and we re-fit our sources assuming these updated transmission curves. We plot the updated photometric redshifts in the right panel of Figure \ref{fig:spec_z_vs_redshift}. Overall, the offset is improved, with $\langle \Delta z \rangle = 0.16$, largely driven by the sources at $z_{\mathrm{spec}} < 9$. A few sources have updated photometric redshifts that push them significantly lower than their spectroscopic redshifts when we include the \citet{asada2025} prescription. As is discussed in \citet{hainline2024c}, DLAs are not observed in all high-redshift galaxies, and the one-size-fits-all approach, which assumes a column density of hydrogen as a function of redshift, will sometimes overcorrect the photometric redshifts. 

The templates introduced in \KHZ{} and used in this study were designed to fit sources at $z = 0 - 22$ in the JADES survey. Multiple studies over the last several years have derived theoretical templates with the goal of replicating the properties of ultra-high-redshift galaxies, especially those with very blue UV slopes (as seen for sources in our analysis in Section \ref{sec:uv_slopes}) not accounted for in standard \texttt{EAZY} template sets. The first set of alternate templates we explored were presented in \citet{larson2023}, which were generated using the Binary Population and Spectral Synthesis \citep[BPASS, ][]{eldridge2017} models and the CLOUDY \citep{ferland2017} software package, and were designed to match the blue UV colors of galaxies at $z > 8$. 

We also compared to the synthetic templates presented in \citet{steinhardt2023} and derived by exploring a range of possible initial mass functions and fitting to galaxies at $z > 12$ in early JWST NIRCam deep field data. They find a prescription for the IMF that relies on an estimate of the temperature of the star-forming gas ($T_g$), and assume that galaxies at $z \sim 15$ should have high star formation rates to allow them to form rapidly, requiring higher values of $T_g$ between 45 -- 60 K. Their final templates are at the extremes of this range, and we fit our objects with both sets.  

We additionally explored the use of templates from \citet{luberto2025} which were created using the public galaxy formation code Accelerated Reionization Era Simulations \citep[\texttt{ARES}; ][]{mirocha2012, mirocha2014}, which starts from dark matter halo growth histories and uses a model for star formation to create realistic galaxy SEDs. The templates were derived by matching \texttt{ARES} models to the evolution of the UV luminosity density to $z \sim 10$ from HST, and are presented both with and without Lyman-$\alpha$ emission. We fit both sets of models to our observed sample.  

Finally, we also fit our objects with two template sets that are currently included within \texttt{EAZY}, the ``\texttt{tweak\_fsps}'' set, and the ``\texttt{agn\_blue\_sfhz}'' set, which are commonly used in the literature for fitting galaxies at a large range of redshifts. For each template set that we fit, we provide the final photometric redshifts and more statistics from the fits in our final output table. We fit all of our sources with the full suite of templates, both without and with the \citet{asada2025} CGM prescription.

\begin{figure*}[t!]
  \centering
  \includegraphics[width=0.48\linewidth]{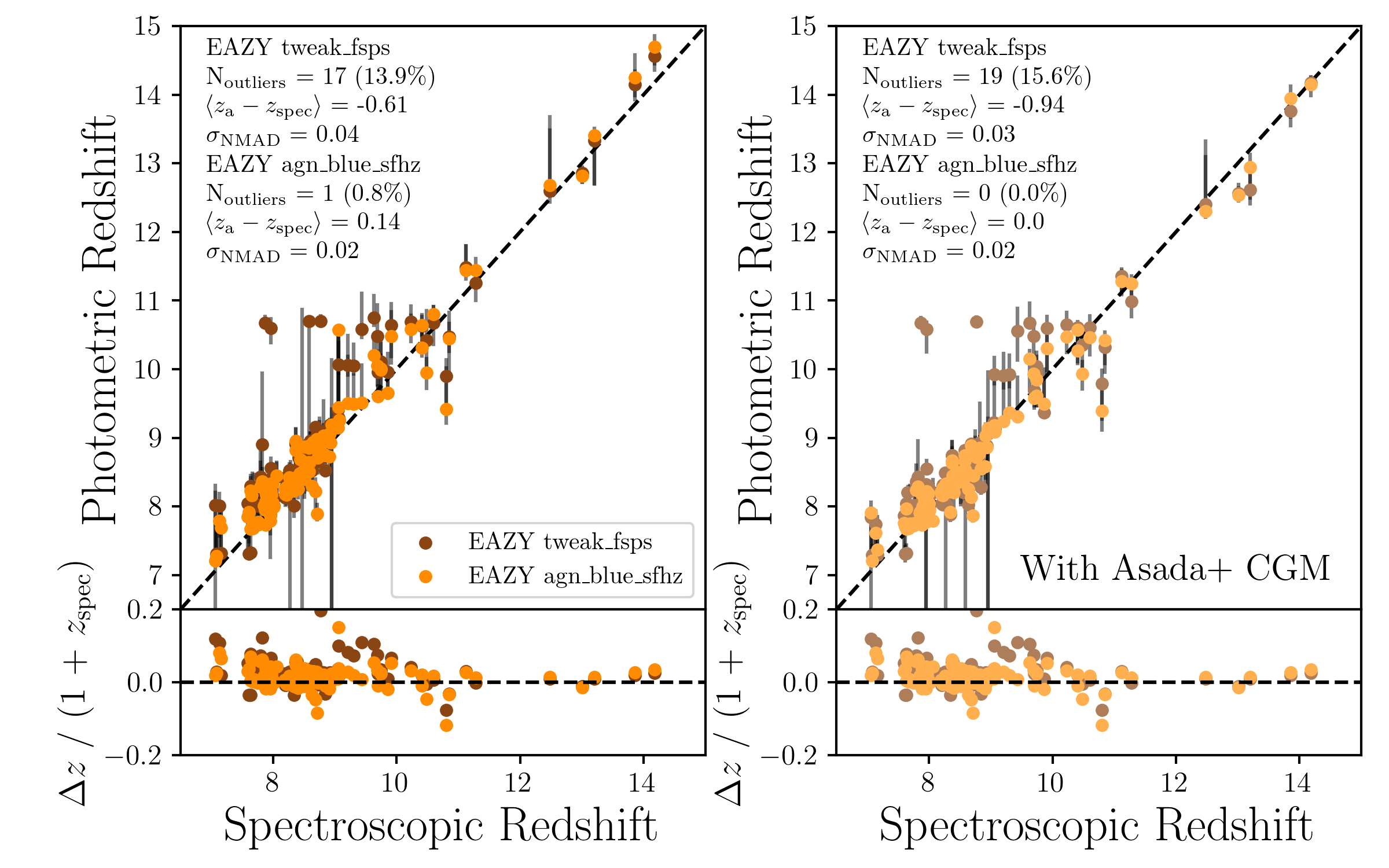}\
  \includegraphics[width=0.48\linewidth]{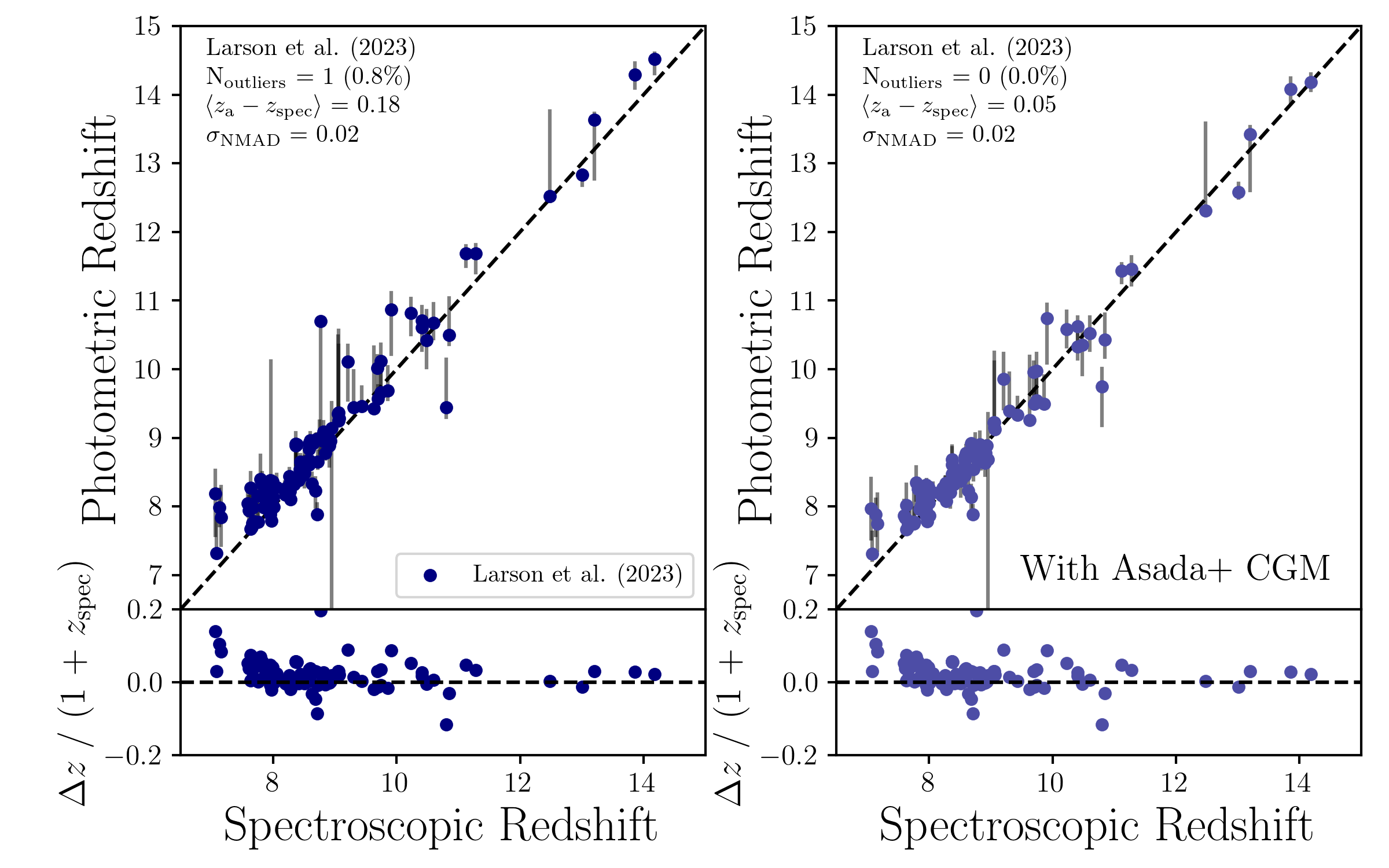}\
  \includegraphics[width=0.48\linewidth]{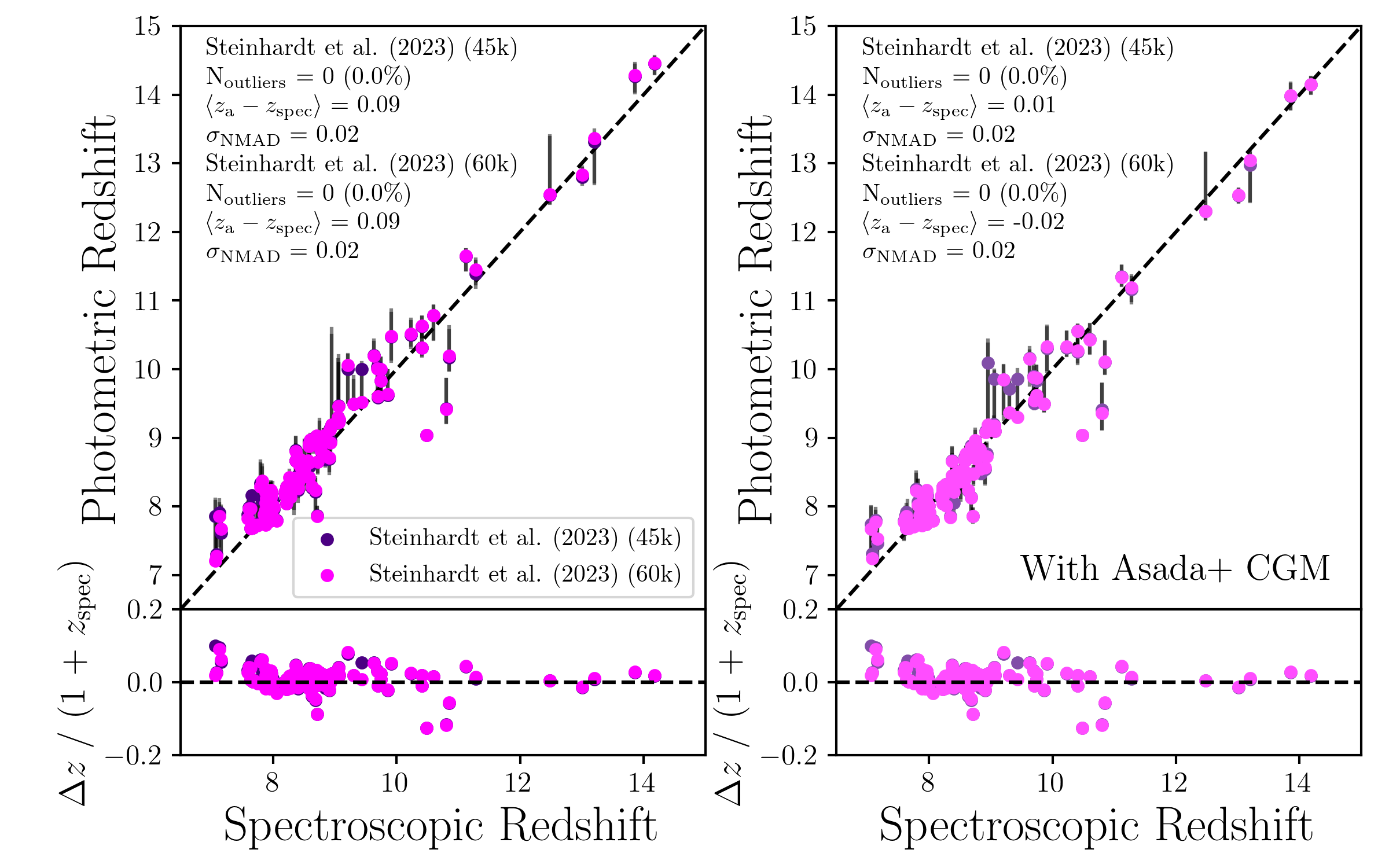}\
  \includegraphics[width=0.48\linewidth]{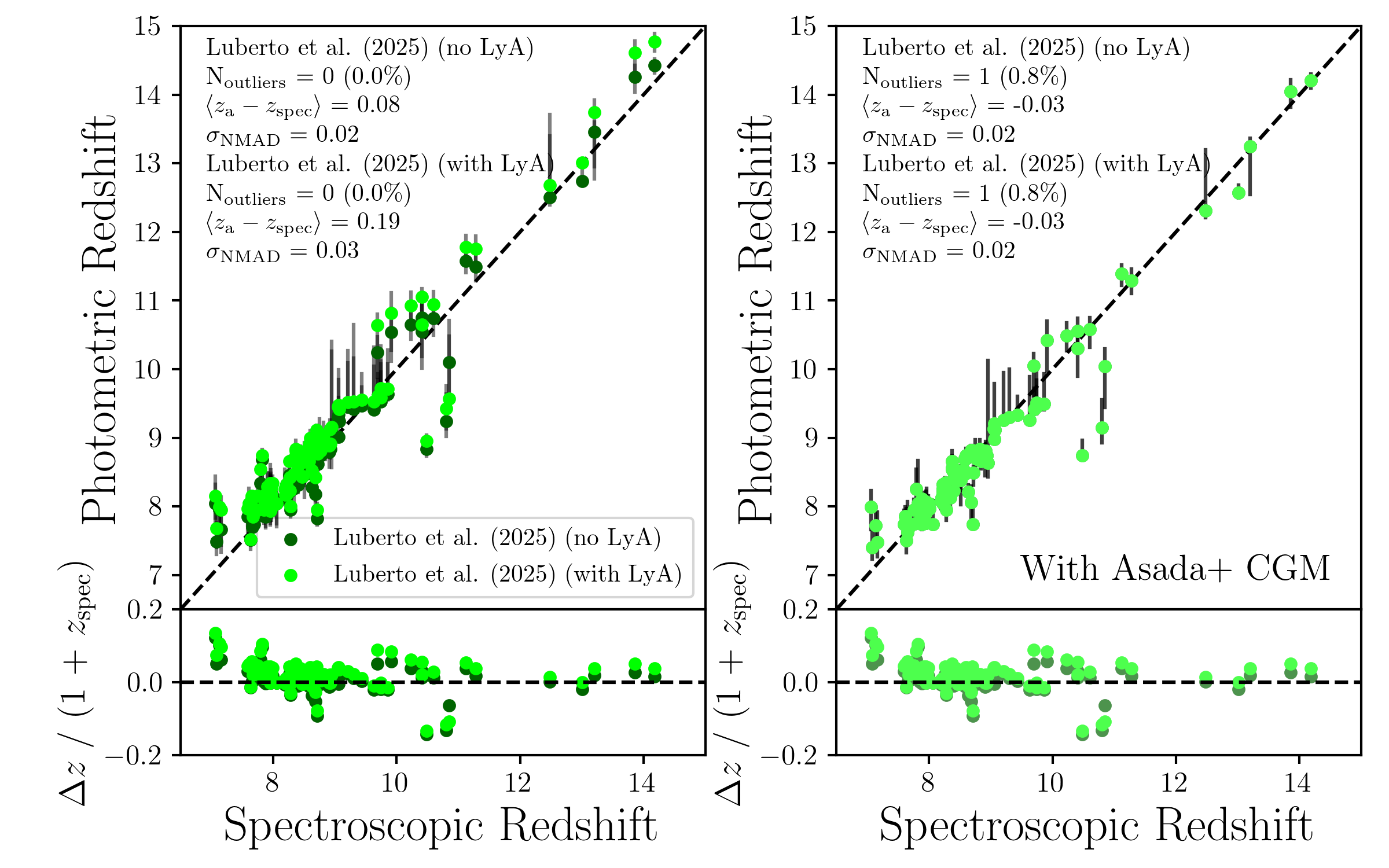}\
  \caption{A comparison between \texttt{EAZY} photometric redshifts and spectroscopic redshifts, with statistics as shown in Figure \ref{fig:spec_z_vs_redshift}, but for multiple template sets: (Top Left) \texttt{EAZY} ``\texttt{tweak\_fsps}'' and ``\texttt{agn\_blue\_sfhz}'' templates. (Top Right) \citet{larson2023b} templates. (Bottom Left) \citet{steinhardt2023} templates. (Bottom Right) \citet{luberto2025} templates.}
  \label{fig:spec_z_vs_redshift_alttemplates}
\end{figure*}

For all of the alternate template sets we explore, we fit the sources with the same photometry and uncertainties, as well as the same photometric offsets, and redshift grid, as we do with our primary template set. Similarly, for all templates, we adopted the minimum-$\chi^2$ redshifts as our final values. In Figure \ref{fig:spec_z_vs_redshift_alttemplates}, we plot all of the photometric redshifts for the full set of templates in individual panels. Like in Figure \ref{fig:spec_z_vs_redshift}, we show the redshifts directly from \texttt{EAZY} in the left panel in each quadrant with darker points, and redshifts from fits with the \citet{asada2025} CGM prescription in the right panel in each quadrant, depicted with lighter points. In each panel, we show the ``z160'' and ``z840'' redshift ranges derived from the P(z) (and estimated from the fit $\chi^2$) as vertical lines. We also provide statistics on the comparison between spectroscopic and photometric redshifts for each template set. 

The first thing to notice is that each of the template sets shows an offset where the photometric redshifts are systematically high as compared to the spectroscopic redshift. The exact magnitude of this offset is different between the template sets, but ranges between $\langle \Delta z \rangle = -0.61 - 0.18$. It is also apparent that the \texttt{EAZY} \texttt{tweak\_fsps} set results in a $\sim 14\%$ outlier fraction, and the largest offset, in large part due to the fact that this set does not contain templates with the blue UV slopes observed in high-redshift galaxies. Putting this template set aside, when comparing $\langle \Delta z \rangle$, all of the template sets perform better than the \KHZ{} templates we use in this study on average, although this varies for individual sources. When we look at the \citet{steinhardt2023} templates, there is very little difference between the 45K and 60K versions with respect to our spectroscopic sample, while the \citet{luberto2025} ARES templates without Lyman-$\alpha$ emission perform significantly better, as might be expected given the redshifts of these objects.

More importantly, the usage of the \citet{asada2025} CGM prescription has a significant impact on the recovered redshifts, and specifically $\langle \Delta z \rangle$, which is much closer to 0.0 for each template set (outside of \texttt{tweak\_fsps}, which worsens), demonstrating the impact that DLA absorption may be having on these high-redshift galaxies. We note that the results for the \citet{luberto2025} template set is identical with and without Lyman-$\alpha$ emission when using the \citet{asada2025} CGM model, as the Lyman-$\alpha$ emission is absorbed entirely at these redshifts, such that the templates are otherwise identical. The success of the \citet{asada2025} model suggests that high-redshift galaxies have significant neutral gas reservoirs, as discussed in \citet{heintz2025}. Notably, \citet{heintz2025b} estimate that one of the highest-redshift spectroscopically confirmed galaxies, JADES-GS-z14-0, has a gas mass of $\log{(M_{\mathrm{gas}}/M_{\odot})} = 9.5 \pm 0.3$ and a gas mass fraction of $0.7 - 0.9$. The effects of this neutral gas needs to be accounted for both when deriving UV luminosity functions from sources using photometric redshifts, and also for designing spectroscopic follow-up observational setups for these objects. 

We find that the usage of the alternate templates from \citet{larson2023b}, \citet{steinhardt2023}, and \citet{luberto2025} are supported for our sample of galaxies at $z > 8$, and we also find promising results for the usage of the stock \texttt{EAZY} ``\texttt{agn\_blue\_sfhz}'' templates, where $\langle \Delta z \rangle$ is consistent with zero with the \citet{asada2025} CGM prescription. In our final output table, we provide the redshifts estimated using each of these templates for the entire sample of $z > 8$ sources, with and without the \citet{asada2025} CGM model, along with some of the additional statistics and uncertainties. From these fits, we should note that a fraction of the sources that have $z_a > 8$ in fits using the \KHZ{} templates are fit with a minimum-$\chi^2$ redshift at $z_{\mathrm{phot}} < 8$ when using alternate templates. We find that 30\% of the sources in our current sample, have $z_a < 8$ when fit with the ``\texttt{tweak\_fsps}'' templates. The percentage is only 17\% when using the ``\texttt{agn\_blue\_sfhz}'' templates, 11\% when using the \citet{larson2023b} templates, 17\% when using the \citet{steinhardt2023} 45K and 60K templates, 12\% when using the \citet{luberto2025} ARES templates without Lyman-$\alpha$, and only 7\% when using the ARES templates with Lyman-$\alpha$. This final comparison is understandable given that the templates used for selecting these sources also have Lyman-$\alpha$ emission, as do the templates used to derive the JADES photometric redshifts from \KHZ{}.

As would be expected based on the results seen in Figure \ref{fig:spec_z_vs_redshift_alttemplates}, the bulk of these outlier sources are between $z_a = 7 - 8$. If we instead look which sources have $z_a < 7$ when fit with these alternate templates, we find that only only 4\% of the sources in our sample are fit at $z_a < 7$ when using the ``\texttt{agn\_blue\_sfhz}'' templates, 3\% when using the \citet{larson2023b} templates, 2\% when using the \citet{steinhardt2023} 45K, 1\% when using the \citet{steinhardt2023} 60K templates, 0.5\% when using the \citet{luberto2025} ARES templates without Lyman-$\alpha$, and only 0.4\% when using the ARES templates with Lyman-$\alpha$. These results demonstrate the validity of the full sample of $z > 8$ sources presented in this study, given the large uncertainty on estimating photometric redshifts for faint, high-redshift galaxies. We also note that a full 21\% of the sources, when fit with the ``\texttt{tweak\_fsps}'' templates, are found at $z_a < 7$, and we do not recommend the usage of these templates in fitting high-redshift galaxies.

Our primary conclusion from this analysis is in support of the suggestions presented in \KHZ{} and \citet{steinhardt2023}, among others, that while galaxies at $z > 8$ should be \textit{initially selected} using templates designed to fit galaxies at all redshifts (and therefore to reduce the number low-redshift interlopers being fit at higher redshifts), the most accurate photometric redshifts for galaxies at $z > 8$ are estimated using updated, bluer template sets which include the effects of CGM absorption. To that end, we provide the photometric redshifts (both with and without the \citet{asada2025} CGM prescription) when using these alternate templates, in our final output table, along with estimates of the photometric redshift uncertainties, in the \texttt{ALTERNATE\_TEMPLATES} hdu as described in Table \ref{tab:catalog_columns}.

\section{Conclusions}

In this paper, we present a sample of \Nsamp{} galaxies with photometric or spectroscopic redshifts at $z > 8$ from across the full JADES DR5 dataset in GOODS-S and GOODS-N. These objects were selected using \texttt{EAZY} fits to the JWST/NIRCam photometry, along with a minimum SNR cut and a requirement that the difference between the best-fit $\chi^2$ and the $\chi^2$ at $z < 7$ is greater than 4. We estimate the properties of these sources, and conclude:

\begin{enumerate}
    \item The galaxies in our sample span a $M_{\mathrm{UV}}$ range of $-16$ to $-21.5$, out to $z \sim 22$, although only one source is at $z_a > 16.5$. There is a distinct drop-off in the number density of sources in our final sample at $z > 14.5$, even though at the depth of the survey these sources would satisfy the SNR requirements. Our highest-redshift source in the sample, JADES-GS+52.96884-27.76382 (ID 385830), at $z \sim 22$, is a bright F200W dropout only detected in four NIRCam filters. Given the lack of filter coverage, there is a possibility of this source being a transient or an emission line galaxy at lower redshift.
    \item We find that the slope of the relationship between $M_{\mathrm{UV}}$ and $\beta$ becomes steeper at higher redshifts, while the normalization of the fit (at $M_{\mathrm{UV}} = -19.0$) is consistent across the redshift bins, in conflict with results from the literature, although our sample has significantly more faint ($M_{\mathrm{UV}} > -18.0$) sources at $z > 11$.
    \item When we explore the morphologies for our full sample, we find that $> 25$\% have evidence for being significantly extended ($\gtrsim 1.0$ kpc), some with multiple knots of star formation, a fraction that is similar at all redshifts within our sample. We observe a number of sources at $z > 13$ with two clumps, or with flattened morphologies, suggesting the presence of merging events.
    \item The $z \sim 15$--18 candidate JADES-GN +189.15982+62.28899 (ID 1019411) is a faint F200W dropout with $z_{\mathrm{phot}} = 16.5$, and we present a 9~ksec JADES DR4 NIRSpec prism spectrum for the source which is consistent with the photometry. We use this very low SNR spectrum to argue against low-redshift solutions where strong emission lines boost the observed photometry.
    \item Using the \Nsamp{} sources with spectroscopic-redshifts in our sample, we explored the accuracy of our photometric redshifts. The estimated photometric redshifts are biased high compared to the spectroscopic redshifts, but we find that \texttt{EAZY} templates designed to fit the blue UV slopes of high-redshift galaxies, and more importantly, accounting for the presence of DLAs in the sources, specifically with the prescription from \citet{asada2025}, provides more redshift accuracy. 
\end{enumerate}

The JWST NIRCam observations collected as part of the JADES DR5 dataset across GOODS-S and GOODS-N offer a uniquely comprehensive view of the distant Universe given the survey area, the number of observed filters and the depth of the coverage, and the imaging and spectroscopic complementarity. While deep spectroscopy is required to confirm the redshifts for our highest-redshift candidates, from their photometric redshifts we observe a Universe that is rapidly changing. The galaxies in our sample have a spread in UV slopes even within the first few hundred million years, indicating the possible presence of dust at early times. A significant fraction have clumpy, elongated morphologies demonstrating the potential role of mergers and the existence of large quantities of neutral gas in and around the sources that needs to be accounted for when estimating redshift. Future spectroscopic campaigns will hopefully help us understand the origin of the relative lower number density of sources at $z > 14.5$, or if this is merely a limitation of the (extremely deep!) JADES imaging. 

The first four years of JWST have offered a host of exciting new galaxies at Cosmic Dawn, and with them, mysteries about their origin and properties. JADES DR5 demonstrates the power of Guaranteed Time Observations to offer a possible deep, wide survey which will enjoy a long legacy. The sources in our $z > 8$ sample presented here represent the culmination of many years of GTO planning, and we are excited to see how they will be analyzed in the future by the astronomical community. 

\bigskip
The authors would like to thank Amber Qureshi for their assistance, and for the anonymous referee who provided comments that greatly improved the manuscript. This research was funded through the JWST/NIRCam contract to the University of Arizona (NAS5-02105). The data were obtained from the Mikulski Archive for Space Telescopes at the Space Telescope Science Institute, which is operated by the Association of Universities for Research in Astronomy, Inc., under NASA contract NAS5-03127 for JWST. These observations are associated with PIDs 1063, 1176, 1180, 1181, 1210, 1264, 1283, 1286, 1287, 1895, 1963, 2079, 2198, 2514, 2516, 2674, 3215, 3577, 3990, 4540, 4762, 5398, 5997, 6434, and 6511. The authors acknowledge the teams of programs 1895, 1963, 2079, 2514, 3215, 3577, 3990, 6434, and 6541 for developing their observing program with a zero-exclusive-access period. The JWST data presented in this manuscript were obtained from the Mikulski Archive for Space Telescopes (MAST) at the Space Telescope Science Institute. The specific observations analyzed can be accessed via doi:10.17909/8tdj-8n28. Additionally, this work made use of the lux supercomputer at UC Santa Cruz, which is funded by NSF MRI grant AST1828315, as well as the High Performance Computing (HPC) resources at the University of Arizona, which is funded by the Office of Research Discovery and Innovation (ORDI), Chief Information Officer (CIO), and University Information Technology Services (UITS). 

DJE is supported as a Simons Investigator and by JWST/NIRCam contract to the University of Arizona. Support for program 3215 was provided by NASA through a grant from the Space Telescope Science Institute, which is operated by the Association of Universities for Research in Astronomy, Inc., under NASA contract NAS 5-03127. LW acknowledges support from the Gavin Boyle Fellowship at the Kavli Institute for Cosmology, Cambridge and from the Kavli Foundation. The Cosmic Dawn Center (DAWN) is funded by the Danish National Research Foundation under grant DNRF140. ST acknowledges support by the Royal Society Research Grant G125142. SA acknowledges grant PID2021-127718NB-I00 funded by the Spanish Ministry of Science and Innovation/State Agency of Research (MICIN/AEI/ 10.13039/501100011033). WMB gratefully acknowledges support from DARK via the DARK fellowship. This work was supported by a research grant (VIL54489) from VILLUM FONDEN. AJB, AJC, and JC acknowledge funding from the ``FirstGalaxies'' Advanced Grant from the European Research Council (ERC) under the European Union’s Horizon 2020 research and innovation programme (Grant agreement No. 789056). SC acknowledges support by European Union’s HE ERC Starting Grant No. 101040227 - WINGS. ECL acknowledges support of an STFC Webb Fellowship (ST/W001438/1). Funding for this research was provided by the Johns Hopkins University, Institute for Data Intensive Engineering and Science (IDIES). TJL gratefully acknowledges support from the Swiss National Science Foundation through a SNSF Mobility Fellowship and from the NASA/JWST Program OASIS (PID 5997). RM and FDE acknowledge support by the Science and Technology Facilities Council (STFC), by the ERC through Advanced Grant 695671 “QUENCH”, and by the UKRI Frontier Research grant RISEandFALL. RM also acknowledges funding from a research professorship from the Royal Society. PGP-G acknowledges support from grant PID2022-139567NB-I00 funded by Spanish Ministerio de Ciencia e Innovaci\'on MCIN/AEI/10.13039/501100011033, FEDER, UE. JAAT acknowledges support from the Simons Foundation and JWST program 3215. Support for program 3215 was provided by NASA through a grant from the Space Telescope Science Institute, which is operated by the Association of Universities for Research in Astronomy, Inc., under NASA contract NAS 5-03127. H\"U acknowledges funding by the European Union (ERC APEX, 101164796). Views and opinions expressed are however those of the authors only and do not necessarily reflect those of the European Union or the European Research Council Executive Agency. Neither the European Union nor the granting authority can be held responsible for them. The research of CCW is supported by NOIRLab, which is managed by the Association of Universities for Research in Astronomy (AURA) under a cooperative agreement with the National Science Foundation. AJC and JW gratefully acknowledge support from the Cosmic Dawn Center through the DAWN Fellowship. 

\vspace{5mm}
\facilities{JWST(NIRCam and NIRSpec), HST(ACS)}

\software{\texttt{astropy} \citep{2013A&A...558A..33A,2018AJ....156..123A}, \texttt{scipy} \citep{2020SciPy-NMeth}, \texttt{EAZY} \citep{brammer2008}.
          }

\bibliography{ms}{}
\bibliographystyle{aasjournal}

\end{document}